\documentclass[preprints,article,accept,moreauthors]{Definitions/mdpi} 

\firstpage{1} 
\pubvolume{1}
\issuenum{1}
\articlenumber{0}
\doinum{}
\pubyear{2023}
\copyrightyear{2023}
\datereceived{} 
\dateaccepted{} 
\datepublished{} 
\hreflink{https://doi.org/} % If needed use \linebreak

\usepackage{color}
       
\Title{Observational Detection of Higher Order Secular Perturbations in Tight Hierarchical Triple Stars}

\TitleCitation{Higher Order Secular Perturbations in Tight Hierarchical Triple Stars}

\Author{Tam\'as Borkovits $^{1,2,3,4,5}$\orcidA{}, Tibor Mitnyan $^2$\orcidB{}}

\AuthorNames{Tam\'as Borkovits, Tibor Mitnyan}

\AuthorCitation{Borkovits, T. \& Mitnyan, T.}
\address{%
$^{1}$ \quad Baja Astronomical Observatory of University of Szeged, H-6500 Baja, Szegedi \'ut, Kt. 766, Hungary\\
$^{2}$ \quad HUN-REN -- SZTE Stellar Astrophysics Research Group, H-6500 Baja, Szegedi \'ut, Kt. 766, Hungary\\
$^{3}$ \quad Konkoly Observatory, HUN-REN Research Centre for Astronomy and Earth Sciences,  H-1121 Budapest, Konkoly Thege Mikl\'os \'ut 15-17, Hungary \\
$^{4}$ \quad ELTE Gothard Astrophysical Observatory, H-9700 Szombathely, Szent Imre h. u. 112, Hungary\\
$^{5}$ \quad HUN-REN -- ELTE Exoplanet Research Group, H-9700 Szombathely, Szent Imre h. u. 112, Hungary}

\corres{Correspondence: borko@electra.bajaobs.hu}

\abstract{
In this work, we search for observational evidence of higher-order secular perturbations in three eclipsing binaries.  These are slightly eccentric binaries, and they form the inner pairs of tight, compact, hierarchical triple star systems. We analyze simultaneously the high precision satellite (\textit{Kepler} and \textit{TESS}) light curves, eclipse timing variations, combined spectral energy distributions (through catalog passband magnitudes) and, where available, radial velocities of KICs~9714358, 5771589 and TIC~219885468. Besides the determination of robust astrophysical and dynamical properties of the three systems, we find evidence that the observed unusual eclipse timing variations of KIC~9714358 are a direct consequence of the octupole-order secular eccentricity perturbations forced by an unusual, resonant behaviour between the lines of the apsides of the inner and outer orbital ellipses. We also show that, despite its evident cyclic eclipse depth variations, KIC~5771589 is an almost perfectly coplanar system (to within $0.3^\circ$), and we explain the rapid eclipse depth variations with the grazing nature of the eclipses. Finally, we find that the inner pair of TIC~219885468 consists of two twin stars and, hence, in this triple there are no octupole order three-body perturbations. Moreover, we show that this triple is also coplanar on the same level as the former one, but due to its deep eclipses, it does not exhibit eclipse depth variations. We intend to follow this work up with further analyses and a quantitative comparison of the theoretical and the observed perturbations.}

\begin{document}
%%%%%%%%%%%%%%%%%%%%%%%%%%%%%%%%%%%%%%%%%%

\section{Introduction}

The new window which was opened up by \textit{Kepler} spacecraft \citep{boruckietal10} to the exotic world of the tight, compact hierarchical triple stellar systems (TCHTs) has now become more widely opened and transparent with the ongoing operation of the \textit{TESS} space telescope \citep{rickeretal15} mission. While \textit{Kepler} had an inevitable primacy in the discovery and principal characterizations of this rare new class of stellar systems, the multiply repeated revisitations of \textit{TESS}, over five years, have allowed us to extend our knowledge about the properties of such systems.  In a wider perspective, our studies of the stellar three-body problem have enhanced our understanding not only quantitatively, but even qualitatively, for the reasons which we discuss briefly in the forthcoming text.

Similar to our previous works \citep[e.~g.][]{borkovitsetal22b} we consider hierarchical triple or multiple (stellar) systems to be `tight', if the orbital period ratio of the outer orbit to the inner orbit remains below $\sim100$ (i.~e., $P_\mathrm{out}/P_\mathrm{in}\lesssim100$), while we call a triple or multiple star system `compact', when the outer period does not exceed, let's say, $1000$\,days.\footnote{In what follows, for simplicity, we will talk about only hierarchical triple star systems, but our discussion can be extended very simply to hierarchical multiple systems as well.} As is widespreadly accepted \citep[see, e.~g.][]{harrington972,zare977,mardlingaarseth01}, the only possible long-term stable configuration amongst three masses, being of the same order of mass, is the hierarchical configuration, i.e., when one of the three bodies is continuously located much farther from the other two than these latter two from each other. In such a system the motion, in general, can be well approximated with two binaries, i.e., two Keplerian two-body motions. The inner binary is formed by the two closer objects, while the outer binary consists of the third, most distant component, and the center of mass of the inner binary. Hence, tightness chiefly characterizes the strength of the third-body perturbation of such systems.  This is so since the magnitude of the gravitational perturbations relative to the pure Keplerian two-body motions primarily depend on the ratio of the semi-major axes or, more strictly, of the instantaneous separations of the outer to the inner binaries.  The former can readily be converted to such a direct observable as the period ratio. 

On the other hand, compactness is directly connected to the physical dimensions of the system as a whole. This characteristic size of the given triple star, naturally, has very crucial astrophysical implications, not only in regard to the early and the late evolutionary phases of such systems \citep[see, e.g.][and references therein]{tokovinin21,offneretal23,toonenetal20}, but how tidal and other non third-body perturbations relate to the gravitational perturbations of the third mass. From our point of view, compactness primarily has some practical significance. This significance has two, somewhat counteracting aspects. First, the largest amplitude, and most interesting, orbital perturbations in tight triple star systems are the so-called long-period or apse-node perturbations -- see below -- which are of primary interest in this paper. The characteristic periods of these perturbations scale with the ratio of $P_\mathrm{out}^2/P_\mathrm{in}$, i.e., these are essentially driven by the product of the measures of the tightness and the compactness.  Hence, the more compact a tight triple is, the shorter the observational window or the length of the necessary dataset for the direct detection and investigation of such effects. On the other hand, however, such triples can be serendipitously discovered and characterized most easily through the medium-period third-body perturbations (see again, below) that drive eclipse timing variations (ETVs) of such eclipsing binaries (EBs) which form the inner binary of a tight hierarchical triple system. The amplitude of such medium-timescale dynamically driven ETVs, however, scales as $P_\mathrm{in}^2/P_\mathrm{out}$ and, in such a way, for the most frequently occurring and shortest period EBs (some hours to few days), the dynamically driven ETVs may remain under the detection limit even in the tightest triples.

This latter fact explains why TCHTs were almost completely unknown before the four-year-long observations of the prime \textit{Kepler} mission. The vast majority of the EBs known before \textit{Kepler} had periods less than 3-4 days, in which case there was no chance of detecting the gravitational third-body perturbations via eclipse timing variations with the accuracies of the highly inhomogeneous ground-based observations.\footnote{Strictly speaking, Earth-based timing observations are less effective in the detection of any short outer period (i.~e., compact) triple star systems irrespective on their tightness. This question was discussed in details in the recent review of \citet{borkovits22}.} The quasi-continuous, almost four-year-long observations of the \textit{Kepler} space telescope has led to the discovery of more than a thousand new EBs with eclipsing periods longer than 5 days. Amongst them, \citet{borkovitsetal15,borkovitsetal16} have identified 46 tight hierarchical triple star candidates through the manifestations of the medium-term ($P_\mathrm{out}$-period) third-body perturbations in their ETVs. (Note, they found 16 further tight triple star candidates amongst the almost 2000 additional, shorter period \textit{Kepler} EBs, but this was mainly a consequence of the unprecedentedly precise observations of the space telescope and/or other effects, such as, e.g., third-body eclipses.) From this sample of the newly discovered 62 tight triple systems, 39 can be ranked as TCHT, i.e., a triple star system which is not only tight, but compact, as well. While these former studies depend primarily on the $P_\mathrm{out}$-timescale, the so-called medium period perturbations, the four-year-long ETVs of most of the eccentric TCHTs also exhibited signals of rapid, dynamically driven apsidal motion. Hence, a simplified, quadrupole approximation of the dynamically forced apsidal motion (and also of the nodal regression) was included into the analytic ETV model. 

The \textit{TESS} spacecraft, \textit{Kepler}'s successor, first reobserved the prime \textit{Kepler}-field almost exactly a decade after \textit{Kepler} began its survey. Since then, the vast majority of the original \textit{Kepler} targets and, hence, the \textit{Kepler} discovered TCHTs have been revisited typically 4--7 times, each for a four-week-long observing session. Thus, currently the lengths of the available datasets are nearly one and a half decades. Naturally, 14\,years is tiny in the context of most interesting astronomical timescales, but in the case of TCHTs, this interval is long enough for the robust detection and quantitative analysis of the `long-period' (or, secular) third-body perturbations. In this paper we have selected a few such TCHTs from the original \textit{Kepler} sample, which are exceptional even amongst the TCHTs. In particular, their special properties make possible the clear detection and investigation of even the higher order secular perturbations with the exclusive use of the high-precision satellite data. In what follows, first we give a brief review in Sect.~\ref{Sect:dynamics} of the analytic theory of the doubly averaged or, secular stellar three-body problem, focusing on the observable quadrupole and octupole perturbation terms in the low mutual inclination domain. Then, in Sect.~\ref{sect:selection} we introduce the three TCHTs that were selected for the current analysis. The details of the observational materials, the data preparation, and the complex photodynamical analysis are described in Sects.~\ref{sect:dataprep} and \ref{sec:dyn_mod}, while the results are discussed in Sect.~\ref{sec:discussion}.

%%%%%%%%%%%%%%%%%%%%%%%%%%%%%%%%%%%%%%%%%%

\section{Dynamics of TCHTs}
\label{Sect:dynamics}

The hierarchical three-body problem is sometimes called the `stellar three-body problem', however, the fields where the hierarchical treatment can be applied are much wider than the domain of triple and multiple stars. For example, the hierarchical three-body treatment can be used for investigations of the perturbations of the Earth-Moon system due to the Sun \citep{brown936a,brown936b,brown936c}, Jupiter's effects on the motion of the main belt asteroids and/or, comets \citep{vonzeipel910,kozai962}, and the perturbations on a binary comprised of the Earth and an artificial satellite due to the Moon \citep{lidov962}.

According to our knowledge, \citet{slavenas927a,slavenas927b} was the first one who investigated the hierarchical three-body problem in the context of a stellar triple ($\lambda$ Tauri, the very first and, for a long-time, the only TCHT). Due to the fact, however, that there were almost no known tight, or even compact, triple stellar systems, the hierarchical or, stellar three-body problem, apart from a very few exceptions \citep[see, e.~g.][]{harrington968,harrington969,soderhjelm975,soderhjelm982,soderhjelm984,mazehshaham979}, has remained beyond of the scope of the scientific community. The situation did change completely, and quickly, when the Kozai mechanism (nowadays called the `Von Zeipel-Lidov-Kozai' -- ZLK -- mechanism) or cycles were essentially `rediscovered' in regard to the formation and evolution of stellar triples during the late 1990's. At that time the first observational evidence was found in triple stellar and planetary systems for the effectiveness of secular third-body perturbations \citep[see, e.~g.][]{beustetal997,innanenetal997,mikkolatanikawa998} and, practically, at the same time, in an epochal paper \citet{kiselevaetal998} proposed a combined mechanism of third-body perturbations with tidal interactions to explain the formation of the close binary systems. In our view, this last work transformed the ZLK mechanism into its current, widespread and acknowledged position.

The mechanism, which is currently known as ZLK theory, is practically nothing more than the secular, doubly averaged theory of the hierarchical or, stellar three-body problem.  This theory describes the long-term dynamical evolution of such a three body system, where the evolution is driven, in its pure form, only by the mutual gravitational interactions of the three bodies. 

If we slightly modify the original classification of \citet{brown936b}, in a hierarchical triple system the perturbations are effective over three different timescales, and they can be classified according to these characteristic timescales as follows:

\begin{itemize}
\item{{\it short-period perturbations,} for which the typical period is in the order of $P_\mathrm{in}$, and the amplitude is related to $(P_\mathrm{in}/P_\mathrm{out})^2$,}
\item{{\it medium-period perturbations,} with a characteristic period of $P_\mathrm{out}$, and amplitude of $P_\mathrm{in}/P_\mathrm{out}$ and,}
\item{{\it long-period perturbations}\footnote{The latter two classes of medium-period and long-period perturbations in Brown's original terminology were named `long-period' and `apse-node type perturbations', respectively, but this would result in some confusion with the terminology of the classical planetary perturbation theories, which can be avoided with the use of this modified terminology.} having period about $P_\mathrm{out}^2/P_\mathrm{in}$, and an amplitude that may reach unity.}
\end{itemize}

It was \citet{harrington968,harrington969} who, for the first time, described the equations of the stellar three-body problem in Hamiltonian formalism for any arbitrary values of outer eccentricities and mutual inclinations and, gave third-order (sometimes called: `octupole') solutions for the long-period or, apse-node time-scale variations of the orbital elements, with the application of the double-averaging method of \citet{vonzeipel916a,vonzeipel916b,vonzeipel917a,vonzeipel917b}. 

While the current discoveries of TCHTs (including the currently analyzed systems, as well) are mainly based on the medium-period perturbations of the ETVs of their inner EBs, the scope of this paper is connected to the long-period perturbations, and thus, in what follows, we concentrate only on the latter. The original ZLK theorem is restricted to the lowest order perturbative terms of the doubly averaged Hamiltonian (which is quadratic in the small parameter $\mathbold{\alpha=a_\mathrm{in}/a_\mathrm{out}}$, hence, it is called the `quadrupole approximation'). At that level, the problem has only one degree of freedom, and thus, it is integrable, and the solution can be given with the use of elliptic functions of Weierstrass $\mathcal{P}$. Note, although the original analytic solution of \citet{kozai962} is strictly valid only if the outer orbit coincides with the invariable plane of the triple (i.e., all the angular momentum of the system is stored in the outer orbit), as was shown, e.g. by \citet{harrington968} and \citet{soderhjelm982}, the same solution may remain valid, with small modifications, even in the case where the orbital angular momentum of the inner orbit is non-negligible, but small enough. On the other hand, however, as was shown by \citet{fordetal00} via numerical integrations and, later analytically by \citet{naozetal13} (amongst others), in the case of an eccentric outer orbit, the octupole order terms may substantially alter the long-term behaviour of a hierarchical triple system.

As was excellently reviewed by \citet{naoz16}, the vast majority of the studies of the doubly averaged stellar three-body problem (being either analytical, numerical or, both) are directed toward the investigation of the very long-timescale evolution of such systems.  Specifically their interests are mainly focused on configurations where large amplitude eccentricity and (mutual) inclination cycles (including prograde-retrograde flip-flops) may occur. Much less effort has been directed toward the short-term (human timescale), directly observable effects of the stellar three-body problem. In this latter context, besides the above mentioned few exceptions, from the point of view of the present work, the paper of \citet{krymolowskimazeh999} is extremely relevant. These authors compared the outcomes of the quadrupole and octupole order perturbation theories of the doubly averaged problem to each other (and also to the results of direct numerical integrations) for different mutual inclinations (from the coplanar case up to $i_\mathrm{mut}=60^\circ$). They found that the octupole perturbations may be very significant even in the coplanar (or, nearly coplanar) case. This is so because, in an exactly coplanar configuration, the long-period quadrupole perturbations, for example, in the inner eccentricity completely disappear. Despite this fact, numerical integrations show that there exist long-period perturbations in the inner eccentricity in the coplanar configurations, and this effect is chiefly determined by the octupole order perturbations.

As those TCHTs that we investigate in this paper are actually nearly coplanar systems, in what follows, first we discuss the octupole effects quantitatively. Following the formulae of \citet{harrington968}, but using partly different notations, the doubly averaged Hamiltonian up to the octupole order takes the form:

\begin{eqnarray}
\mathcal{H}(\Delta\Omega^\mathrm{dyn}=\pi)&=&\beta_2\left[\left(2+3e_\mathrm{in}^2\right)(3I^2-1)+15e_\mathrm{in}^2\left(1-I^2\right)\cos2\omega_\mathrm{in}^\mathrm{dyn}\right] \nonumber\\
&&+\beta_3e_\mathrm{in}e_\mathrm{out}\left[\frac{1}{2}\left(1+\frac{3}{4}e_\mathrm{in}^2\right)\left(1+11I-5I^2-15I^3\right)\cos(\omega_\mathrm{in}^\mathrm{dyn}-\omega_\mathrm{out}^\mathrm{dyn})\right. \nonumber \\
&&+\frac{1}{2}\left(1+\frac{3}{4}e_\mathrm{in}^2\right)\left(1-11I-5I^2+15I^3\right)\cos(\omega_\mathrm{in}^\mathrm{dyn}+\omega_\mathrm{out}^\mathrm{dyn}) \nonumber \\
&&-\frac{35}{8}e_\mathrm{in}^2\left(1+I-I^2-I^3\right)\cos(3\omega_\mathrm{in}^\mathrm{dyn}-\omega_\mathrm{out}^\mathrm{dyn}) \nonumber \\
&&\left.-\frac{35}{8}e_\mathrm{in}^2\left(1-I-I^2+I^3\right)\cos(3\omega_\mathrm{in}^\mathrm{dyn}+\omega_\mathrm{out}^\mathrm{dyn})\right],
\label{Eq:secHamilton}
\end{eqnarray}
where $e_\mathrm{in,out}$ and $\omega_\mathrm{in,out}^\mathrm{dyn}$ denote the eccentricities and dynamical arguments of periastron of the inner and outer orbits, respectively. These latter quantities (sometimes denoted as the Delaunay variables $g_\mathrm{in,out}$) -- should not be confused with the observable arguments of periastrons (the latter of which will hereafter be denoted as $\omega_\mathrm{in,out}$) -- give the angular distances of the pericenter points of the orbits from the ascending nodes of the intersections of the orbital planes. Furthermore, $I=\cos i_\mathrm{mut}$ stands for the cosine of the mutual (or, relative) inclination of the two orbital planes. Finally,
\begin{eqnarray}
\beta_2&=&\frac{G^2}{16}\frac{m_\mathrm{A}^7}{m_\mathrm{AB}^3}\frac{m_\mathrm{Aa}^7}{(m_\mathrm{Aa}m_\mathrm{Ab})^3}\frac{L_\mathrm{in}^4}{L_\mathrm{out}^3G_\mathrm{out}^3} \nonumber \\
&=&\frac{2\pi}{30}A_\mathrm{G}L_\mathrm{in}, \\
\beta_3&=&-\frac{15}{4}\frac{G^2}{16}\frac{m_\mathrm{A}^9}{m_\mathrm{AB}^4}\frac{m_\mathrm{B}^9(m_\mathrm{Aa}-m_\mathrm{Ab})}{(m_\mathrm{Aa}m_\mathrm{Ab})^5}\frac{L_\mathrm{in}^6}{L_\mathrm{out}^3G_\mathrm{out}^5} \nonumber \\
&=&-4\pi A_\mathrm{G}^\mathrm{oct}L_\mathrm{in},
\end{eqnarray}
where we have introduced the parameters $A_\mathrm{G}$ and $A_\mathrm{G}^\mathrm{oct}$ which are more closely related to observable quantities, and can be expressed as
\begin{eqnarray}
A_\mathrm{G}&=&\frac{15}{8}\frac{q_\mathrm{out}}{1+q_\mathrm{out}}\frac{P_\mathrm{in}}{P_\mathrm{out}^2}\left(1-e_\mathrm{out}^2\right)^{-3/2}, \\
A_\mathrm{G}^\mathrm{oct}&=&\frac{1-q_\mathrm{in}}{1-q_\mathrm{out}}\left(\frac{1}{1+q_\mathrm{out}}\right)^{1/3}\left(\frac{P_\mathrm{in}}{P_\mathrm{out}}\right)^{2/3}\frac{A_\mathrm{G}}{1-e_\mathrm{out}^2}.
\end{eqnarray}
Additional, newly introduced quantities in the equations above are the gravitational constant $G$, the individual masses of the inner, close binary stars, as $m_{Aa}$, $m_{Ab}$, the total mass of the inner pair, $m_\mathrm{A}=m_\mathrm{Aa}+m_\mathrm{Ab}$, the individual mass of the third component, $m_\mathrm{B}$, and the total mass of the triple system $m_\mathrm{AB}=m_\mathrm{A}+m_\mathrm{B}$. Moreover, the mass ratios are denoted by $q_\mathrm{in,out}$ for the inner and outer binaries, respectively. 

In regard to the amplitude-like quantity of $A_\mathrm{G}^\mathrm{oct}$, as one can see, it disappears for $q_\mathrm{in}=1$, i~e., when the inner pair consists of two equal mass stars.\footnote{Note, that, in general, for equal mass inner stars all the higher, even order perturbative terms disappear, as well.} It is common also to introduce the quantity of
\begin{equation}
\epsilon=e_\mathrm{out}\frac{A_\mathrm{G}^\mathrm{oct}}{A_\mathrm{G}}=\frac{1-q_\mathrm{in}}{1-q_\mathrm{out}}\left(\frac{1}{1+q_\mathrm{out}}\right)^{1/3}\left(\frac{P_\mathrm{in}}{P_\mathrm{out}}\right)^{2/3}\frac{e_\mathrm{out}}{1-e_\mathrm{out}^2},
\label{Eq:epsilon_def}
\end{equation}
which gives the strength of the octupole order terms relative to the quadrupole ones.

Finally, note, that we use the usual Delaunay action variables as
\begin{eqnarray}
L_\mathrm{in}&=&\frac{m_\mathrm{Aa}m_\mathrm{Ab}}{m_\mathrm{A}}\sqrt{Gm_\mathrm{A}a_\mathrm{in}}, \\
L_\mathrm{out}&=&\frac{m_\mathrm{A}m_\mathrm{B}}{m_\mathrm{AB}}\sqrt{Gm_\mathrm{AB}a_\mathrm{out}}, \\
G_\mathrm{in,out}&=&L_\mathrm{in,out}\sqrt{1-e_\mathrm{in,out}^2}, \\
H_\mathrm{in,out}&=&G_\mathrm{in,out}\cos i_\mathrm{in,out}^\mathrm{dyn},
\end{eqnarray}
while the conjugate angular variables are the mean anomalies ($l_\mathrm{in,out}$), dynamical arguments of pericenters ($\omega_\mathrm{in}^\mathrm{dyn}$ -- or, traditionally, $g_\mathrm{in,out}$), and dynamical longitudes of the nodes ($\Omega_\mathrm{in,out}^\mathrm{dyn}$ -- or, $h_\mathrm{in,out}$) of the inner and outer orbits, respectively. One should also keep in mind that, as was noted by \citet{naozetal13}, although Eq.~(\ref{Eq:secHamilton}) formally does not contain the nodes ($\Omega_\mathrm{in,out}^\mathrm{dyn}$), which would suggest the constancy of the conjugate variables $H_\mathrm{in,out}$, this arises only from an incorrect application of the elimination of the nodes. For a correct treatment, one should take it into account that, in the original Hamiltonian, both the inner and outer dynamical nodes are present, but only through the sines and cosines of their differences, which is a constant $\Delta\Omega_\mathrm{dyn}=\Omega_\mathrm{in}^\mathrm{dyn}-\Omega_\mathrm{in}^\mathrm{out}=\pi$ and, therefore, $\sin\Delta\Omega^\mathrm{dyn}=0$, and $\cos\Delta\Omega^\mathrm{dyn}=1$. If one substitutes these values into the original Hamiltonian, it leads to the usual form of Eq.~(\ref{Eq:secHamilton}). Hence, strictly speaking, from the simplified (Eq.~\ref{Eq:secHamilton}) form of the Hamiltonian, the constancy of $H_\mathrm{in,out}$ does not follow. This fact, however, does not affect the calculation of the perturbation equations of the other elements through the usual, formal way, and hence, in what follows, we do not take it into account. 

The perturbation equations that are interesting for us, take the following form:
\begin{eqnarray}
\frac{\mathrm{d}e_\mathrm{in}}{\mathrm{d}\tau}&=&-\frac{1-e_\mathrm{in}^2}{e_\mathrm{in}}\frac{1}{G_\mathrm{in}}\frac{\mathrm{d}G_\mathrm{in}}{\mathrm{d}\tau} \\
&=&2\pi A_\mathrm{G}e_\mathrm{in}\left(1-e_\mathrm{in}^2\right)^{1/2}\left(1-I^2\right)\sin2\omega_\mathrm{in}^\mathrm{dyn}\nonumber \\
&&+2\pi A_\mathrm{G}^\mathrm{oct}e_\mathrm{out}\left(1-e_\mathrm{in}^2\right)^{1/2}\left\{\left(1+\frac{3}{4}e_\mathrm{in}^2\right)\left[\left(1+11I-5I^2-15I^3\right)\sin(\omega_\mathrm{in}^\mathrm{dyn}-\omega_\mathrm{out}^\mathrm{dyn})\right.\right. \nonumber \\
&&\left.+\left(1-11I-5I^2+15I^3\right)\sin(\omega_\mathrm{in}^\mathrm{dyn}+\omega_\mathrm{out}^\mathrm{dyn})\right] \nonumber \\
&&-\frac{105}{4}e_\mathrm{in}^2\left[\left(1+I-I^2-I^3\right)\sin(3\omega_\mathrm{in}^\mathrm{dyn}-\omega_\mathrm{out}^\mathrm{dyn})\right. \nonumber \\
&&\left.\left.+\left(1-I-I^2+I^3\right)\sin(3\omega_\mathrm{in}^\mathrm{dyn}+\omega_\mathrm{out}^\mathrm{dyn})\right]\right\}, 
\label{Eq:dote_in}
\end{eqnarray}
\begin{eqnarray}
\frac{\mathrm{d}\omega_\mathrm{in}^\mathrm{dyn}}{\mathrm{d}\tau}=-\frac{\partial\mathcal{H}}{\partial{G_\mathrm{in}}}&=&2\pi\frac{A_\mathrm{G}}{\left(1-e_\mathrm{in}^2\right)^{1/2}}\left[I^2-\frac{1}{5}\left(1-e_\mathrm{in}^2\right)+\frac{2+3e_\mathrm{in}^2}{5}\frac{G_\mathrm{in}}{G_\mathrm{out}}I\right. \nonumber \\
&&\left.+\left(1-e_\mathrm{in}^2-I^2-e_\mathrm{in}^2\frac{G_\mathrm{in}}{G_\mathrm{out}}I\right)\cos2\omega_\mathrm{in}^\mathrm{dyn}\right] \nonumber\\
&&+2\pi\frac{A^\mathrm{oct}_\mathrm{G}}{\left(1-e_\mathrm{in}^2\right)^{1/2}}e_\mathrm{in}e_\mathrm{out}\left\{\left[\frac{1-e_\mathrm{in}^2}{e_\mathrm{in}^2}\left(1+\frac{9}{4}e_\mathrm{in}^2\right)\left(1+11I-5I^2-15I^3\right)\right.\right. \nonumber \\
&&\left.+\left(1+\frac{3}{4}e_\mathrm{in}^2\right)\left(I+\frac{G_\mathrm{in}}{G_\mathrm{out}}\right)\left(11-10I-45I^2\right)\right]\cos(\omega_\mathrm{in}^\mathrm{dyn}-\omega_\mathrm{out}^\mathrm{dyn}). \nonumber \\
&&+\left[\frac{1-e_\mathrm{in}^2}{e_\mathrm{in}^2}\left(1+\frac{9}{4}e_\mathrm{in}^2\right)\left(1-11I-5I^2+15I^3\right)\right. \nonumber \\
&&\left.-\left(1+\frac{3}{4}e_\mathrm{in}^2\right)\left(I+\frac{G_\mathrm{in}}{G_\mathrm{out}}\right)\left(11+10I-45I^2\right)\right]\cos(\omega_\mathrm{in}^\mathrm{dyn}+\omega_\mathrm{out}^\mathrm{dyn}) \nonumber \\
&&-\frac{35}{4}e_\mathrm{in}^2\left[3\frac{1-e_\mathrm{in}^2}{e_\mathrm{in}^2}\left(1+I-I^2-I^3\right)\right. \nonumber \\
&&\left.+\left(I+\frac{G_\mathrm{in}}{G_\mathrm{out}}\right)\left(1-2I-3I^2\right)\right]\cos(3\omega_\mathrm{in}^\mathrm{dyn}-\omega_\mathrm{out}^\mathrm{dyn}) \nonumber \\
&&-\frac{35}{4}e_\mathrm{in}^2\left[3\frac{1-e_\mathrm{in}^2}{e_\mathrm{in}^2}\left(1-I-I^2+I^3\right)\right. \nonumber \\
&&\left.\left.-\left(I+\frac{G_\mathrm{in}}{G_\mathrm{out}}\right)\left(1+2I-3I^2\right)\right]\cos(3\omega_\mathrm{in}^\mathrm{dyn}+\omega_\mathrm{out}^\mathrm{dyn})\right\},
\label{Eq:dotom_indyn}
\end{eqnarray}
\begin{eqnarray}
\frac{\mathrm{d}\Omega_\mathrm{in}^\mathrm{dyn}}{\mathrm{d}\tau}&=&-\frac{\partial\mathcal{H}}{\partial{H_\mathrm{in}}} \nonumber \\
&=&-\frac{2\pi}{5}\frac{A_\mathrm{G}}{\left(1-e_\mathrm{in}^2\right)^{1/2}}\frac{C}{G_\mathrm{out}}I\left(2+3e_\mathrm{in}^2-5e_\mathrm{in}^2\cos2\omega_\mathrm{in}^\mathrm{dyn}\right) \nonumber\\
&&-2\pi\frac{A_\mathrm{G}^\mathrm{oct}}{\left(1-e_\mathrm{in}^2\right)^{1/2}}\frac{C}{G_\mathrm{out}}e_\mathrm{in}e_\mathrm{out}\left\{\left(1+\frac{3}{4}e_\mathrm{in}^2\right)\left[\left(11-10I-45I^2\right)\cos(\omega_\mathrm{in}^\mathrm{dyn}-\omega_\mathrm{out}^\mathrm{dyn})\right.\right. \nonumber  \nonumber \\
&&-\left.\left(11+10I-45I^2\right)\cos(\omega_\mathrm{in}^\mathrm{dyn}+\omega_\mathrm{out}^\mathrm{dyn})\right] \nonumber \\
&&-\frac{35}{4}e_\mathrm{in}^2\left[\left(1-2I-3I^2\right)\cos(3\omega_\mathrm{in}^\mathrm{dyn}-\omega_\mathrm{out}^\mathrm{dyn})\right. \nonumber \\
&&\left.\left.-\left(1+2I-3I^2\right)\cos(3\omega_\mathrm{in}^\mathrm{dyn}+\omega_\mathrm{out}^\mathrm{dyn})\right]\right\}, 
\label{Eq:dotom_indyn}
\end{eqnarray}
and, regarding the perturbations in the observable arguments of periastron:
\begin{eqnarray}
\frac{\mathrm{d}\omega_\mathrm{in}}{\mathrm{d}\tau}&=&\frac{\mathrm{d}\omega_\mathrm{in}^\mathrm{dyn}}{\mathrm{d}\tau}+\frac{\mathrm{d}\Omega_\mathrm{in,out}^\mathrm{dyn}}{\mathrm{d}\tau}\cos i_\mathrm{in}^\mathrm{dyn}-\frac{\mathrm{d}\Omega_\mathrm{in}}{\mathrm{d}\tau}\cos i_\mathrm{in} \\
&=&\frac{2\pi}{5}A_\mathrm{G}\left(1-e_\mathrm{in}^2\right)^{1/2}\left[3I^2-1+5\left(1-I^2\right)\cos2\omega_\mathrm{in}^\mathrm{dyn}\right] \nonumber \\
&&+2\pi A_\mathrm{G}^\mathrm{oct}\left(1-e_\mathrm{in}^2\right)^{1/2}\frac{e_\mathrm{out}}{e_\mathrm{in}}\left\{\left(1+\frac{9}{4}e_\mathrm{in}^2\right)\left[\left(1+11I-5I^2-15I^3\right)\cos(\omega_\mathrm{in}^\mathrm{dyn}-\omega_\mathrm{out}^\mathrm{dyn})\right.\right. \nonumber \\
&&\left.+\left(1-11I-5I^2+15I^3\right)\cos(\omega_\mathrm{in}^\mathrm{dyn}+\omega_\mathrm{out}^\mathrm{dyn})\right] \nonumber \\
&&-\frac{105}{4}e_\mathrm{in}^2\left[\left(1+I-I^2-I^3\right)\cos(3\omega_\mathrm{in}^\mathrm{dyn}-\omega_\mathrm{out}^\mathrm{dyn})\right. \nonumber \\
&&\left.\left.+\left(1-I-I^2+I^3\right)\cos(3\omega_\mathrm{in}^\mathrm{dyn}+\omega_\mathrm{out}^\mathrm{dyn})\right]\right\}-\frac{\mathrm{d}\Omega_\mathrm{in}}{\mathrm{d}\tau}\cos i_\mathrm{in} \label{Eq:dotom_in} \\
\frac{\mathrm{d}\omega_\mathrm{out}}{\mathrm{d}\tau}&=&\frac{\mathrm{d}\omega_\mathrm{out}^\mathrm{dyn}}{\mathrm{d}\tau}+\frac{\mathrm{d}\Omega_\mathrm{in,out}}{\mathrm{d}\tau}\cos i_\mathrm{out}^\mathrm{dyn}-\frac{\mathrm{d}\Omega_\mathrm{out}}{\mathrm{d}\tau}\cos i_\mathrm{out} \\
&=&\frac{2\pi}{10}\frac{A_\mathrm{G}}{\left(1-e_\mathrm{in}^2\right)^{1/2}}\frac{G_\mathrm{in}}{G_\mathrm{out}}\left[\left(2+3e_\mathrm{in}^2\right)\left(3I^2-1\right)+15e_\mathrm{in}^2\left(1-I^2\right)\cos2\omega_\mathrm{in}^\mathrm{dyn}\right] \nonumber \\
&&+2\pi\frac{A_\mathrm{G}^\mathrm{oct}}{\left(1-e_\mathrm{in}^2\right)^{1/2}}\frac{G_\mathrm{in}}{G_\mathrm{out}}e_\mathrm{in}\frac{1+4e_\mathrm{out}^2}{e_\mathrm{out}}\left\{\left[\left(1+11I-5I^2-15I^3\right)\cos(\omega_\mathrm{in}^\mathrm{dyn}-\omega_\mathrm{out}^\mathrm{dyn})\right.\right. \nonumber \\
&&\left.+\left(1-11I-5I^2+15I^3\right)\cos(\omega_\mathrm{in}^\mathrm{dyn}+\omega_\mathrm{out}^\mathrm{dyn})\right] \nonumber \\
&&-\frac{35}{4}e_\mathrm{in}^2\left[\left(1+I-I^2-I^3\right)\cos(3\omega_\mathrm{in}^\mathrm{dyn}-\omega_\mathrm{out}^\mathrm{dyn})\right. \nonumber \\
&&\left.\left.+\left(1-I-I^2+I^3\right)\cos(3\omega_\mathrm{in}^\mathrm{dyn}+\omega_\mathrm{out}^\mathrm{dyn})\right]\right\}-\frac{\mathrm{d}\Omega_\mathrm{out}}{\mathrm{d}\tau}\cos i_\mathrm{out}.
\label{Eq:dotom_out}
\end{eqnarray}

In what follows, we consider only the prograde, coplanar scenario, i.e., when $i_\mathrm{mut}=0^\circ$. This restriction can be justified by the fact that, as will be shown in Sect.~\ref{sec:discussion}, the mutual inclination in none of the three considered systems exceed $i_\mathrm{mut}=0.5^\circ$. In the coplanar case, when $I=1$, Eqs.~(\ref{Eq:dote_in}--\ref{Eq:dotom_out}) become much simpler. Before we give these simpler expressions, we note that, for coplanar orbits, the dynamical arguments of periastron, and the dynamical nodes lose their meanings. The quantities that continue to remain meaningful in this situation are the dynamical longitudes of periastron ($\varpi_\mathrm{in,out}^\mathrm{dyn}=\omega_\mathrm{in,out}^\mathrm{dyn}+\Omega_\mathrm{in,out}^\mathrm{dyn}$), and also the angle between the directions of the two periastron points, i.e.,
$\omega_\mathrm{in}^\mathrm{dyn}-\omega_\mathrm{out}^\mathrm{dyn}=\varpi_\mathrm{in}^\mathrm{dyn}-\varpi_\mathrm{out}^\mathrm{dyn}-\pi$, where in the last expression we took into account the fact that $\Delta\Omega^\mathrm{dyn}=\pi$. Furthermore, we notice that in the coplanar case $\omega_\mathrm{in,out}=\varpi_\mathrm{in,out}^\mathrm{dyn}$. Thus, as can be easily seen, for the prograde, coplanar case, the perturbation equations up to the octupole order have the following forms:
\begin{eqnarray}
\frac{\mathrm{d}e_\mathrm{in}}{\mathrm{d}\tau}&=&16\pi A_\mathrm{G}^\mathrm{oct}e_\mathrm{out}\left(1-e_\mathrm{in}^2\right)^{1/2}\left(1+\frac{3}{4}e_\mathrm{in}^2\right)\sin(\omega_\mathrm{in}-\omega_\mathrm{out}), 
\label{Eq:dote_in_coplanar} \\
\frac{\mathrm{d}\omega_\mathrm{in}}{\mathrm{d}\tau}&=&\frac{4\pi}{5}A_\mathrm{G}\left(1-e_\mathrm{in}^2\right)^{1/2} \nonumber \\
&&+16\pi A_\mathrm{G}^\mathrm{oct}\left(1-e_\mathrm{in}^2\right)^{1/2}\frac{e_\mathrm{out}}{e_\mathrm{in}}\left(1+\frac{9}{4}e_\mathrm{in}^2\right)\cos(\omega_\mathrm{in}-\omega_\mathrm{out}) \label{Eq:dotom_in_coplanar} \\
\frac{\mathrm{d}\omega_\mathrm{out}}{\mathrm{d}\tau}&=&\frac{4\pi}{5}\frac{A_\mathrm{G}}{\left(1-e_\mathrm{in}^2\right)^{1/2}}\frac{G_\mathrm{in}}{G_\mathrm{out}}\left(1+\frac{3}{2}e_\mathrm{in}^2\right) \nonumber \\
&&+16\pi\frac{A_\mathrm{G}^\mathrm{oct}}{\left(1-e_\mathrm{in}^2\right)^{1/2}}\frac{G_\mathrm{in}}{G_\mathrm{out}}e_\mathrm{in}\frac{1+4e_\mathrm{out}^2}{e_\mathrm{out}}\cos(\omega_\mathrm{in}-\omega_\mathrm{out})
\label{Eq:dotom_out_coplanar}
\end{eqnarray}
and, finally,
\begin{eqnarray}
\frac{\mathrm{d}(\omega_\mathrm{in}-\omega_\mathrm{out})}{\mathrm{d}\tau}&=&\frac{4\pi}{5}A_\mathrm{G}\left(1-e_\mathrm{in}^2\right)^{1/2}\left(1-\frac{G_\mathrm{in}}{G_\mathrm{out}}\frac{1+\frac{3}{2}e_\mathrm{in}^2}{1-e_\mathrm{in}^2}\right) \nonumber \\
&&+16\pi A_\mathrm{G}^\mathrm{oct}\left(1-e_\mathrm{in}^2\right)^{1/2}e_\mathrm{in}e_\mathrm{out}\nonumber \\
&&\times\left(\frac{1+\frac{9}{4}e_\mathrm{in}^2}{e_\mathrm{in}^2}-\frac{G_\mathrm{in}}{G_\mathrm{out}}\frac{1}{1-e_\mathrm{in}^2}\frac{1+4e_\mathrm{out}^2}{e_\mathrm{out}^2}\right)\cos(\omega_\mathrm{in}-\omega_\mathrm{out}). 
\label{Eq:dotom_in-om_out_coplanar} 
\end{eqnarray}

As can be easily seen, in this scenario there are no quadrupole level long-period perturbations in the inner eccentricity, and the observable dynamical apsidal motion rates of both orbits remain constant (again, at the quadrupole level). In contrast to this, the octupole order perturbations of these elements depend on  trigonometric functions of the angle between the two periastron directions. One should keep in mind, that this angle, in general, varies at a lower rate than the individual apsidal lines themselves and, hence, one expect a longer period cyclic, octupole variation, than the usual (quadrupole) dynamically forced apsidal motion period.

A detailed analytic and/or numerical analysis of these perturbation equations are beyond the scope of the current paper. Our aim remains only to identify the consequences of these orbital perturbations directly from the high-precision observations of selected TCHTs, and then, to determine accurate dynamical (and astrophysical) parameters of these systems, and finally, to compare our findings at least qualitatively with the predictions of the theory, discussed briefly above. A more detailed study including the quantitative comparison of the theoretical and the observed perturbations will be published later.

\section{Selected systems}
\label{sect:selection}

We selected two TCHTs from the well-observed primary \textit{Kepler}-sample and, moreover, a third system, which serves as a counterexample in the sense that, although this system is very similar to the previous two in several aspects, the octupole order perturbations are almost nulled out due to a near unit mass ratio of the inner binary. Unfortunately, we did not find such an illustrative TCHT amongst the \textit{Kepler} systems and, hence, we took one from the \textit{TESS} northern continuous viewing zone (NCVZ) sample. The main catalog data for the three systems are collected in Table~\ref{tbl:mags}, where, in addition to the J2000.0 coordinates and the catalog magnitudes from near ultraviolet to infrared passbands (i.e., form GALEX NUV to Wise W4 magnitudes) we also tabulate the effective temperatures ($T_\mathrm{eff}$), photogeometric distances, metallicities ([$M/H$]), interstellar reddenings ($E(B-V)$) and proper motions ($\mu_{\alpha,\delta}$) from different catalogs. Finally, we also give the specific $RUWE$ (renormailzed unit weight error) parameters, which was introduced in \textit{Gaia} DR2 as an indicator of the quality of the astrometric solutions. This parameter is relevant in the context of the current study because it was found that a value greater than $\gtrsim1.4$ could indicate the multiplicity of the source \citep[see, e.~g.][]{stassuntorres21}. In what follows, we discuss briefly the available basic information on these three TCHTs.

\begin{table}
\centering
\caption{Main properties of the three systems from different catalogs}
\begin{tabular}{lccc}
\hline
\hline
Parameter & KIC~9714358 & KIC~5771589 & TIC~219885468 \\
\hline
RA (J2000) & $19:34:09.667$ & $18:58:20.526$ & $17:28:57.721$ \\  
Dec (J2000)& $+46:26:14.02$ & $+41:00:34.46$ & $+70:42:03.09$ \\  
%$T^a$ & $10.9412\pm0.0069$  & $14.0809\pm0.0292$ & $12.6336\pm0.0165$ \\
$G^b$ & $14.9892\pm0.0007$  & $11.8349\pm0.0020$ & $13.0798\pm0.0002$ \\
$G_{\rm BP}^b$ & $15.4996\pm0.0031$ & $12.0314\pm0.0005$ & $13.3429\pm0.0001$ \\
$G_{\rm RP}^b$ & $14.3166\pm0.0024$ & $11.2839\pm0.0004$ & $12.6629\pm0.0005$ \\
B$^a$ & $16.336 \pm 0.067$ & $12.512\pm0.013$ & $13.729\pm0.200$ \\
V$^a$ & $15.422 \pm 0.1114$ & $11.943\pm0.027$ & $13.200\pm0.069$ \\
g$'$  & $15.632\pm0.004^c$ & $12.228\pm0.001^d$ & $13.442\pm0.196^e$ \\
r$'$  & $14.989\pm0.004^c$ & $11.849\pm0.001^d$ & $13.015\pm0.173^e$ \\
i$'$  & $14.696\pm0.002^c$ & $11.691\pm0.001^d$ & $12.905\pm0.205^e$ \\
J$^f$ & $13.513 \pm 0.023$ & $10.787\pm0.020$ & $12.233\pm0.021$ \\
H$^f$ & $13.001 \pm 0.026$ & $10.511\pm0.021$ & $11.970\pm0.023$ \\
K$^f$ & $12.884 \pm 0.025$ & $10.453\pm0.018$ & $11.942\pm0.022$ \\
W1$^g$ & $12.801\pm0.023$  & $10.379\pm0.023$ & $11.885\pm0.023$ \\
W2$^g$ & $12.794\pm0.023$  & $10.408\pm0.020$ & $11.909\pm0.021$ \\
W3$^g$ & $12.619\pm0.368$  & $10.340\pm0.049$ & $11.973\pm0.125$ \\
W4$^g$ & $ 9.198$          & $8.842$        & $9.577$ \\
NUV    & $22.713\pm0.324$  &                & $17.232\pm0.032$   \\
$T_{\rm eff}$ (K)$^a$ & $4764\pm109$ & $6231$ & $6350\pm133$ \\
Distance (pc)$^h$ & $ 577\pm7 $ & $870\pm100$ & $1111\pm13$ \\ 
$[M/H]^a$ & $-0.462\pm0.013$ & $0.083\pm0.011$ & $-$ \\ 
$E(B-V)^a$ & $0.050$ & $-$ & $0.030\pm0.007$ \\
$\mu_\alpha$ (mas ~${\rm yr}^{-1}$)$^b$ & $-8.77\pm0.02$ & $-3.02\pm0.15$ & $1.02\pm0.01$ \\ 
$\mu_\delta$ (mas ~${\rm yr}^{-1}$)$^b$ & $3.14\pm0.02$ & $1.19\pm0.17$ & $-0.64\pm0.02$ \\ 
RUWE$^b$ & $1.03356$ & $11.34329$ & $0.94989$ \\
\hline
\label{tbl:mags}  % Table 1
\end{tabular}

\textit{Notes.}  (a) TESS Input Catalog (TIC v8.2) \citep{TIC8}. (b) Gaia EDR3 \citep{GaiaEDR3}. (c) PanSTARRS, \citep{PanSTARRS}. (d) The Kepler-INT survey \citep{kisdr2}. (e) AAVSO Photometric All Sky Survey (APASS) DR9, \citep{APASS}, \url{http://vizier.u-strasbg.fr/viz-bin/VizieR?-source=II/336/apass9}. (f) 2MASS catalog \citep{2MASS}.  (g) WISE point source catalog \citep{WISE}. (h) Photogeometric distances from \citet{bailer-jonesetal21}. \\
Note also, that for the SED analysis in Sect.~\ref{sec:dyn_mod} the uncertainties of the passband magnitudes were set to $\sigma_\mathrm{mag}=\mathrm{max}(\sigma_\mathrm{catalog},0.030)$ to avoid the strong overdominance of the extremely accurate Gaia magnitudes over the other measurements.
\end{table}

\textit{KIC 9714359 = KOI-6073} ($P_\mathrm{in}=6.48$\,d; $P_\mathrm{out}=103.8$\,d; $P_\mathrm{out}/P_\mathrm{in}=16.0$, $\epsilon\approx0.016$) is the compact, tight triple star system which chiefly inspired this study. This target was identified first as an EB in the first release of the Kepler EB catalog \citep{prsaetal11}. The hierarchical triple star nature of this source was reported first by \citet{rappaportetal13} analyzing the ETVs of the first 13 quarters of data of \textit{Kepler} EBs, who also pointed out that the ETV is dominated almost exclusively by the dynamical, third-body perturbation effects over the classic light-travel time effect (LTTE). Then the system was included in the sample of 26 eccentric EBs which were comprehensively reanalyzed by \citet{borkovitsetal15} with the use of their most complex analytic ETV model. Despite the extreme tightness of the triple, they found a quite satisfactory analytical, coplanar ETV solution with a slightly eccentric inner orbit, and moderately eccentric outer orbit ($e_\mathrm{in}=0.015\pm0.001$ and $e_\mathrm{out}=0.30\pm0.01$, respectively), and with a dynamically forced apsidal motion period of $P_\mathrm{apse, in}=30.6$\,yr. The model also indicated inner and outer mass ratios of $q_\mathrm{in}=0.45\pm0.01$ and $q_\mathrm{out}=0.21\pm0.01$, respectively.

Further studies, however, resulted in slightly discrepant values for some of the parameters above. First, \citet{kjurkchievaetal17} found a substantially ($\sim 5-6$ times) larger inner eccentricity of $e_\mathrm{in}=0.085$ from the \textit{Kepler} light curves, and, moreover, based on astrophysical reasons, they estimated a lower inner mass ratio ($q_\mathrm{in}=0.33$). Then, \citet{windemuthetal19} also reanalyzed the \textit{Kepler} light curves, adding SED information and theoretical \texttt{PARSEC} \citep{PARSEC} evolutionary tracks to them in an automated way. While their eccentricity of $e_\mathrm{in}=0.016\pm0.002$ was really in good agreement with that of the dynamically inferred value of \citet{borkovitsetal15}, they obtained an even much more extreme inner mass ratio ($q_\mathrm{in}=0.234$) and, correspondingly, they found that the system was in a pre-MS state with an age of $\log\tau\approx7.44\pm0.04$.

\begin{figure}
\includegraphics[width=7 cm]{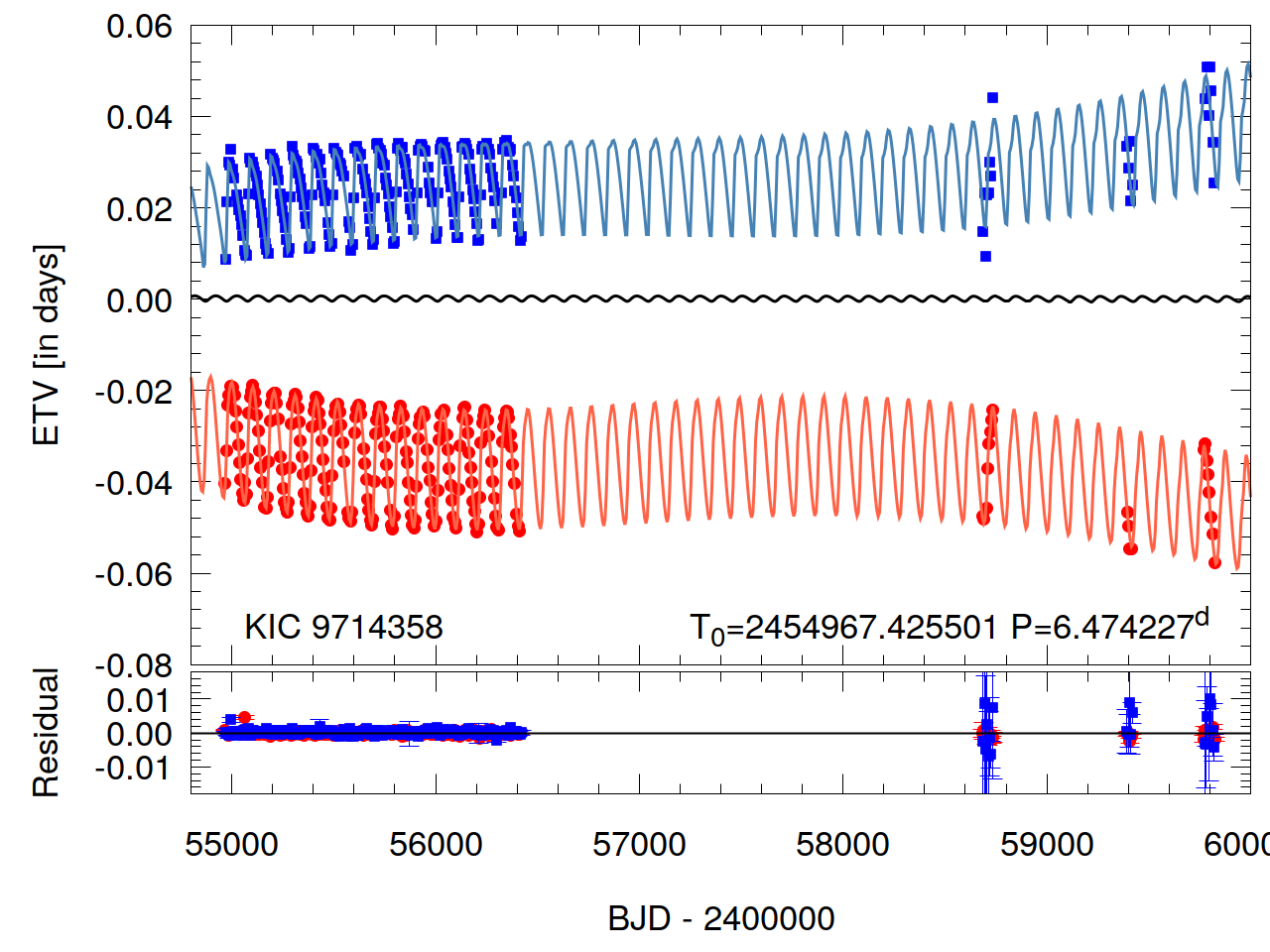}\includegraphics[width=7 cm]{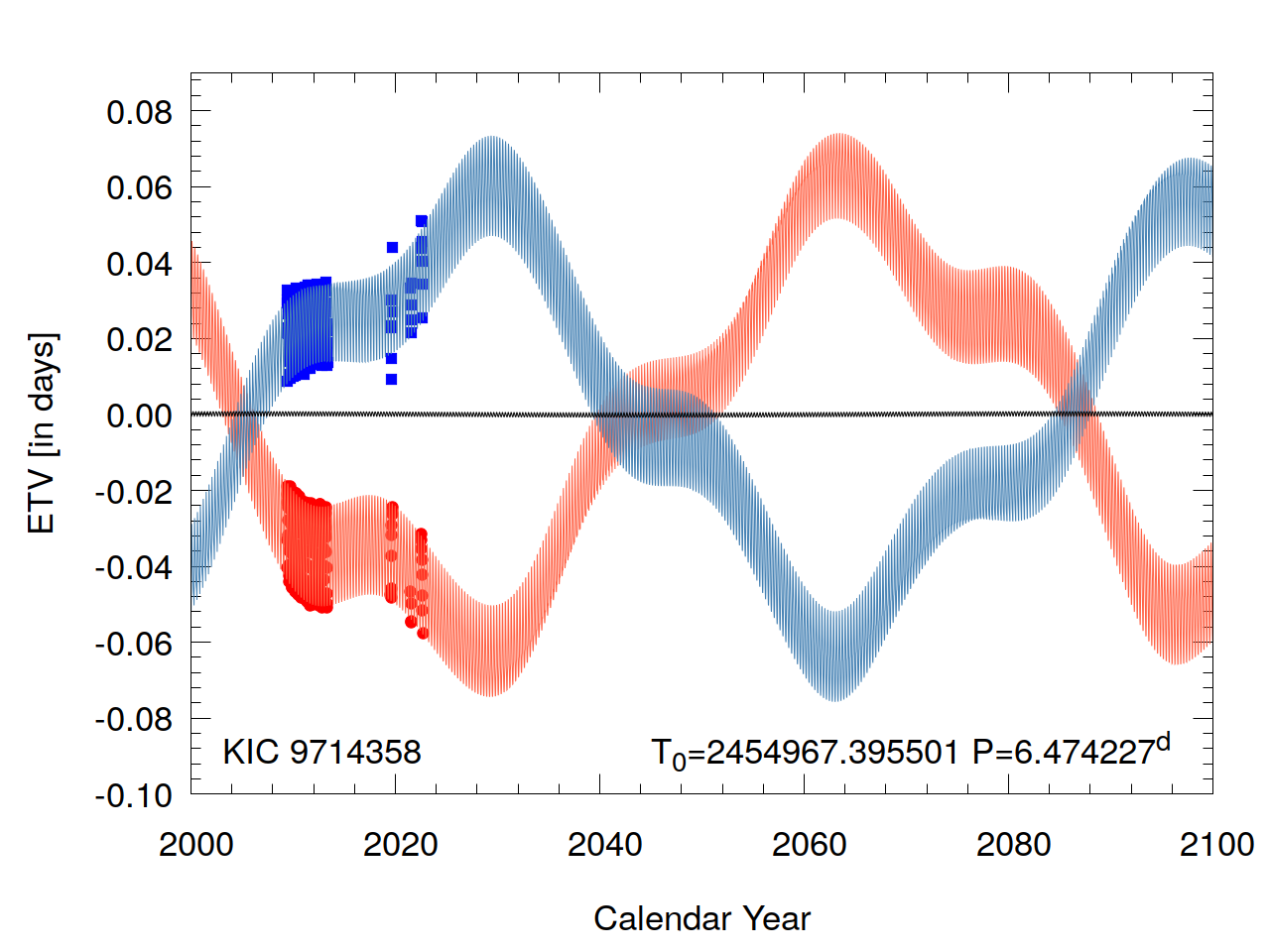}
\includegraphics[width=7 cm]{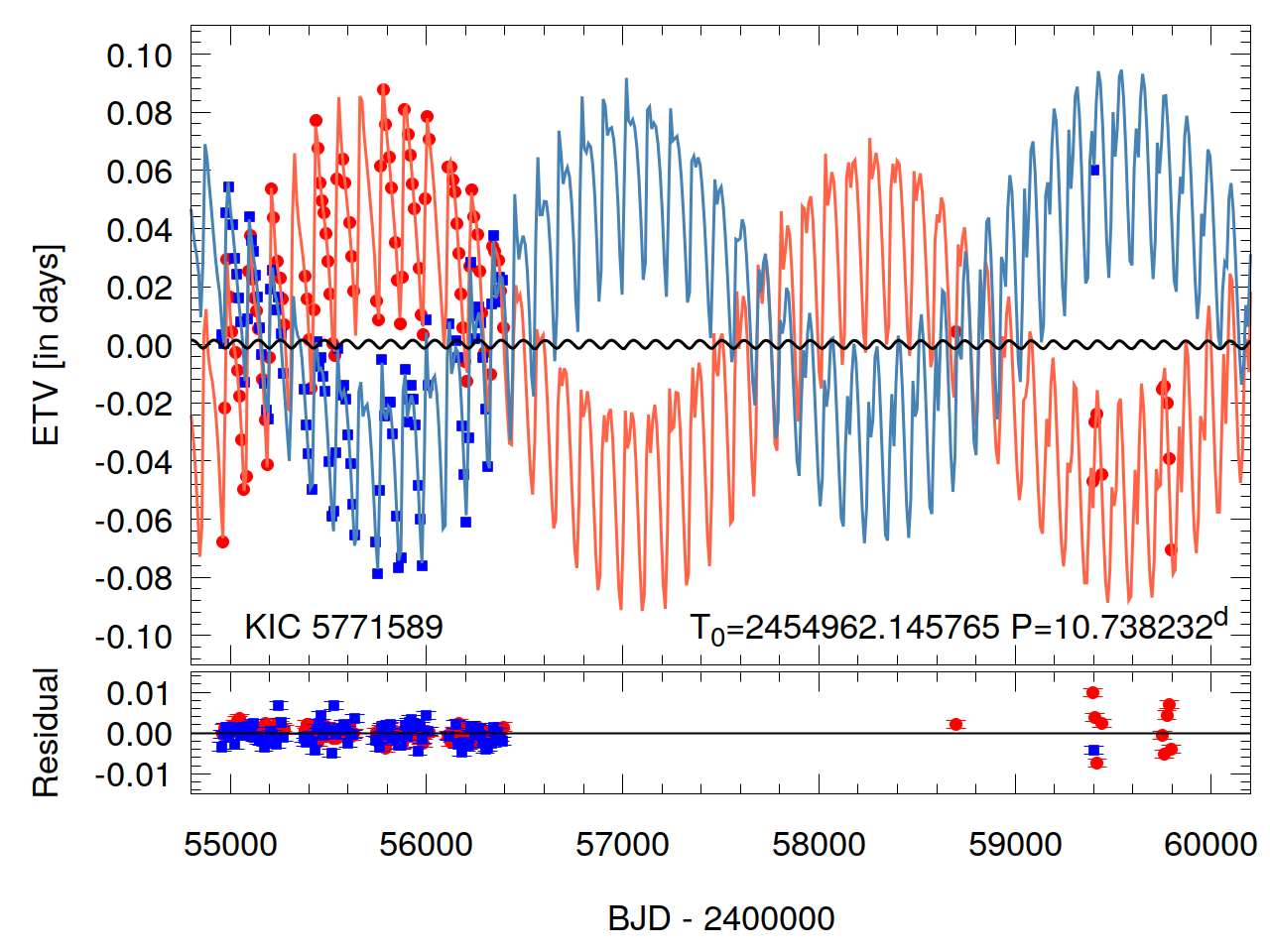}\includegraphics[width=7 cm]{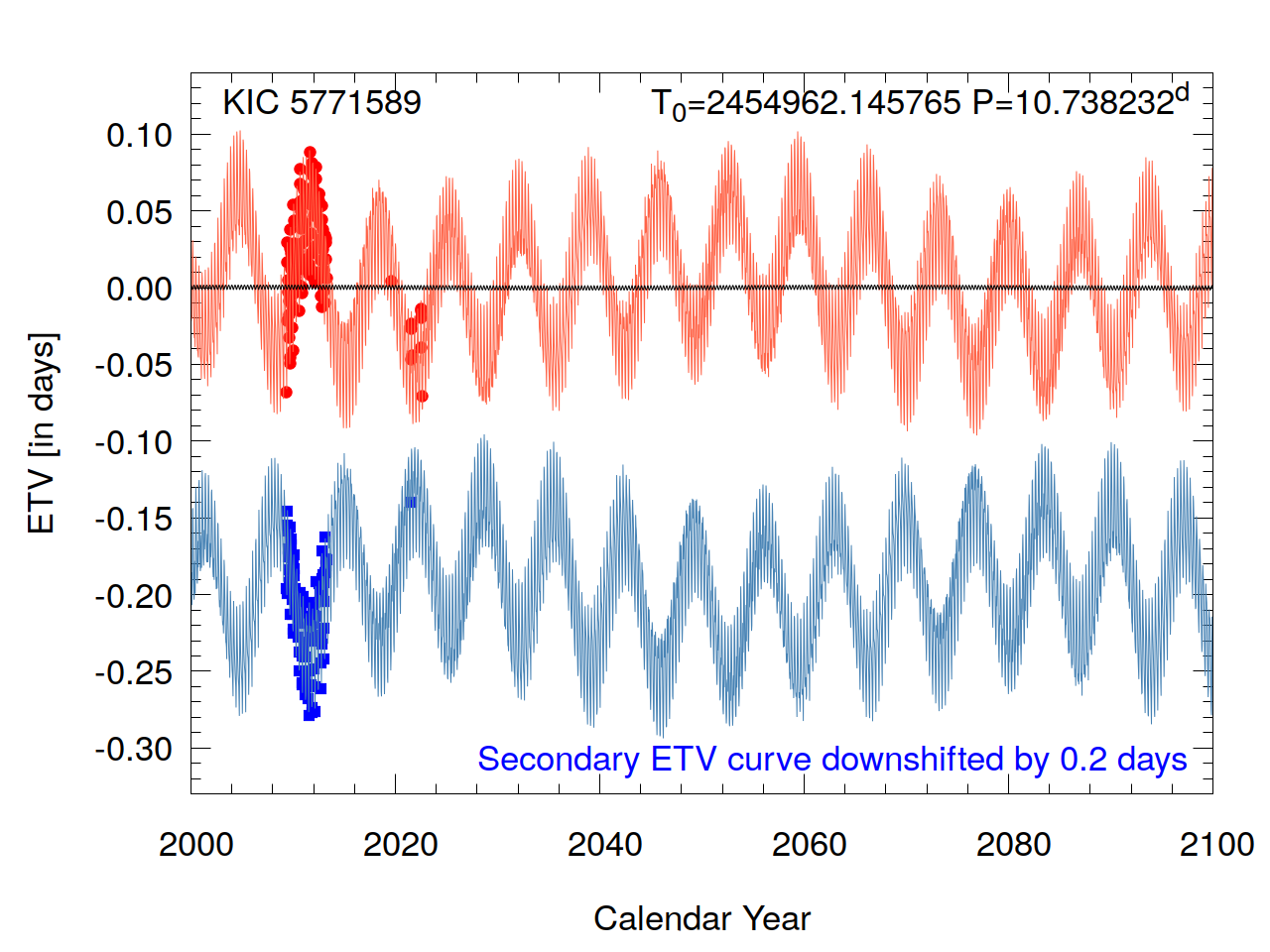}
\includegraphics[width=7 cm]{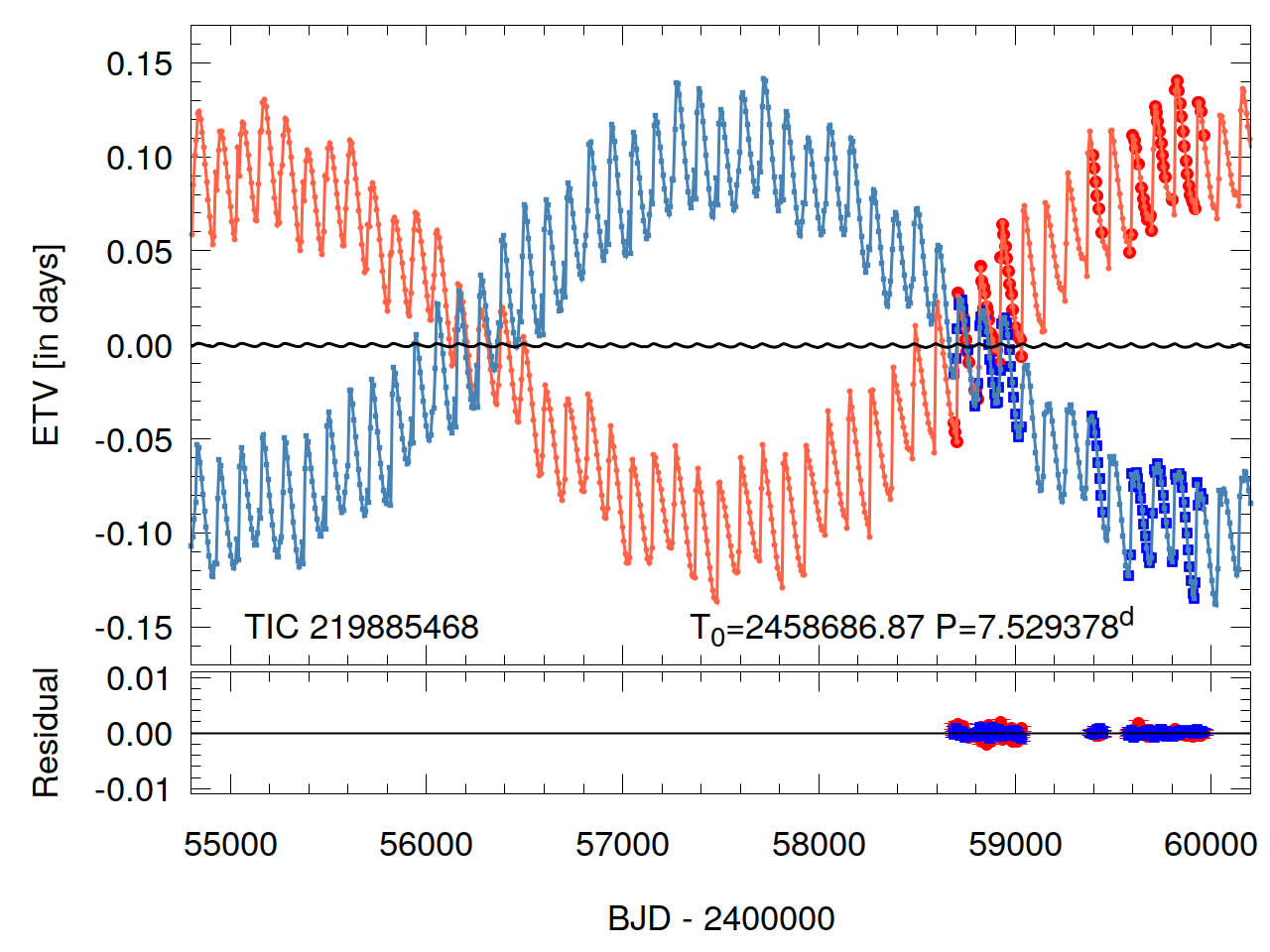}\includegraphics[width=7 cm]{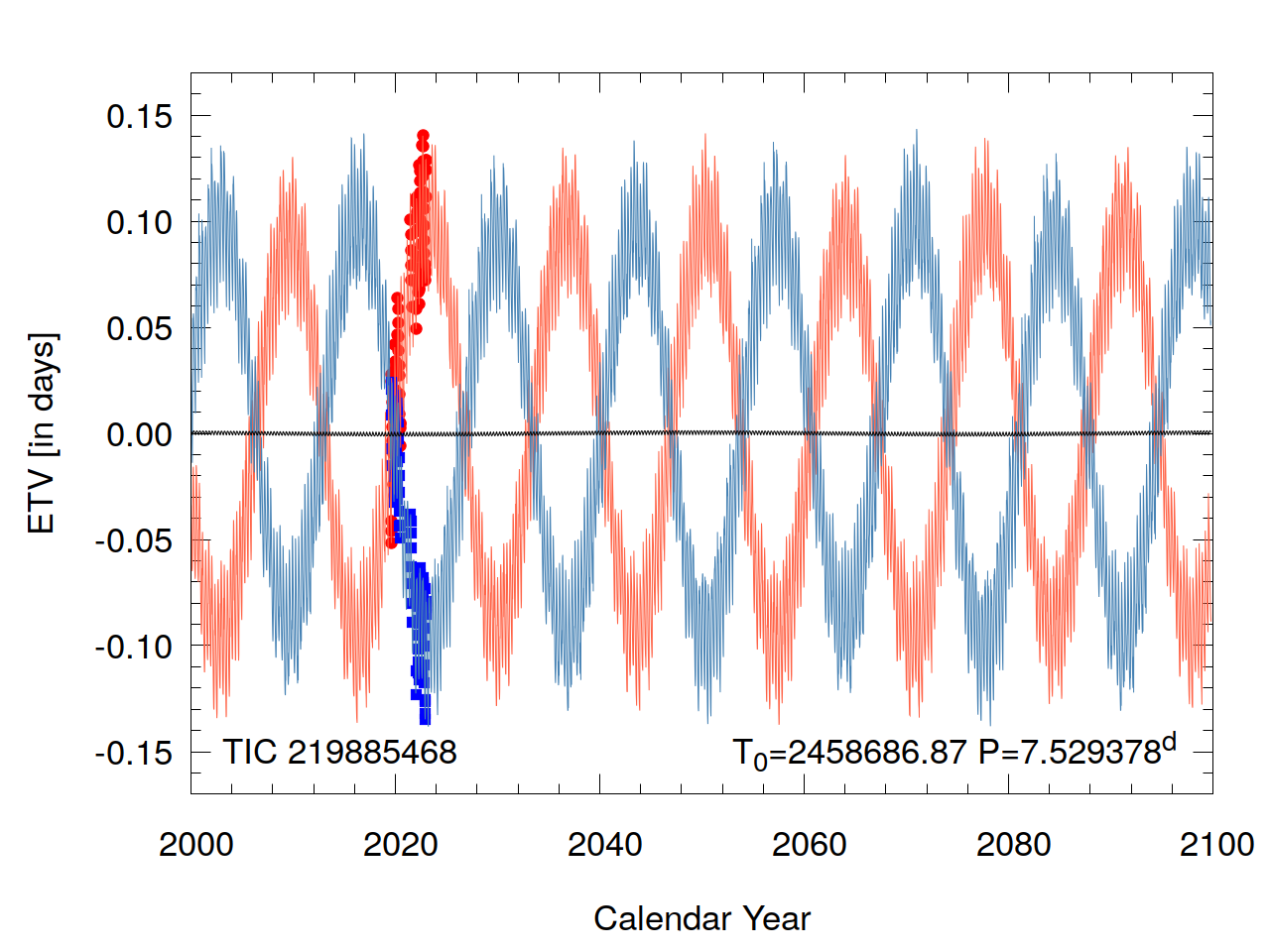}
\caption{Observed and modelled ETVs of KICs~9714358 (upper), 5771578 (middle) and TIC 219885468 (bottom). The left-hand panels cover the time interval approximately from the beginning of the \textit{Kepler} observations to the end of the second northern ecliptic hemisphere scan of \textit{TESS}, while the right-hand panels show longer, 100\,yr-long intervals, which offer direct predictions for future observations. Larger red and blue symbols represent primary and secondary ETV points determined from the observations, while the smaller points connected with straight lines stand for ETV points calculated according to the best-fitting complex photodynamical models. Finally, the black, almost horizontal lines represent the LTTE contributions to the curves. Note, in the middle right panel we downshifted the originally overlapping secondary ETV curve of KIC~5771589 by 0.2\,days, for a better view of the small effect of variations in the inner eccentricity forced by the octupole perturbations. See text for further details.}
\label{fig:ETVs}
\end{figure}

While these discrepancies, noted above, might be interesting in and of themselves, what makes this system especially important for us is the unusual behavior of the ETV curves during the \textit{TESS} observations. KIC~9714358 was reobserved by \textit{TESS} in Sectors 14, 15, 40, 54 and 55 in the lower cadence full frame image (FFI) mode. As one can readily see in the upper left panel of Fig.~\ref{fig:ETVs} the new ETV points, determined from \textit{TESS}-observed eclipses, do not follow the former trend, and the long-term behaviour of the red (primary) and blue (secondary) ETV curves can no longer be described with two antiphased cosine curves, as one would expect for an EB with low eccentricity and (almost constant) apsidal motion rate (the latter of which is a clear prediction of the quadrupole-order hierarchical three-body problem). Hence, our immediate suspicion was that what we are seeing is nothing more than a direct, observational manifestation of the octupole-order perturbation effects. We will return to this question in Sect.~\ref{sec:discussion}.

\begin{figure}
\includegraphics[width=14 cm]{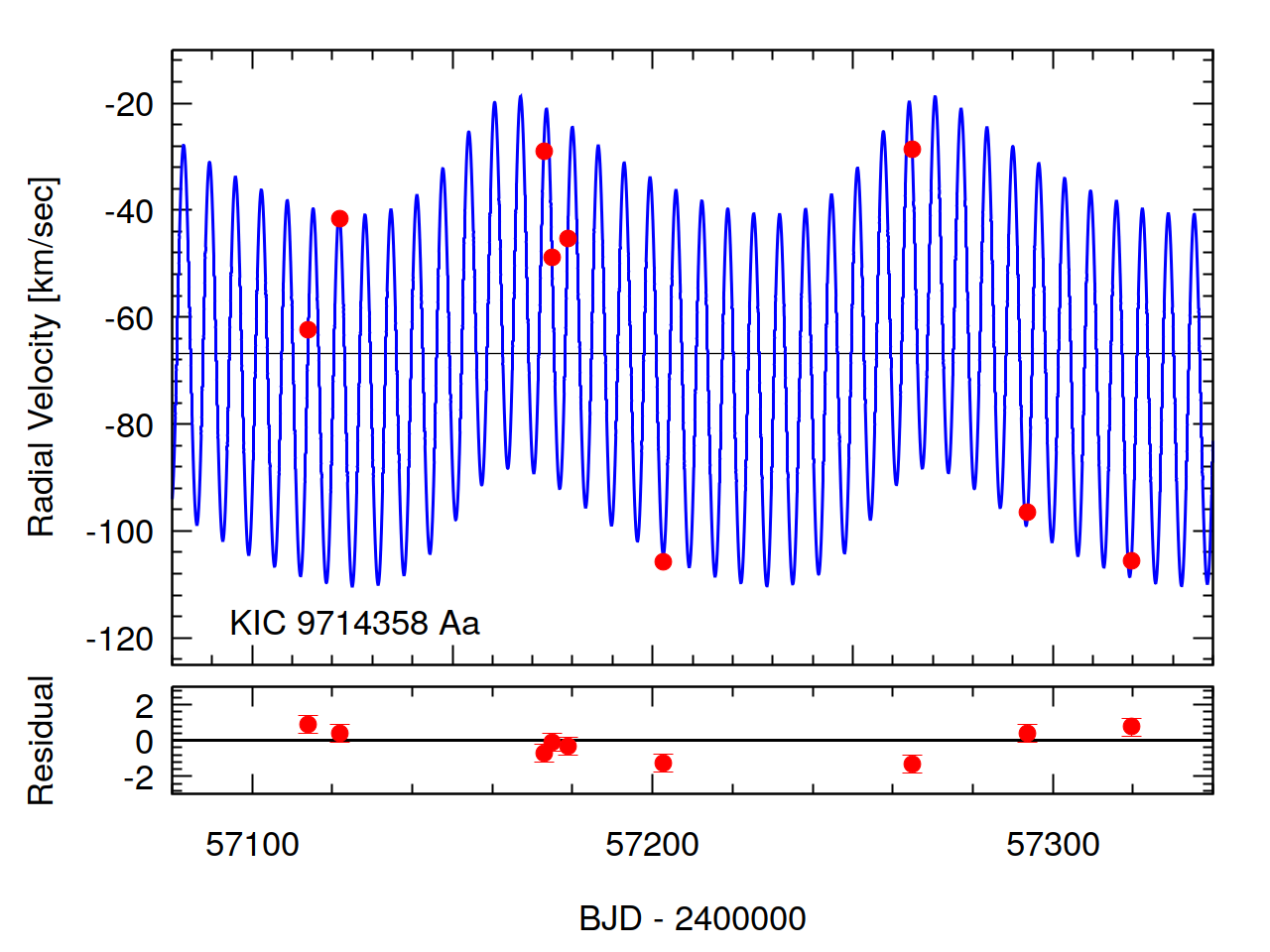}
\includegraphics[width=7 cm]{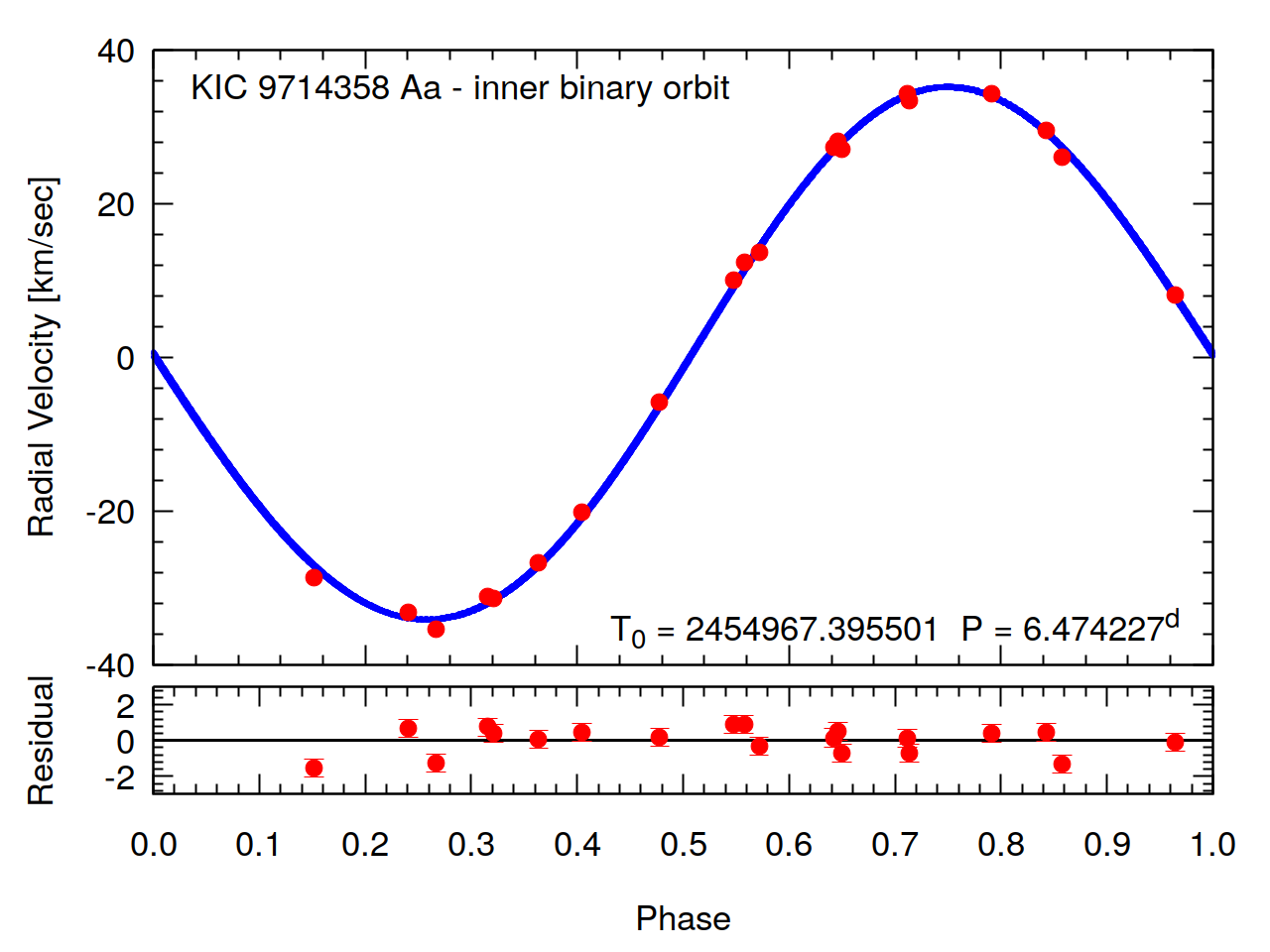}\includegraphics[width=7 cm]{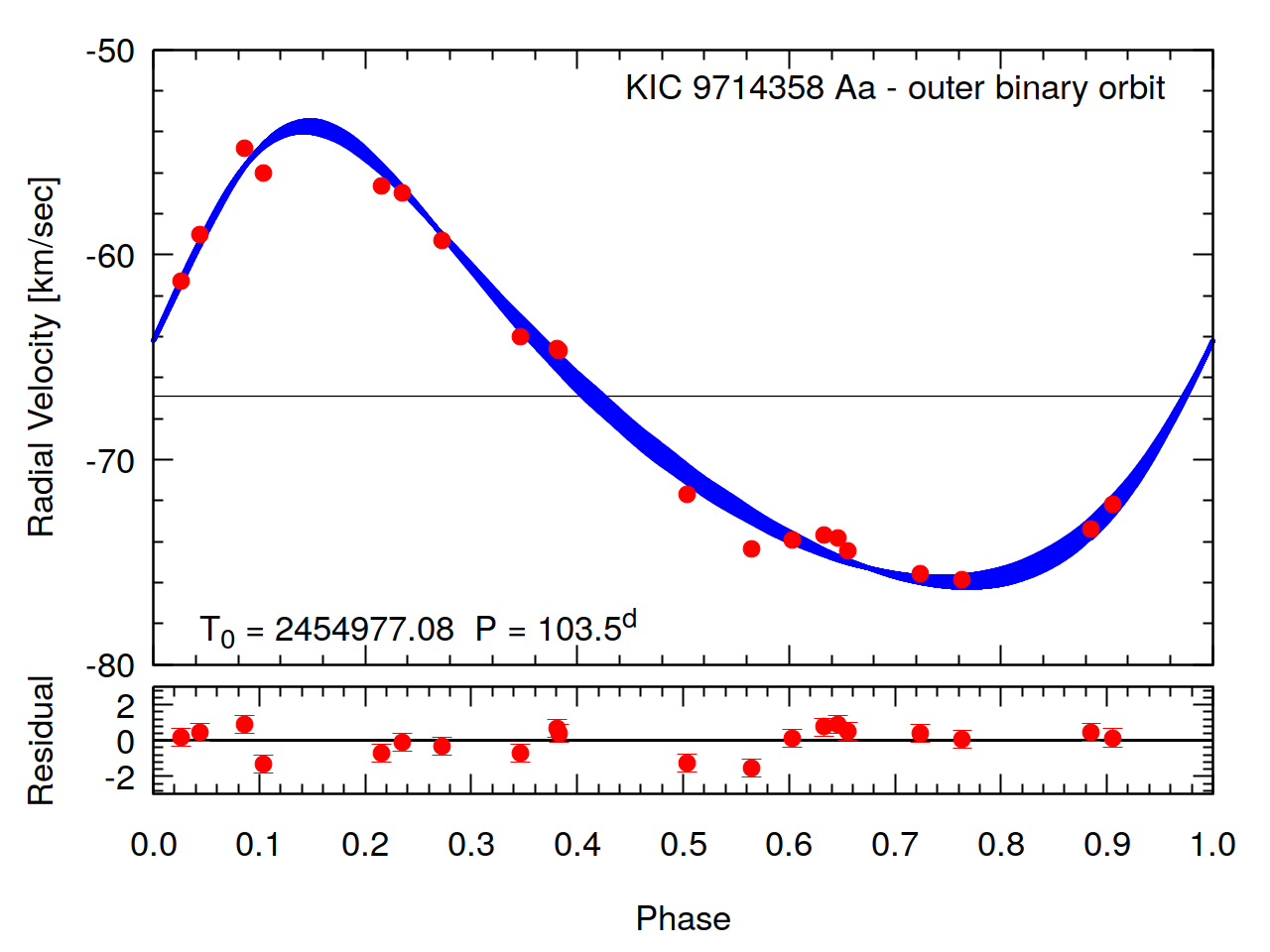}
\caption{APOGEE-2 RVs of the primary component of KIC~9714358 (red dots) together with the best photodynamical model (blue curves). {\it Upper panel}: A characteristic section of the observed RV points and the corresponding model fit in the time domain. Both the inner and the outer orbits are well visible. {\it Lower left panel}: Phase folded RV data according to the inner period, after the removal of the signals of the outer orbit. {\it Lower right panel}: Phase folded RV data according to the outer period, after the removal of the effects of the inner orbital motion. The varying shape of the folded solution, caused primarily by the rapid apsidal motion of the outer orbit is readily visible. Note, the thin horizontal black lines at $V=-66.91\,\mathrm{km\,s}^{-1}$ denote the systemic radial velocity of the triple star.}
\label{fig:K9714358RVs}
\end{figure}

Finally, note that this is the only system in our sample for which radial velocity (RV) data are available from the literature. Data release 16 of the APOGEE-2 project \citep{jonssonetal20} tabulates 16 individual RV observations, which we used for our analysis (see Fig.~\ref{fig:K9714358RVs}).

\textit{KIC 5771589 = KOI-6625} ($P_\mathrm{in}=10.79$\,d; $P_\mathrm{out}=113.1$\,d; $P_\mathrm{out}/P_\mathrm{in}=10.5$, $\epsilon\approx0.004$). This system was identified as an EB, and then as a triple star candidate in the same sequence as for the previous target. It was also included in the detailed analytical ETV studies of \citet{borkovitsetal15,borkovitsetal16}. This is the second tightest known triple in the whole studied \textit{Kepler} sample, and the only one for which the ETV curves, determined purely from the $\sim4$\,yr-long original \textit{Kepler} observations, cover more than half of an apsidal cycle.\footnote{Note that in the case of the most compact (but less tight) \textit{Kepler} triple, KOI-126, the full apsidal motion period is shorter than the length of the \textit{Kepler} dataset \citep{carteretal11,yenawineetal21}, but in that case no precise eclipse times and, hence, no ETVs, are available due to the very shallow, and frequently missing eclipses.} The analytical ETV studies suggested a large outer mass ratio ($q_\mathrm{out}=0.98\pm0.06$), which  was strongly supported by the very shallow eclipses ($\sim0.2-0.4\%$). 

\begin{figure}
\includegraphics[width=14 cm]{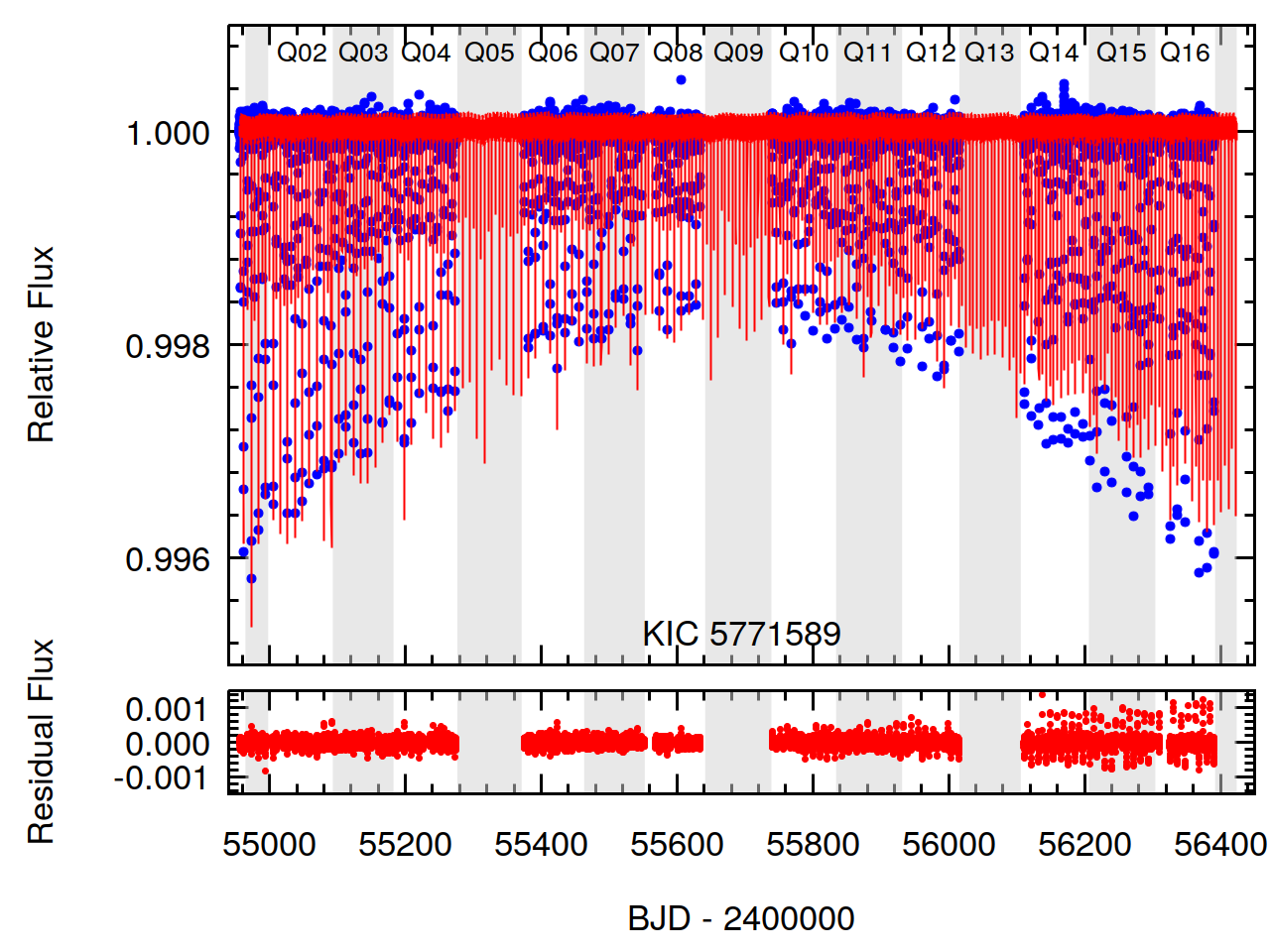}
\includegraphics[width=7 cm]{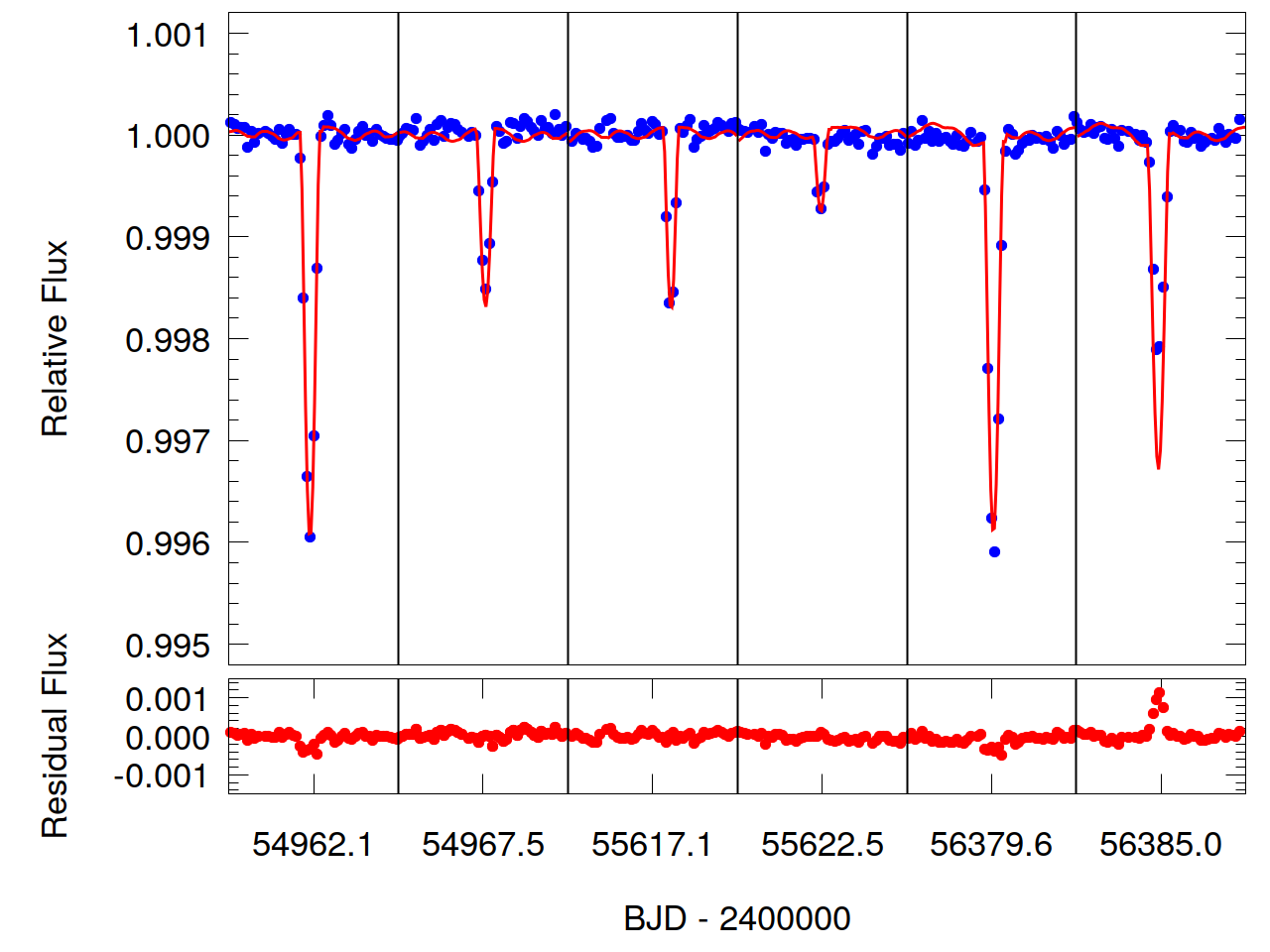}\includegraphics[width=7 cm]{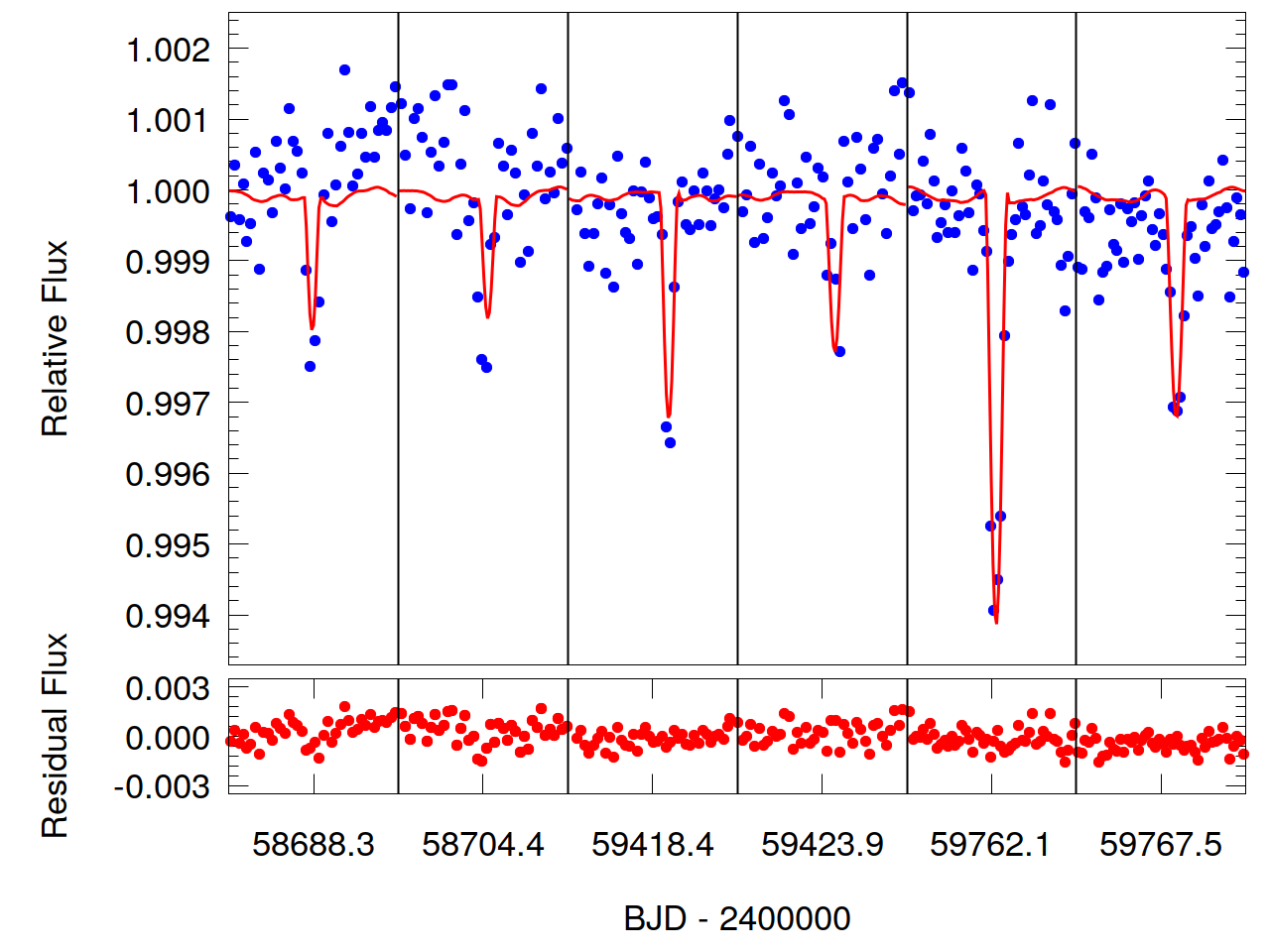}
\caption{{\it Upper panel}: The $\sim4$-year-long \textit{Kepler} light curve of KIC 5771589 (blue dots) with the best-fitted photodynamical model (red). Alternating gray-white stripes denote the consecutive observing quarters. Note, Q5, Q9 and Q13 data are missing, as in these quarters the stellar light had fallen onto a bad CCD camera. {\it Lower left panel}: Consecutive primary and secondary eclipses from the beginning, mid-time, and near to the end of the prime \textit{Kepler} mission. {\it Lower right panel:} Characteristic primary and secondary eclipses from the 2019, 2021 and 2022 visits of \textit{TESS}. Each of the sections is 1\,day long.}
\label{fig:K5771589lcs}
\end{figure}

The most remarkable feature of the \textit{Kepler} light curve is, however, the varying depths of the eclipses (see Fig.~\ref{fig:K5771589lcs}), implying orbital plane precession forced by the not-exactly coplanar tertiary. According to our knowledge, there are only two EBs in the whole primary \textit{Kepler} sample which exhibit a clear reversal of the trend of the eclipse depth variations (EDVs) during the four-year-long observations. The other such target is the triply eclipsing triple star system KIC~6964043 \citep{borkovitsetal22b}, which displayed an opposite trend in its EDV compared to KIC~5771589. In the former case the eclipse depths increased first and then decreased, while in the currently considered KIC~5771589 the eclipses became shallower and shallower during the first half of the \textit{Kepler} observations, and then they started to increase again. These two situations are not symmetric with respect to each other. This is because an EB can reach the maximum of its eclipse depths in two different ways during a nodal precession cycle. The EB may either pass through the exact edge-on view (i.e. $i=90^\circ$ -- which, as was pointed out by \citealt{borkovitsetal22b}, really did happen in KIC~6964043) without reaching an extremum in its observable inclination or, it may reach its nearest position to an edge-on view (i.e. $|\cos i|$ has a minimum, but not zero value or, of course, the inclination itself has an extremum). In contrast to this, when eclipse depths reach a minimum value, one can be certain that the visible inclination has reached its extremum farthest from $i=90^\circ$\footnote{Note, this does not mean that `at the other end' of the precession cycle (i.e. a half precession period later) it cannot reach another extremum which is even farther from the edge-on view.}. In such a way, the current light curve behaviour offers more strict constraints on the orbital configuration.

KIC~5771589 was reobserved with the \textit{TESS} space telescope in Sectors 14, 26, 40, 41, 53 and 54. In contrast to KIC~9714358, for this target 2-min cadence observations are also available for all sectors. Unfortunately, despite the relative brightness of the system, due to the very shallow eclipse depths, the \textit{TESS} data have much lower quality. Despite the low S/N ratios, however, very shallow eclipses can still be identified at the expected times, but in the Year 2 data they are almost lost in the noise. By contrast, the eclipses can be seen clearly at the end of the Year 4 observations, indicating that the eclipse depths must have increased again in between the Year 2 and Year 4 \textit{TESS} observations (see the lower right panel of Fig.~\ref{fig:K5771589lcs}).

\textit{TIC 219885468} has $P_\mathrm{in}=7.54$\,d; $P_\mathrm{out}=110.8$\,d; $P_\mathrm{out}/P_\mathrm{in}=14.7$, and $\epsilon\approx0.0004$. This system was not observed with the \textit{Kepler} spacecraft, but instead, it is located near to the northern ecliptic pole, in the NCVZ of the \textit{TESS} space telescope. But, we did not choose it primarily for its relatively long data train, but rather for didactic reasons. In particular, we wanted to illustrate that, despite the clear similarity to the above-mentioned KIC~97154358 (in both their inner and outer periods), due to the twin-nature of the inner binary star, and hence, in the absence of octupole order perturbations, the short-term orbital evolution of the two triples differ significantly.

\begin{figure}
\includegraphics[width=14cm]{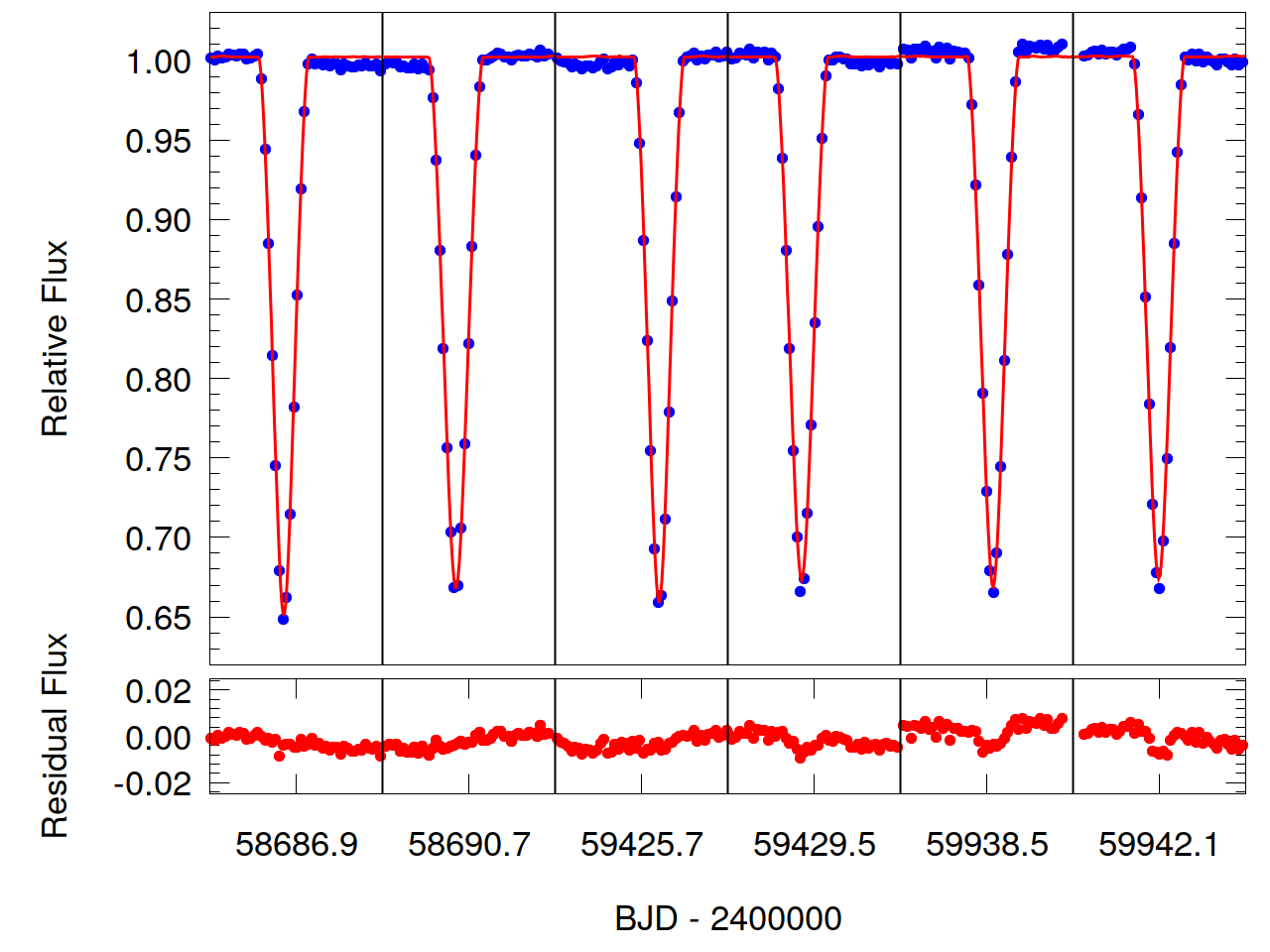}
\caption{Primary and secondary eclipses from the beginning of Year 2 and Year 4, as well as from the end of Year 5 \textit{TESS}-observations of TIC 219885468. Each of the sections is 1\,day long.}
\label{fig:T219885468lcs}
\end{figure}

TIC~219885468, a previously unknown EB, was identified as a likely triple star candidate during our ongoing study of \textit{TESS} EBs in the NCVZ which surveys the ETVs of these objects to identify triple star candidates through their eclipse timing variations (Mitnyan et al., to be submitted soon). The ETVs of this EB display typical timing variations that are dominated by third-body perturbations, including medium-term effects as well as dynamically forced rapid apsidal motion (see bottom row of Fig.~\ref{fig:ETVs}). The primary and secondary eclipses look very similar in depth, which indicates that the inner mass ratio should be near unity--and this does null out the octupole perturbation terms. Moreover, no eclipse depth variations are observed during the $\sim1300$\,days of \textit{TESS} observations, which suggests a strong coplanarity (see, Fig.~\ref{fig:T219885468lcs}).

\begin{figure}
\includegraphics[width=7 cm]{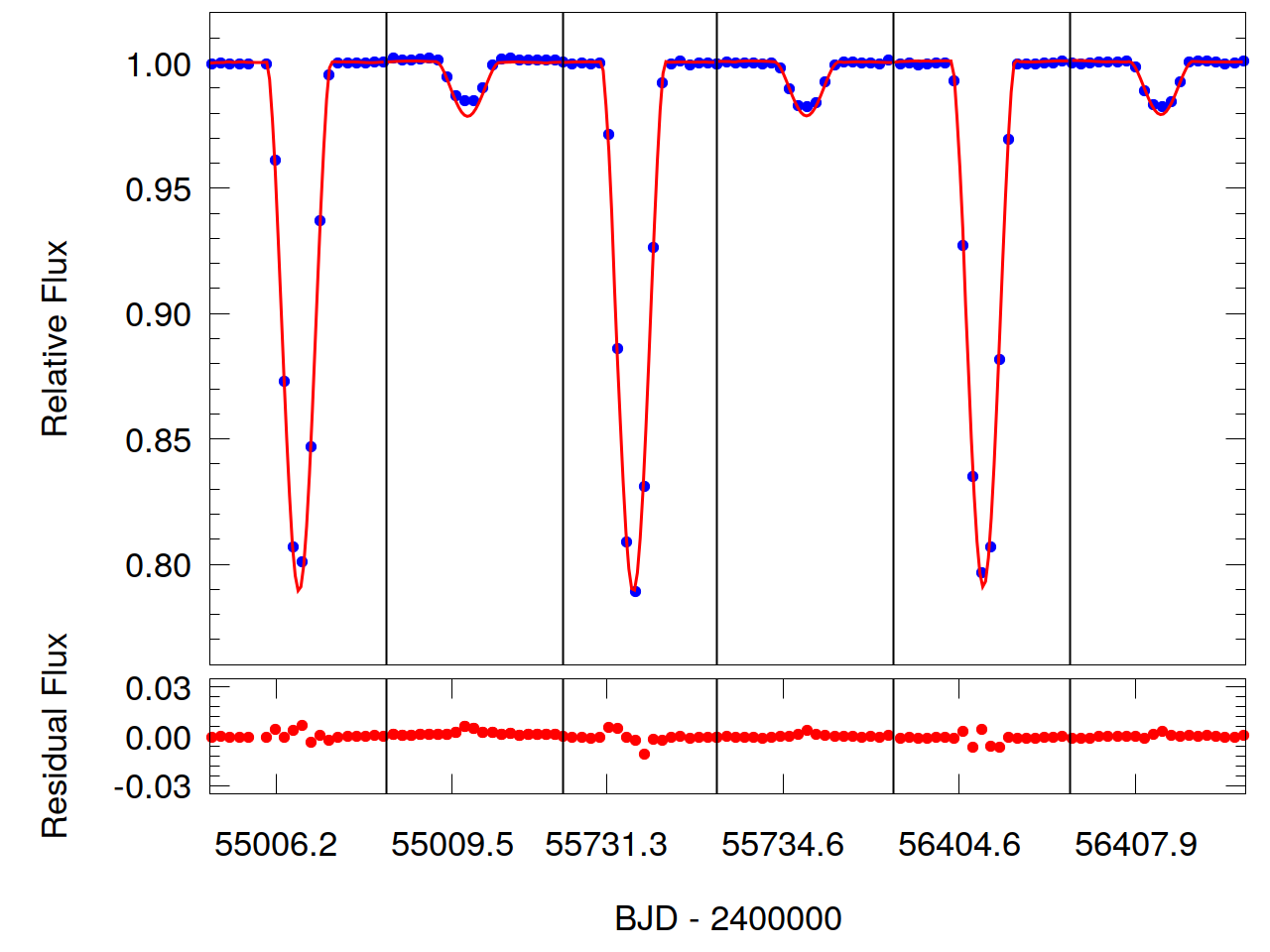}\includegraphics[width=7 cm]{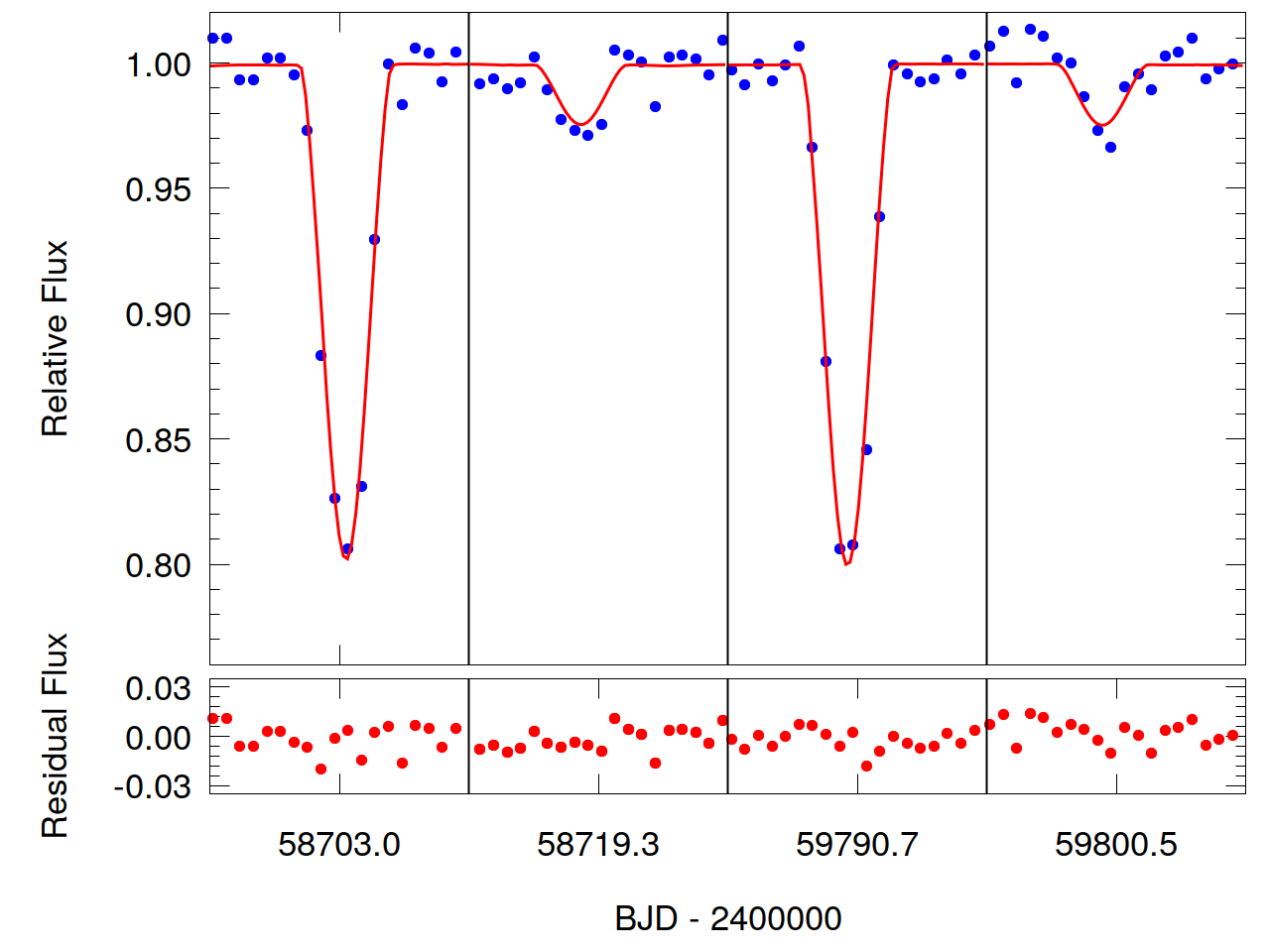}
\caption{Some satellite-observed primary and secondary eclipses of KIC~9714358 (blue dots) with the best-fitting photodynamical model (red curves). {\it Left panel}: Consecutive primary and secondary eclipses from the beginning, mid-time, and near to the end of the prime \textit{Kepler} mission. {\it Right panel:} Characteristic primary and secondary eclipses from the 2019 and 2022 visits of \textit{TESS}. Each of the sections is 0.4\,day long.}
\label{fig:K9714358lcs}
\end{figure}

\section{Observational data, and its preparation for the analysis}
\label{sect:dataprep}

Two of our three targets were observed nearly continuously in long cadence (LC; time resolution of 29.4 min) mode during the prime \textit{Kepler} mission. Furthermore, for KIC~5771589 one month of short cadence (SC; time resolution of 58.9 sec) data in Quarter 4.1 is also available. For our analysis we used those preprocessed LC datasets which can be publicly downloaded from the Villanova website\footnote{\url{http://keplerebs.villanova.edu/}} of the third revision of the Kepler Eclipsing Binary Catalog \citep{kirketal16,abdul-masihetal16}. Quarters 4, 12 and 13 data of KIC~9714358, however, are not available at this site, hence, these were downloaded directly from the MAST database\footnote{\url{https://archive.stsci.edu/missions-and-data/kepler}}.

Regarding the \textit{TESS} observations, we downloaded the FFIs of all three targets for all sectors, when they were observed. (As is well known, Year 2 FFIs were obtained with 30-min cadence time, Year 4 ones with 10 minutes, while Year 5 FFIs -- available only for TIC~219885468 among our targets -- have 200-sec cadence time.) We then processed these FFIs using a convolution-based differential photometric pipeline implemented in the {\sc FITSH} package \citep{pal12}. Finally, we detrended these raw light curves, which were affected by differing amounts of scattered light and other effects, with smoothing polynomials of various degrees to each light curve section.

After the detrending processes, as a first step, we determined mid-eclipse times for each individual eclipse, in the very same manner as was described in \citet{borkovitsetal16}.\footnote{For \textit{Kepler} data, naturally, we did not recalculate the times of minima, as was done already for the \citet{borkovitsetal15,borkovitsetal16} papers.} Then, preparing the dataset for the light curve analysis section of our complex photodynamical treatment, first, we binned the Year 4-5 data into 1800 sec bins for a homogeneous sampling (not only with respect to Year 2 \textit{TESS} data, but also to the \textit{Kepler} LC data). Then, in order to reduce computational costs, and also to give higher statistical weight to the eclipses (which carry most of the relevant astrophysical and dynamical information amongst the light curve points), we dropped out the larger parts of the out-of-eclipse light curves and, kept only small light-curve sections around the primary and secondary eclipses.

In the case of KIC~9714358 we downloaded the 16 heliocentric RV data points, published in APOGEE DR 16 \citep{jonssonetal20}, from the Vizier site\footnote{\url{http://vizier.cfa.harvard.edu/viz-bin/VizieR?-source=III/284}} and also used them.

Finally, note that, similar to our previous works, we included a composite spectral energy distribution (SED) analysis into our joint photodynamical runs (see below) for which we mainly used those catalog magnitudes which are tabulated in Table~\ref{tbl:mags}.

\section{Photodynamical modeling}
\label{sec:dyn_mod}
%=================================================================================

We have carried out complex photodynamical analyses of our three targets with the software package {\sc Lightcurvefactory} \citep[see, e.g.][and references therein]{borkovitsetal19a,borkovitsetal20a}. This code contains (i) a built-in numerical integrator to calculate the gravitationally (three-body effects), tidally, and (optionally) relativistically perturbed Jacobian coordinates and velocities of the three stars in the system; (ii) emulators for multiband light curves, ETV curves, and RV curves, (iii) built-in, tabulated \texttt{PARSEC} grids {\citep{PARSEC} for interpolating fundamental stellar parameters (e.g. radii and effective temperatures) as well as a composite SED model for the three stars, and (iv) an MCMC-based search routine for fitting the parameters. The latter applies our own implementation of the generic Metropolis-Hastings algorithm \citep[see e.g.][]{ford05}. The use of this software package and the consecutive steps of the entire analysis process have been previously explained in detail in several former papers \citep[see, e.~g.][]{borkovitsetal19a,borkovitsetal19b,borkovitsetal20a,borkovitsetal20b}. In this regard we note that, despite the fact that {\sc Lightcurvefactory} has a built-in subroutine for correcting the emulated light curves for finite cadence times \citep[see~][for the description of the process]{borkovitsetal19a}, due to the relatively long cadence times of 1800~sec, we did not applied such corrections. This decision can be justified a posteriori by the fact, that applying cadence time corrections in the case of KIC 9714358, resulted in parameters that remained within the $\pm1\sigma$ domains of the uncorrected results (tabulated in Table~\ref{tab: syntheticfit_KIC9714358}) -- even in the case of the most sensitive parameters (the inclinations).  However, since the analysis runs with the cadence time corrections result in computation times about four-five times longer (on average), we declined to use this feature routinely.

In the absence of an RV curve for each of the three stars in any of our triple systems (or even for two of the stars), we combined a composite SED analysis and the use of precalculated \texttt{PARSEC} grids as proxies to determine the masses of the constituent stars. Naturally, such a photodynamical model solution is no longer astrophysically model-independent.  Nonetheless, in a previous work \citep{borkovitsetal22a} we have shown that for compact, dynamically interacting triple stars (as is the case for the currently investigated systems), such a solution may result in stellar masses and radii within 5-10\% of the results of an astrophysical model-independent solution, the latter of which is based on dynamical masses from  RV data.  The geometric and dynamical parameters (i.e. the orbital elements and also the mass ratios), as well as dimensionless quantities such as the fractional radii and temperature ratios of the stars remain practically unchanged between the two kinds of solutions.

In the case of these model-dependent runs, the adjusted parameters were as follows:
\begin{itemize}
\item[(i)] Stars: Three stellar mass related parameters: the mass of the most massive component (being either $m_\mathrm{Aa}$, the primary of the inner pair or, $m_\mathrm{B}$, the tertiary star), and the inner and outer mass ratios ($q_\mathrm{in,out}$). Moreover, the metallicity of the system ([$M/H$]), the (logarithmic) age of the three coeval stars ($\log\tau$), and the interstellar reddening $E(B-V)$ for the given triple were also varied. Additionally,  the `extra light' contamination, $\ell_4$ parameter was also freely adjusted for two of the three systems.
\item[(ii)] Orbits: Three of six orbital-element related parameters of the inner, and six parameters of the outer orbits, i.e., the eccentricity vector components of both orbits $(e\sin\omega)_\mathrm{in,out}$, $(e\cos\omega)_\mathrm{in,out}$, the inclinations relative to the plane of the sky ($i_\mathrm{in,out}$), and moreover, three other parameters for the outer orbit, including the period ($P_\mathrm{out}$), the longitude of the node relative to the inner binary's node ($\Omega_\mathrm{out}$), and the periastron passage times ($\tau_\mathrm{out}$) were adjusted. 
\end{itemize} 

A couple of other parameters were {\it constrained} instead of being adjusted or held constant during our analyses, as follows:
\begin{itemize}
\item[(i)] Stars: The radii and temperatures of the three stars were calculated with the use of three-linear interpolations from the precomputed 3D (metallicity; logarithmic age; stellar mass) \texttt{PARSEC} grids. Additionally, the distance of the system (which is necessary for the SED fitting) was calculated a posteriori at the end of each trial step, by minimizing the value of $\chi^2_\mathrm{SED}$.
\item[(ii)] Atmospheric parameters of the stars: we handled them in a similar manner as in our previous photodynamical studies. We utilized a logarithmic limb-darkening law \citep{klinglesmithsobieski70} for which the passband-dependent linear and non-linear coefficients were interpolated at each trial step via the tables from the original version of the {\tt Phoebe} software \citep{Phoebe}. We set the gravity darkening exponents for all late type stars to $\beta=0.32$ in accordance with the classic model of \citet{lucy67} valid for convective stars and held them constant. Note, however, that the choice of this parameter has only minor consequences, since the stars in the present study are close to spheroids.
\item[(iii)] Orbits: The orbital period of the inner binary ($P_\mathrm{in}$) and its orbital phase (through the time of an arbitrary primary eclipse or, more strictly, the time of the inferior conjunction of the secondary star -- $\mathcal{T}^\mathrm{inf}_\mathrm{in}$) were constrained internally through the ETV curves. Finally, in the case of KIC~9714358, the systemic radial velocity ($V_\gamma$) was constrained in a similar way, as was done with the distances, described above, i.e. this parameter was calculated a posteriori with a minimization of the value of $\chi^2_\mathrm{RV}$.
\end{itemize} 

 The median values of the orbital and physical parameters, as well as some derived quantities, of the three triple systems, computed from the MCMC posteriors and their $1\sigma$ uncertainties are tabulated in Tables~\ref{tab: syntheticfit_KIC9714358} -- \ref{tab: syntheticfit_TIC219885468}. Furthermore, the observed vs.~model lightcurves are plotted in Figs.~\ref{fig:K9714358lcs}, \ref{fig:K5771589lcs}, \ref{fig:T219885468lcs}, while the observed vs.~model ETV curves are shown in Fig.~\ref{fig:ETVs}. In the case of KIC~9714358 we also plot the APOGEE-2 RV data vs. the best photodynamical model, both in the time and the phase domains in Fig.~\ref{fig:K9714358RVs}.

While the majority of the tabulated parameters are self-explanatory, here we add some notes on the apsidal and nodal motion related parameters. We give the theoretically estimated apsidal motion periods both in the observational ($P_\mathrm{apse}$) and the dynamical ($P_\mathrm{apse}^\mathrm{dyn}$) frame of references. For their calculation we do not use exclusively the theoretical (quadrupole) model of the point-mass third-body perturbations, but the tidal and general relativistic effects are also taken into account. In this regard, we also tabulate the individual contributions of the third-body, tidal and relativistic effects to the apsidal advance rate ($\Delta\omega_\mathrm{3b,tide,GR}$). Moreover, besides the apsidal motion parameters, we also give the theoretical, quadrupole, nodal regression period ($P_\mathrm{node}^\mathrm{dyn}$), i.e., the time needed for a $360^\circ$ regression of the parameters $\Omega_\mathrm{in,out}^\mathrm{dyn}$. Further details of these derivations are discussed in Section 6.2 of \citet{kostovetal21}.

\begin{table}
{\small
 \centering
\caption{Orbital and astrophysical parameters of KIC\,97143586 from the joint photodynamical lightcurve, ETV, RV, SED and \texttt{PARSEC} isochrone solution. Meanings of most of the parameters are explained in the text, with the exceptions of $i_\mathrm{inv}$ and $\Omega_\mathrm{inv}$, which quantities give the position of the invariable plane with respect to the tangent plane of the sky. The osculating orbital elements are given for epoch $t_0=2\,454\,953.0$.}
 \label{tab: syntheticfit_KIC9714358}
\begin{tabular}{@{}llll}
%\hline
% & \multicolumn{3}{c}{KIC\,6964043} \\
\hline
\multicolumn{4}{c}{orbital elements} \\
\hline
   & \multicolumn{3}{c}{subsystem}  \\
   & \multicolumn{2}{c}{Aa--Ab} & A--B  \\
  \hline
 % $t_0$ [BJD - 2400000]& \multicolumn{3}{c}{$54953.0$} \\
  $P$ [days] & \multicolumn{2}{c}{$6.47075_{-0.00019}^{+0.00019}$} & $104.083_{-0.010}^{+0.010}$ \\
  $a$ [R$_\odot$] & \multicolumn{2}{c}{$15.32_{-0.10}^{+0.21}$} & $105.6_{-0.7}^{+1.4}$ \\
  $e$ & \multicolumn{2}{c}{$0.02862_{-0.00034}^{+0.00038}$} & $0.2524_{-0.0020}^{+0.0019}$ \\
  $\omega$ [deg]& \multicolumn{2}{c}{$120.35_{-0.51}^{+0.52}$} & $102.06_{-0.61}^{+0.65}$ \\ 
  $i$ [deg] & \multicolumn{2}{c}{$87.594_{-0.036}^{+0.083}$} & $87.531_{-0.040}^{+0.086}$ \\
  $\mathcal{T}_0^\mathrm{inf}$ [BJD - 2400000]& \multicolumn{2}{c}{$54967.3844_{-0.0004}^{+0.0004}$} &  ... \\
  $\tau$ [BJD - 2400000]& \multicolumn{2}{c}{$54964.7250_{-0.0095}^{+0.0097}$} & $54977.087_{-0.083}^{+0.082}$ \\
  $\Omega$ [deg] & \multicolumn{2}{c}{$0.0$} & $-0.011_{-0.041}^{+0.026}$ \\
  $i_\mathrm{mut}$ [deg] & \multicolumn{3}{c}{$0.074_{-0.026}^{+0.041}$} \\
  $\varpi^\mathrm{dyn}$ [deg]& \multicolumn{2}{c}{$303.4_{-0.5}^{+0.5}$} & $282.1_{-0.6}^{+0.6}$ \\
  $i^\mathrm{dyn}$ [deg] & \multicolumn{2}{c}{$0.055_{-0.020}^{+0.030}$} & $0.019_{-0.007}^{+0.010}$ \\
  $\Omega^\mathrm{dyn}$ [deg] & \multicolumn{2}{c}{$189_{-24}^{+20}$} & $9_{-24}^{+20}$ \\
  $i_\mathrm{inv}$ [deg] & \multicolumn{3}{c}{$87.547_{-0.037}^{+0.083}$} \\
  $\Omega_\mathrm{inv}$ [deg] & \multicolumn{3}{c}{$-0.008_{-0.030}^{+0.019}$} \\
  \hline
  mass ratio $[q=m_\mathrm{sec}/m_\mathrm{pri}]$ & \multicolumn{2}{c}{$0.408_{-0.008}^{+0.004}$} & $0.265_{-0.003}^{+0.003}$ \\
  RV amplitude $K_\mathrm{pri}$ [km\,s$^{-1}$] & \multicolumn{2}{c}{$34.69_{-0.15}^{+0.16}$} & $11.12_{-0.12}^{+0.13}$ \\ 
  RV amplitude $K_\mathrm{sec}$ [km\,s$^{-1}$] & \multicolumn{2}{c}{$85.08_{-0.76}^{+1.73}$} & $41.92_{-0.31}^{+0.59}$ \\ 
  $V_\gamma$ [km\,s$^{-1}$] & \multicolumn{3}{c}{$-66.91_{-0.02}^{+0.02}$} \\
  \hline
  \multicolumn{4}{c}{Apsidal and nodal motion related parameters} \\
  \hline
$P_\mathrm{apse}$ [year] & \multicolumn{2}{c}{$26.43_{-0.20}^{+0.17}$} & $76.84_{-0.47}^{+0.61}$ \\ 
$P_\mathrm{apse}^\mathrm{dyn}$ [year] & \multicolumn{2}{c}{$11.27_{-0.07}^{+0.06}$} & $15.63_{-0.07}^{+0.07}$ \\ 
$P_\mathrm{node}^\mathrm{dyn}$ [year] & \multicolumn{3}{c}{$19.63_{-0.10}^{+0.10}$} \\
$\Delta\omega_\mathrm{3b}$ [arcsec/cycle] & \multicolumn{2}{c}{$2037_{-11}^{+13}$} & $23622_{-104}^{+100}$ \\ 
$\Delta\omega_\mathrm{GR}$ [arcsec/cycle] & \multicolumn{2}{c}{$0.622_{-0.008}^{+0.017}$} & $0.122_{-0.002}^{+0.003}$ \\ 
$\Delta\omega_\mathrm{tide}$ [arcsec/cycle] & \multicolumn{2}{c}{$0.182_{-0.018}^{+0.007}$} & $0.00072_{-0.00007}^{+0.00003}$  \\ 
  \hline  
\multicolumn{4}{c}{stellar parameters} \\
\hline
   & Aa & Ab &  B \\
  \hline
 \multicolumn{4}{c}{Relative quantities} \\
  \hline
 fractional radius [$R/a$]  & $0.0488_{-0.0004}^{+0.0003}$ & $0.0321_{-0.0011}^{+0.0002}$ & $0.00447_{-0.00014}^{+0.00004}$ \\
 temperature relative to $(T_\mathrm{eff})_\mathrm{Aa}$ & $1$ & $0.6118_{-0.0056}^{+0.0079}$ & $0.6040_{-0.0056}^{+0.0075}$ \\
 fractional flux [in \textit{Kepler}-band] & $0.9135_{-0.0293}^{+0.0118}$ & $0.0365_{-0.0059}^{+0.0018}$ & $0.0314_{-0.0040}^{+0.0016}$ \\
 fractional flux [in \textit{TESS}-band] & $0.8934_{-0.0221}^{+0.0201}$ & $0.0435_{-0.0070}^{+0.0018}$ & $0.0353_{-0.0041}^{+0.0018}$ \\
 \hline
 \multicolumn{4}{c}{Physical Quantities} \\
  \hline 
 $m$ [M$_\odot$] & $0.818_{-0.018}^{+0.039}$ & $0.334_{-0.005}^{+0.009}$ & $0.306_{-0.007}^{+0.011}$ \\
 $R$ [R$_\odot$] & $0.746_{-0.006}^{+0.010}$ & $0.491_{-0.009}^{+0.005}$ & $0.471_{-0.010}^{+0.005}$ \\
 $T_\mathrm{eff}$ [K]& $5312_{-118}^{+82}$ & $3237_{-31}^{+56}$ & $3197_{-32}^{+55}$ \\
 $L_\mathrm{bol}$ [L$_\odot$] & $0.397_{-0.036}^{+0.034}$ & $0.023_{-0.001}^{+0.001}$ & $0.021_{-0.002}^{+0.001}$ \\
 $M_\mathrm{bol}$ & $5.77_{-0.09}^{+0.10}$ & $8.84_{-0.05}^{+0.07}$ & $8.99_{-0.04}^{+0.10}$ \\
 $M_V           $ & $5.94_{-0.11}^{+0.14}$ & $11.02_{-0.15}^{+0.21}$ & $11.25_{-0.12}^{+0.25}$ \\
 $\log g$ [dex] & $4.605_{-0.009}^{+0.008}$ & $4.575_{-0.006}^{+0.032}$ & $4.572_{-0.006}^{+0.031}$ \\
 \hline
\multicolumn{4}{c}{Global system parameters} \\
  \hline
$\log$(age) [dex] &\multicolumn{3}{c}{$7.666_{-0.123}^{+0.048}$} \\
$[M/H]$  [dex]    &\multicolumn{3}{c}{$-0.129_{-0.045}^{+0.058}$} \\
$E(B-V)$ [mag]    &\multicolumn{3}{c}{$0.087_{-0.046}^{+0.047}$} \\
extra light $\ell_4$ [in \textit{Kepler}-band] & \multicolumn{3}{c}{$0.017_{-0.012}^{+0.032}$} \\
extra light $\ell_4$ [in \textit{TESS}-band] & \multicolumn{3}{c}{$0.030_{-0.018}^{+0.023}$} \\
$(M_V)_\mathrm{tot}$  &\multicolumn{3}{c}{$5.92_{-0.11}^{+0.14}$} \\
distance [pc]           &\multicolumn{3}{c}{$666_{-11}^{+10}$} \\  
\hline
\end{tabular}}
\end{table}

\begin{table}
{\small
 \centering
\caption{Orbital and astrophysical parameters of KIC\,5771589 from the joint photodynamical lightcurve, ETV, SED and \texttt{PARSEC} isochrone solution. The osculating orbital elements are given for epoch $t_0=2\,454\,953.0$.}
 \label{tab: syntheticfit_KIC5771589}
\begin{tabular}{@{}llll}
%\hline
% & \multicolumn{3}{c}{KIC\,6964043} \\
\hline
\multicolumn{4}{c}{orbital elements} \\
\hline
   & \multicolumn{3}{c}{subsystem}  \\
   & \multicolumn{2}{c}{Aa--Ab} & A--B  \\
  \hline
 % $t_0$ [BJD - 2400000]& \multicolumn{3}{c}{$54953.0$} \\
  $P$ [days] & \multicolumn{2}{c}{$10.6791_{-0.0033}^{+0.0024}$} & $113.872_{-0.020}^{+0.023}$ \\
  $a$ [R$_\odot$] & \multicolumn{2}{c}{$24.20_{-0.10}^{+0.10}$} & $137.5_{-0.6}^{+0.5}$ \\
  $e$ & \multicolumn{2}{c}{$0.00358_{-0.00025}^{+0.00042}$} & $0.1615_{-0.0018}^{+0.0018}$ \\
  $\omega$ [deg]& \multicolumn{2}{c}{$176.3_{-3.4}^{+3.4}$} & $91.12_{-0.99}^{+1.07}$ \\ 
  $i$ [deg] & \multicolumn{2}{c}{$86.048_{-0.085}^{+0.073}$} & $86.154_{-0.084}^{+0.074}$ \\
  $\mathcal{T}_0^\mathrm{inf}$ [BJD - 2400000]& \multicolumn{2}{c}{$54962.0758_{-0.0044}^{+0.0029}$} &  ... \\
  $\tau$ [BJD - 2400000]& \multicolumn{2}{c}{$54959.308_{-0.101}^{+0.101}$} & $54974.152_{-0.174}^{+0.186}$ \\
  $\Omega$ [deg] & \multicolumn{2}{c}{$0.0$} & $0.266_{-0.025}^{+0.026}$ \\
  $i_\mathrm{mut}$ [deg] & \multicolumn{3}{c}{$0.286_{-0.025}^{+0.026}$} \\
  $\varpi^\mathrm{dyn}$ [deg]& \multicolumn{2}{c}{$356.3_{-3.4}^{+3.4}$} & $271.1_{-1.0}^{+1.1}$ \\
  $i^\mathrm{dyn}$ [deg] & \multicolumn{2}{c}{$0.235_{-0.021}^{+0.021}$} & $0.051_{-0.005}^{+0.010}$ \\
  $\Omega^\mathrm{dyn}$ [deg] & \multicolumn{2}{c}{$68.0_{-2.1}^{+2.0}$} & $248.0_{-2.1}^{+2.0}$ \\
  $i_\mathrm{inv}$ [deg] & \multicolumn{3}{c}{$86.135_{-0.084}^{+0.073}$} \\
  $\Omega_\mathrm{inv}$ [deg] & \multicolumn{3}{c}{$0.219_{-0.021}^{+0.021}$} \\
  \hline
  mass ratio $[q=m_\mathrm{sec}/m_\mathrm{pri}]$ & \multicolumn{2}{c}{$0.776_{-0.011}^{+0.010}$} & $0.612_{-0.003}^{+0.003}$ \\
  RV amplitude $K_\mathrm{pri}$ [km\,s$^{-1}$] & \multicolumn{2}{c}{$49.98_{-0.46}^{+0.50}$} & $23.47_{-0.14}^{+0.13}$ \\ 
  RV amplitude $K_\mathrm{sec}$ [km\,s$^{-1}$] & \multicolumn{2}{c}{$64.47_{-0.45}^{+0.53}$} & $38.31_{-0.17}^{+0.19}$ \\ 
%  $V_\gamma$ [km\,s$^{-1}$] & \multicolumn{3}{c}{$-66.91_{-0.02}^{+0.02}$} \\
  \hline
  \multicolumn{4}{c}{Apsidal and nodal motion related parameters} \\
  \hline
$P_\mathrm{apse}$ [year] & \multicolumn{2}{c}{$11.216_{-0.036}^{+0.036}$} & $51.73_{-0.15}^{+0.15}$ \\ 
$P_\mathrm{apse}^\mathrm{dyn}$ [year] & \multicolumn{2}{c}{$5.059_{-0.014}^{+0.015}$} & $7.823_{-0.016}^{+0.018}$ \\ 
$P_\mathrm{node}^\mathrm{dyn}$ [year] & \multicolumn{3}{c}{$9.217_{-0.024}^{+0.023}$} \\
$\Delta\omega_\mathrm{3b}$ [arcsec/cycle] & \multicolumn{2}{c}{$7489_{-23}^{+22}$} & $51651_{-113}^{+101}$ \\ 
$\Delta\omega_\mathrm{GR}$ [arcsec/cycle] & \multicolumn{2}{c}{$0.569_{-0.005}^{+0.005}$} & $0.166_{-0.001}^{+0.001}$ \\ 
$\Delta\omega_\mathrm{tide}$ [arcsec/cycle] & \multicolumn{2}{c}{$0.095_{-0.010}^{+0.014}$} & $0.00054_{-0.00008}^{+0.00010}$  \\ 
  \hline  
\multicolumn{4}{c}{stellar parameters} \\
\hline
   & Aa & Ab &  B \\
  \hline
 \multicolumn{4}{c}{Relative quantities} \\
  \hline
 fractional radius [$R/a$]  & $0.0475_{-0.0012}^{+0.0016}$ & $0.0286_{-0.0003}^{+0.0004}$ & $0.0168_{-0.0020}^{+0.0010}$ \\
 temperature relative to $(T_\mathrm{eff})_\mathrm{Aa}$ & $1$ & $0.8475_{-0.0081}^{+0.0080}$ & $0.9176_{-0.0108}^{+0.0207}$ \\
 fractional flux [in \textit{Kepler}-band] & $0.2230_{-0.0176}^{+0.0200}$ & $0.0419_{-0.0037}^{+0.0046}$ & $0.6397_{-0.0526}^{+0.0467}$ \\
 fractional flux [in \textit{TESS}-band] & $0.2332_{-0.0207}^{+0.0230}$ & $0.0472_{-0.0048}^{+0.0068}$ & $0.6954_{-0.0479}^{+0.0272}$ \\
 \hline
 \multicolumn{4}{c}{Physical Quantities} \\
  \hline 
 $m$ [M$_\odot$] & $0.938_{-0.013}^{+0.014}$ & $0.727_{-0.010}^{+0.013}$ & $1.017_{-0.013}^{+0.013}$ \\
 $R$ [R$_\odot$] & $1.150_{-0.032}^{+0.040}$ & $0.694_{-0.009}^{+0.012}$ & $2.303_{-0.259}^{+0.146}$ \\
 $T_\mathrm{eff}$ [K]& $6260_{-77}^{+42}$ & $5295_{-70}^{+67}$ & $5746_{-42}^{+72}$ \\
 $L_\mathrm{bol}$ [L$_\odot$] & $1.811_{-0.147}^{+0.173}$ & $0.340_{-0.021}^{+0.024}$ & $5.209_{-0.946}^{+0.531}$ \\
 $M_\mathrm{bol}$ & $4.12_{-0.10}^{+0.09}$ & $5.94_{-0.07}^{+0.07}$ & $2.98_{-0.11}^{+0.22}$ \\
 $M_V           $ & $4.17_{-0.10}^{+0.09}$ & $6.12_{-0.09}^{+0.09}$ & $3.08_{-0.10}^{+0.20}$ \\
 $\log g$ [dex] & $4.289_{-0.026}^{+0.020}$ & $4.617_{-0.007}^{+0.006}$ & $3.718_{-0.052}^{+0.111}$ \\
 \hline
\multicolumn{4}{c}{Global system parameters} \\
  \hline
$\log$(age) [dex] &\multicolumn{3}{c}{$9.832_{-0.018}^{+0.020}$} \\
$[M/H]$  [dex]    &\multicolumn{3}{c}{$-0.504_{-0.014}^{+0.138}$} \\
$E(B-V)$ [mag]    &\multicolumn{3}{c}{$0.004_{-0.003}^{+0.010}$} \\
extra light $\ell_4$ [in \textit{Kepler}-band] & \multicolumn{3}{c}{$0.100_{-0.051}^{+0.038}$} \\
extra light $\ell_4$ [in \textit{TESS}-band] & \multicolumn{3}{c}{$0.023_{-0.018}^{+0.025}$} \\
$(M_V)_\mathrm{tot}$  &\multicolumn{3}{c}{$2.70_{-0.09}^{+0.16}$} \\
distance [pc]           &\multicolumn{3}{c}{$709_{-55}^{+33}$} \\  
\hline
\end{tabular}}
\end{table}

\begin{table}
{\small
 \centering
\caption{Orbital and astrophysical parameters of TIC\,219885468 from the joint photodynamical lightcurve, ETV, SED and \texttt{PARSEC} isochrone solution. The osculating orbital elements are given for epoch $t_0=2\,458\,683.0$.}
 \label{tab: syntheticfit_TIC219885468}
\begin{tabular}{@{}llll}
%\hline
% & \multicolumn{3}{c}{KIC\,6964043} \\
\hline
\multicolumn{4}{c}{orbital elements} \\
\hline
   & \multicolumn{3}{c}{subsystem}  \\
   & \multicolumn{2}{c}{Aa--Ab} & A--B  \\
  \hline
 % $t_0$ [BJD - 2400000]& \multicolumn{3}{c}{$54953.0$} \\
  $P$ [days] & \multicolumn{2}{c}{$7.51281_{-0.00043}^{+0.00040}$} & $111.5498_{-0.0071}^{+0.0063}$ \\
  $a$ [R$_\odot$] & \multicolumn{2}{c}{$21.52_{-0.14}^{+0.21}$} & $142.6_{-0.9}^{+1.4}$ \\
  $e$ & \multicolumn{2}{c}{$0.04233_{-0.00058}^{+0.00060}$} & $0.3903_{-0.0007}^{+0.0007}$ \\
  $\omega$ [deg]& \multicolumn{2}{c}{$259.33_{-0.22}^{+0.23}$} & $256.02_{-0.64}^{+0.69}$ \\ 
  $i$ [deg] & \multicolumn{2}{c}{$88.227_{-0.051}^{+0.050}$} & $88.125_{-0.218}^{+0.206}$ \\
  $\mathcal{T}_0^\mathrm{inf}$ [BJD - 2400000]& \multicolumn{2}{c}{$58686.8073_{-0.0004}^{+0.0004}$} &  ... \\
  $\tau$ [BJD - 2400000]& \multicolumn{2}{c}{$54964.7250_{-0.0095}^{+0.0097}$} & $58704.054_{-0.050}^{+0.052}$ \\
  $\Omega$ [deg] & \multicolumn{2}{c}{$0.0$} & $0.073_{-0.192}^{+0.191}$ \\
  $i_\mathrm{mut}$ [deg] & \multicolumn{3}{c}{$0.228_{-0.089}^{+0.164}$} \\
  $\varpi^\mathrm{dyn}$ [deg]& \multicolumn{2}{c}{$79.33_{-0.22}^{+0.23}$} & $76.01_{-0.64}^{+0.69}$ \\
  $i^\mathrm{dyn}$ [deg] & \multicolumn{2}{c}{$0.166_{-0.065}^{+0.119}$} & $0.063_{-0.025}^{+0.045}$ \\
  $\Omega^\mathrm{dyn}$ [deg] & \multicolumn{2}{c}{$145_{-72}^{+59}$} & $325_{-72}^{+59}$ \\
  $i_\mathrm{inv}$ [deg] & \multicolumn{3}{c}{$88.15_{-0.17}^{+0.16}$} \\
  $\Omega_\mathrm{inv}$ [deg] & \multicolumn{3}{c}{$0.053_{-0.14}^{+0.14}$} \\
  \hline
  mass ratio $[q=m_\mathrm{sec}/m_\mathrm{pri}]$ & \multicolumn{2}{c}{$0.988_{-0.017}^{+0.016}$} & $0.319_{-0.002}^{+0.002}$ \\
  RV amplitude $K_\mathrm{pri}$ [km\,s$^{-1}$] & \multicolumn{2}{c}{$72.26_{-1.14}^{+0.90}$} & $17.00_{-0.14}^{+0.19}$ \\ 
  RV amplitude $K_\mathrm{sec}$ [km\,s$^{-1}$] & \multicolumn{2}{c}{$73.15_{-0.83}^{+0.57}$} & $53.23_{-0.36}^{+0.54}$ \\ 
%  $V_\gamma$ [km\,s$^{-1}$] & \multicolumn{3}{c}{$-66.91_{-0.02}^{+0.02}$} \\
  \hline
  \multicolumn{4}{c}{Apsidal and nodal motion related parameters} \\
  \hline
$P_\mathrm{apse}$ [year] & \multicolumn{2}{c}{$19.489_{-0.087}^{+0.087}$} & $51.24_{-0.10}^{+0.10}$ \\ 
$P_\mathrm{apse}^\mathrm{dyn}$ [year] & \multicolumn{2}{c}{$8.178_{-0.033}^{+0.034}$} & $11.051_{-0.038}^{+0.041}$ \\ 
$P_\mathrm{node}^\mathrm{dyn}$ [year] & \multicolumn{3}{c}{$14.090_{-0.068}^{+0.064}$} \\
$\Delta\omega_\mathrm{3b}$ [arcsec/cycle] & \multicolumn{2}{c}{$3258_{-13}^{+13}$} & $35816_{-131}^{+123}$ \\ 
$\Delta\omega_\mathrm{GR}$ [arcsec/cycle] & \multicolumn{2}{c}{$0.911_{-0.012}^{+0.018}$} & $0.214_{-0.003}^{+0.004}$ \\ 
$\Delta\omega_\mathrm{tide}$ [arcsec/cycle] & \multicolumn{2}{c}{$0.997_{-0.027}^{+0.027}$} & $0.0047_{-0.0001}^{+0.0001}$  \\ 
  \hline  
\multicolumn{4}{c}{stellar parameters} \\
\hline
   & Aa & Ab &  B \\
  \hline
 \multicolumn{4}{c}{Relative quantities} \\
  \hline
 fractional radius [$R/a$]  & $0.0651_{-0.0013}^{+0.0015}$ & $0.0633_{-0.0014}^{+0.0013}$ & $0.00487_{-0.00004}^{+0.00007}$ \\
 temperature relative to $(T_\mathrm{eff})_\mathrm{Aa}$ & $1$ & $0.9989_{-0.0016}^{+0.0014}$ & $0.7625_{-0.0051}^{+0.0044}$ \\
 fractional flux [in \textit{TESS}-band] & $0.4915_{-0.0195}^{+0.0233}$ & $0.4628_{-0.0214}^{+0.0202}$ & $0.0449_{-0.0021}^{+0.0025}$ \\
 \hline
 \multicolumn{4}{c}{Physical Quantities} \\
  \hline 
 $m$ [M$_\odot$] & $1.189_{-0.021}^{+0.033}$ & $1.178_{-0.033}^{+0.038}$ & $0.755_{-0.015}^{+0.024}$ \\
 $R$ [R$_\odot$] & $1.407_{-0.032}^{+0.024}$ & $1.367_{-0.042}^{+0.031}$ & $0.694_{-0.009}^{+0.018}$ \\
 $T_\mathrm{eff}$ [K]& $6405_{-36}^{+46}$ & $6399_{-36}^{+41}$ & $4885_{-50}^{+53}$ \\
 $L_\mathrm{bol}$ [L$_\odot$] & $3.009_{-0.174}^{+0.115}$ & $2.814_{-0.212}^{+0.190}$ & $0.246_{-0.016}^{+0.023}$ \\
 $M_\mathrm{bol}$ & $3.57_{-0.04}^{+0.06}$ & $3.65_{-0.07}^{+0.08}$ & $6.29_{-0.10}^{+0.07}$ \\
 $M_V           $ & $3.57_{-0.04}^{+0.06}$ & $3.65_{-0.07}^{+0.09}$ & $6.64_{-0.12}^{+0.10}$ \\
 $\log g$ [dex] & $4.221_{-0.020}^{+0.015}$ & $4.241_{-0.016}^{+0.013}$ & $4.632_{-0.008}^{+0.003}$ \\
 \hline
\multicolumn{4}{c}{Global system parameters} \\
  \hline
$\log$(age) [dex] &\multicolumn{3}{c}{$9.498_{-0.068}^{+0.044}$} \\
$[M/H]$  [dex]    &\multicolumn{3}{c}{$-0.089_{-0.023}^{+0.054}$} \\
$E(B-V)$ [mag]    &\multicolumn{3}{c}{$0.029_{-0.012}^{+0.020}$} \\
$(M_V)_\mathrm{tot}$  &\multicolumn{3}{c}{$2.84_{-0.05}^{+0.04}$} \\
distance [pc]           &\multicolumn{3}{c}{$1152_{-13}^{+22}$} \\  
\hline
\end{tabular}}
\end{table}

%%%%%%%%%%%%%%%%%%%%%%%%%%%%%%%%%%%%%%%%%%
\section{Discussion}
\label{sec:discussion}

In what follows we discuss our results for the three investigated systems individually. Though we are primarily interested in the inferred dynamical behaviour of our triples, first we consider briefly the astrophysical implications of our results and then discuss the dynamics of the systems in detail.

\subsection{KIC~9714358}

In accord with the previous results of \citet{windemuthetal19}, we found this triple to be very young. Our inferred age of $\tau\approx46$\,Myr implies that the two less massive components of the triple must still be in the pre-Main Sequence contraction stage. This conclusion is imposed by the fact that the secondary is too large, while at the same time is too cool relative to the primary. To be more specific, the surface brightness (and, hence, indirectly, the temperature ratio) and the ratio of the stellar radii are strongly constrained through the shape, durations and depth ratios of the primary and secondary eclipses, and our {\sc Lightcurvefactory} code was unable to find any reliable coeval \texttt{PARSEC} isochrone MS solutions for the two stars. Moreover, in the present situation the inner mass ratio ($q_\mathrm{in}=0.406\pm0.006$) is also very robustly constrained dynamically both from the ETV curve (where, however, the inner mass ratio occurs only at the octupole levels of both the medium- and long-term perturbations) and also from the RV curve of the primary component of the EB. Strictly speaking, as is well-known, in a single-lined spectroscopic binary (SB1) the amplitude of the RV curve constrains directly only the spectroscopic mass function, which can easily be written in the following form:
\begin{equation}
f(m_\mathrm{Ab})=m_\mathrm{Aa}\sin^3i\frac{q^3}{(1+q)^2}.
\end{equation}
This coupled with the known (inner) inclination of $i_\mathrm{in}=87.61^\circ\pm0.06^\circ$, and the primary's mass (at least to within a few percent uncertainty, as in the current situation), of $m_\mathrm{Aa}=0.83\pm0.03\,\mathrm{M}_\odot$, uniquely gives the (inner) mass ratio $q_{\rm in} = 0.406 \pm 0.006$. Thus, this very robust mass ratio clearly excludes any post-MS solutions and, hence, insofar as we assume that the triple (or, at least the two inner stars) have formed together at the same time, and without any substantial interactions (in particular, mass exchange) during their prior evolution, we can conclude that this system is at the very beginning of its life.

Comparing our results with former survey results, our SED solution clearly supports a hotter primary component with $T_\mathrm{Aa}=5\,300\pm100$\,K, than the one listed in either the TIC v8.2 catalog ($4\,764\pm109$\,K, see in Table~\ref{tbl:mags}) or inferred from the APOGEE-2 spectra ($4\,807\pm95$\,K, \citealt{jonssonetal20}). In this context some further caution is needed because our inferred (photometric) distance was found to be $d=666\pm11$\,pc, which substantially exceeds the Gaia EDR3 value of $d_\mathrm{EDR3}=577\pm7$\,pc, \citep{bailer-jonesetal21}. One should keep in mind, however, that Gaia EDR3 parallaxes and, hence, the calculated distances have not yet been corrected for multiplicity. Thus, naturally, due to this discrepancy, the astrophysical implications of our results should be considered only with some caution.

On the other hand, these discrepancies do not influence the validity of our dynamical results. Besides the robust inner and outer mass ratios (the latter being $q_\mathrm{out}=0.265\pm0.003$), and the eccentricities of the two orbits ($e_\mathrm{in}=0.0286\pm0.0003$ and $e_\mathrm{out}=0.252\pm0.002$), here we mention the complete absence of any eclipse depth variations not only during the four years of the \textit{Kepler} data, but also during the extended 13-year-long interval of the \textit{Kepler} and \textit{TESS} observations.  This latter point very strongly supports our findings about the nearly exact coplanarity of the inner and outer orbital planes, which was found to be $i_\mathrm{mut}=0.07^\circ\pm0.03^\circ$. (It will be shown in the next subsection that even a few tenths of a degree departure from exact coplanarity may lead to a robust detection of EDVs for such a tight system.) 

The most interesting dynamical feature of the current system is, however, the unusual behaviour of the lines of the apsides of both orbits, which, in our interpretation, is a clear manifestation of the octupole order perturbation effects and, it has clearly observable consequences.

The theoretical observable apsidal motion period of the inner EB of KIC~9714358, according to the quadrupole-level perturbation theory, should be $P_\mathrm{apse,in}^\mathrm{theo}=26.4\pm0.2$\,yr (see the middle of Table~\ref{tab: syntheticfit_KIC9714358}). As one can readily see in the upper left panel of Fig.~\ref{fig:ETVs}, the $\sim$$5\,000$-day-long observational dataset (which is longer than half the duration of the predicted full apsidal cycle) contradicts the quadrupole theory. In the left panel of Fig.~\ref{fig:K9714358elements} we plot the variations of the observable and dynamical arguments of periastron according to our best-fitted solution, extending the numerical integration of {\sc Lightcurvefactory} to the end of the current century. As one can see, the most characteristic effect is that the lines of the apsides of the inner and outer orbits (red and blue curves) revolve with the same speed, resulting in similar inner and outer apsidal motion periods of $P_\mathrm{apse}^\mathrm{meas}\approx78$\,yr (which is quite close to the theoretical quadrupole outer apsidal motion period, $P_\mathrm{apse,out}^\mathrm{theo}=76.8\pm0.6$\,yr).  However, they do so in such a manner that, while the outer orbital ellipse rotates with a constant rate, the major axis (i.e. the apsidal line) of the inner orbit oscillates or, librates, around the direction of the outer apsidal line with a period of $P_\mathrm{apse}^\mathrm{osc}\sim P_\mathrm{apse}^\mathrm{meas}/2.5$ and, with a half-amplitude of $\sim15^\circ$. (In other words, the difference of $\omega_\mathrm{in}-\omega_\mathrm{out}$ oscillates around $0^\circ$). Naturally, the same is true for the dynamical arguments of periastrons with the difference here being that $\omega^\mathrm{dyn}_\mathrm{in}-\omega^\mathrm{dyn}_\mathrm{out}$ oscillates around $180^\circ$, as is shown nicely in the right panel of Fig.~\ref{fig:K9714358elements}.

\begin{figure}
\includegraphics[width=7 cm]{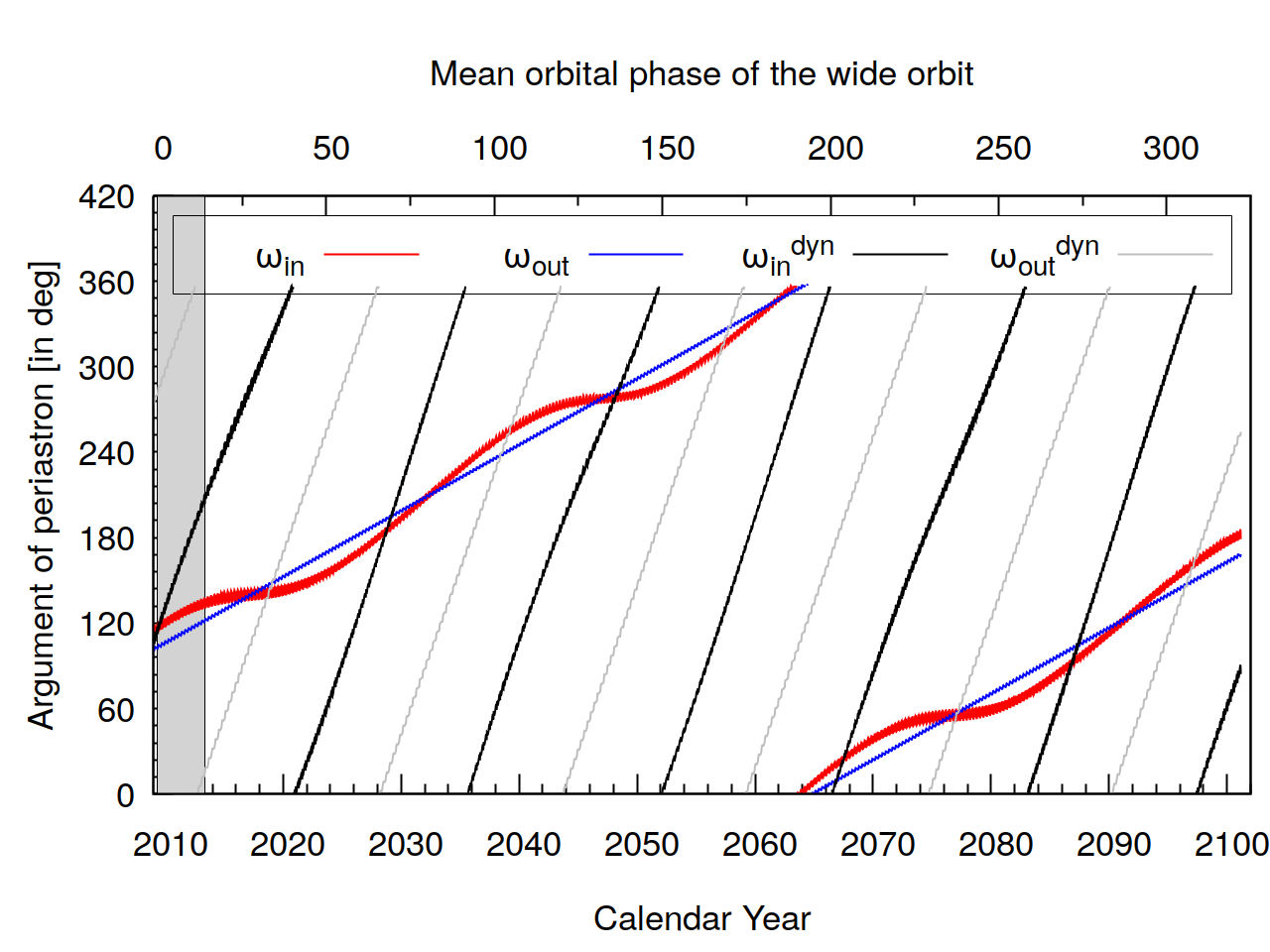}\includegraphics[width=7 cm]{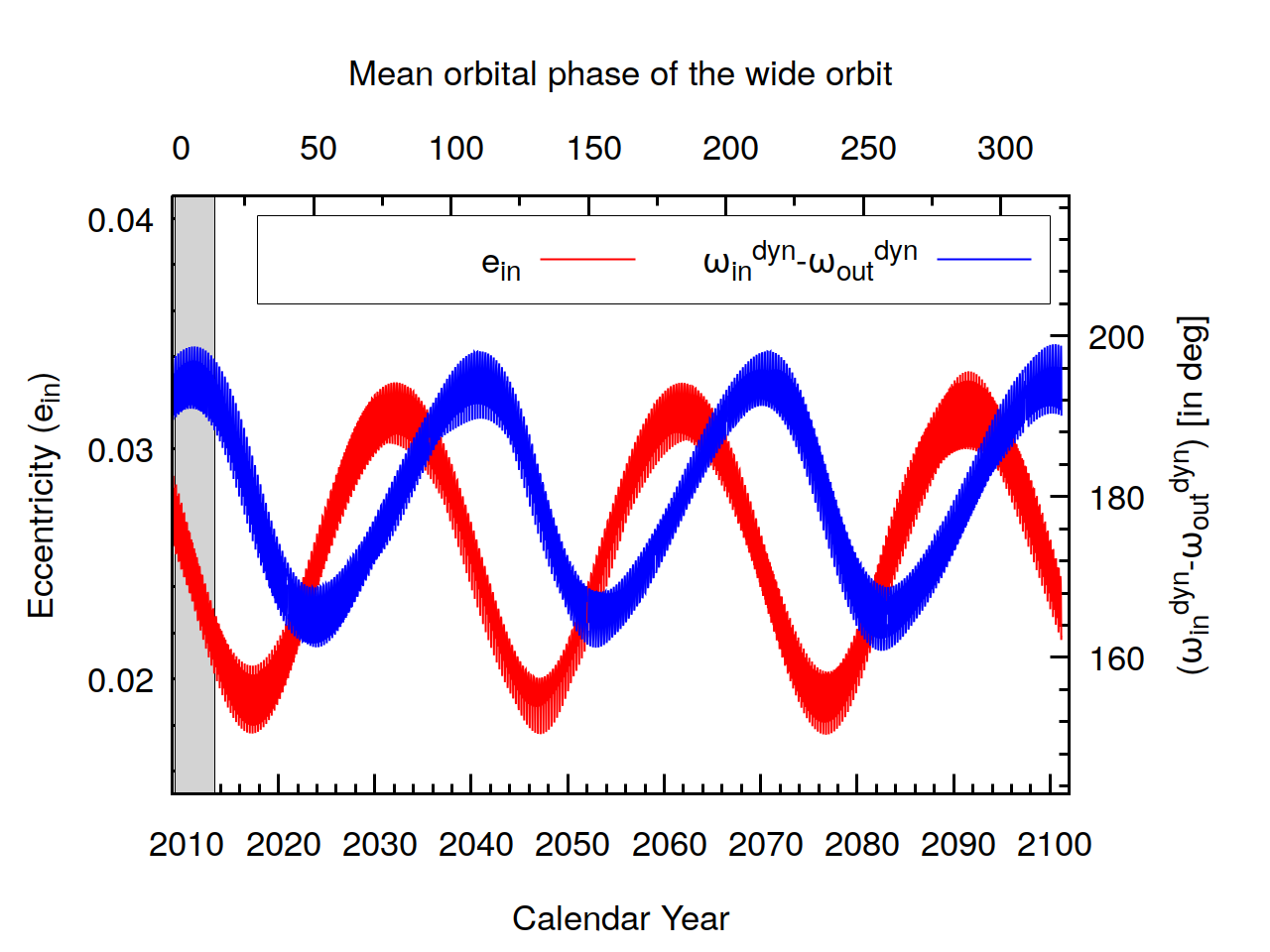}
\caption{The evolution of the orbits in KIC~9714358 from its discovery to the end of the current century. {\it Left panel:} The variations of the observable and the dynamical arguments of periastron of the inner and outer orbits ($\omega_\mathrm{in, out}$ and $\omega^\mathrm{dyn}_\mathrm{in, out}$, respectively). {\it Right panel:} The cyclic variation of the inner eccentricity, and the very similar variations of the difference of the dynamical arguments of periastrons of the two orbits. The gray shaded area represents the interval of the \textit{Kepler} observations. See text for further details.}
\label{fig:K9714358elements}
\end{figure}  

In the right panel of Fig.~\ref{fig:K9714358elements}, besides the difference of $\omega_\mathrm{in}^\mathrm{dyn}-\omega_\mathrm{out}^\mathrm{dyn}$, we also plot the variations of the inner eccentricity ($e_\mathrm{in}$) at the same times. As one can see, the inner eccentricity oscillates with the very same period and, with a $0.25$ phase offset relative to the apsidal oscillations. This is in nice accord with the octupole order long-timescale perturbation equation for the eccentricity (see above, in Eq.~\ref{Eq:dote_in_coplanar}). And, one can conclude that the currently observable uneven variations in the ETV curves of KIC~9714358, i.e. the momentarily rapidly diverging primary and secondary ETV curves, are the direct manifestations of the current growth of the eccentricity of the binary orbit due to the octupole-order terms. 

At this point, however, one should note an important caveat. As was mentioned above, our analysis has shown that this triple system is quite young, and its components (or, at least, the less massive ones) are most probably in pre-MS stages. Such components may potentially exhibit strong spot activity and may rotate non-synchronously (and perhaps even in a non-aligned way). Both effects can affect the ETV curves either indirectly, through the spot-induced light curve distortions which may mimic rapid quasi-periodic and anticorrelated variations in the primary and secondary ETV curves \citep{kalimerisetal02,tranetal13,balajietal15} or, directly influencing the apsidal motion rate of the inner binary \citep{shakura985,hegedusnuspl986}. In the current situation, however, we are convinced that the peculiar behaviour of the ETVs cannot be explained with these latter effects. First of all, in all the known cases, the time-scales of the quasi-periodic spot-induced variations range from a few days to a few months \citep[see, again, the plots in][]{tranetal13,balajietal15} and, moreover, their amplitudes do not exceed 5-10 minutes. Moreover, as these ETVs are virtual and caused only by light curve distortions which affect the determinations of the mid-eclipse times, these distortions would appear in the residuals of our model light curves and, hence, may become directly detectable.\footnote{At this point, naturally, we are talking about the residuals of the eclipsing sections of the observed vs model light curves, as the majority of the non-eclipse light curve sections were dropped out from our analysis.} On the other hand, regarding the possibility of the non-synchronous rotation, this may actually affect the apsidal motion rate, and, hence, the apsidal motion period, but would not change the eccentricity of the inner pair. It is the latter which is the main source of that bump in the ETVs, which now looks evident in the \textit{TESS} measured eclipse times of KIC~9714358.

Here we note, that KIC~9714358 is scheduled to be reobserved by \textit{TESS} in Sectors 74, 75 (January -- February, 2024) and 81 (July/August, 2024), when the offset between the primary and secondary eclipses is expected to be near the next maximum (see upper right panel of Fig.~\ref{fig:ETVs}).  Thus, these findings and the corresponding predictions of our model will be verifiable very soon.

\begin{figure}
\includegraphics[width=5.5 cm]{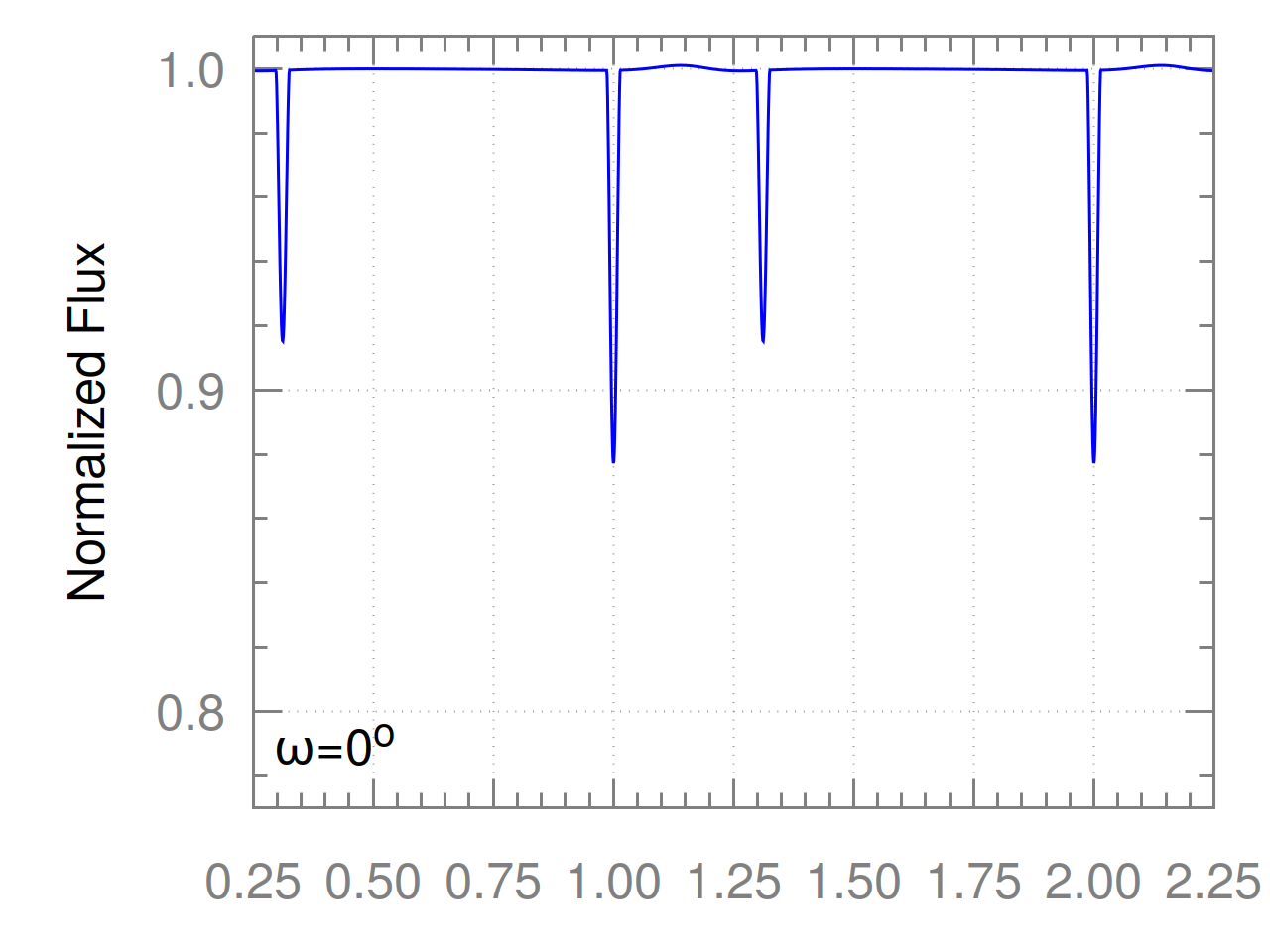}\includegraphics[width=5.5 cm]{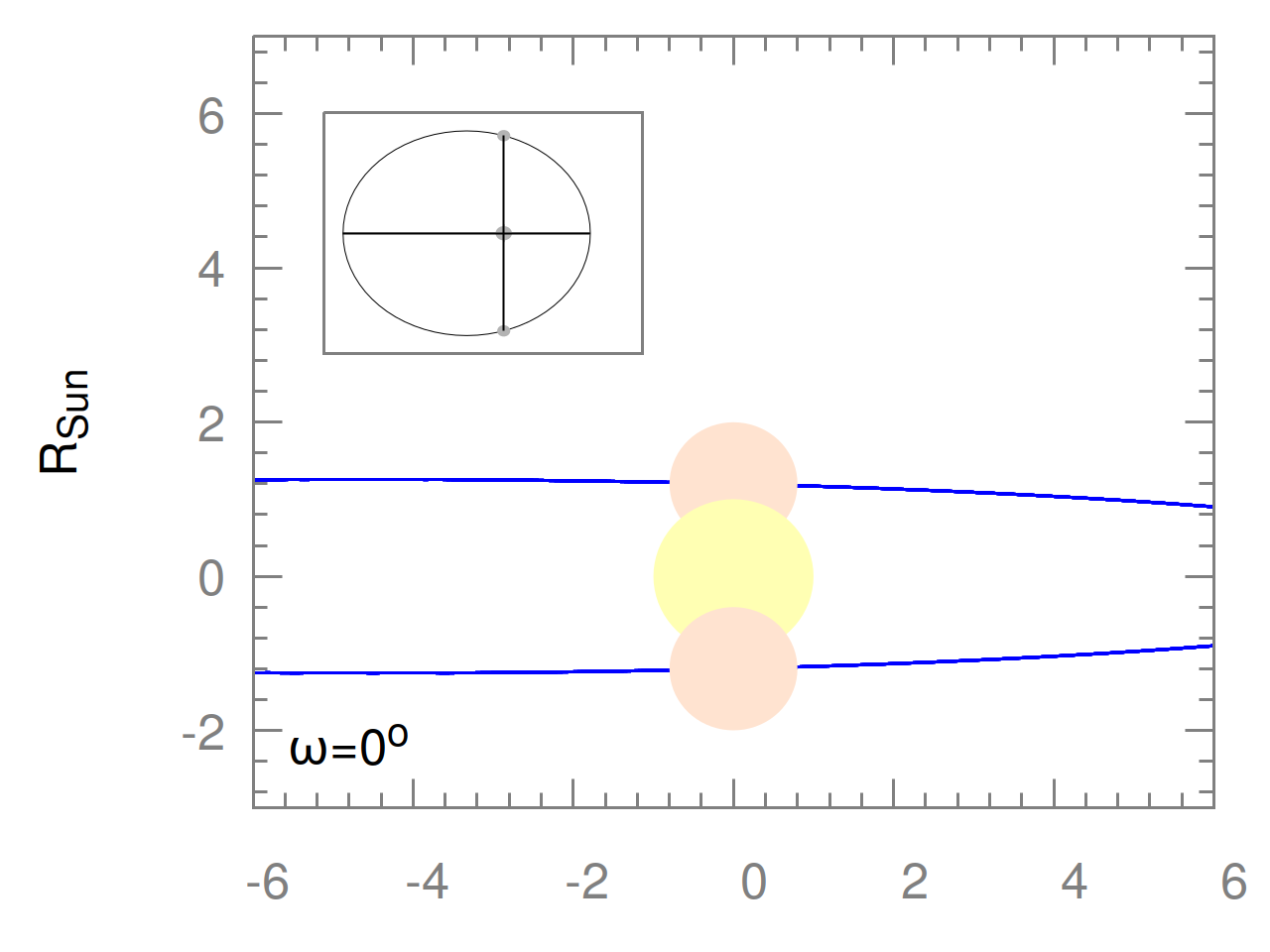}
\includegraphics[width=5.5 cm]{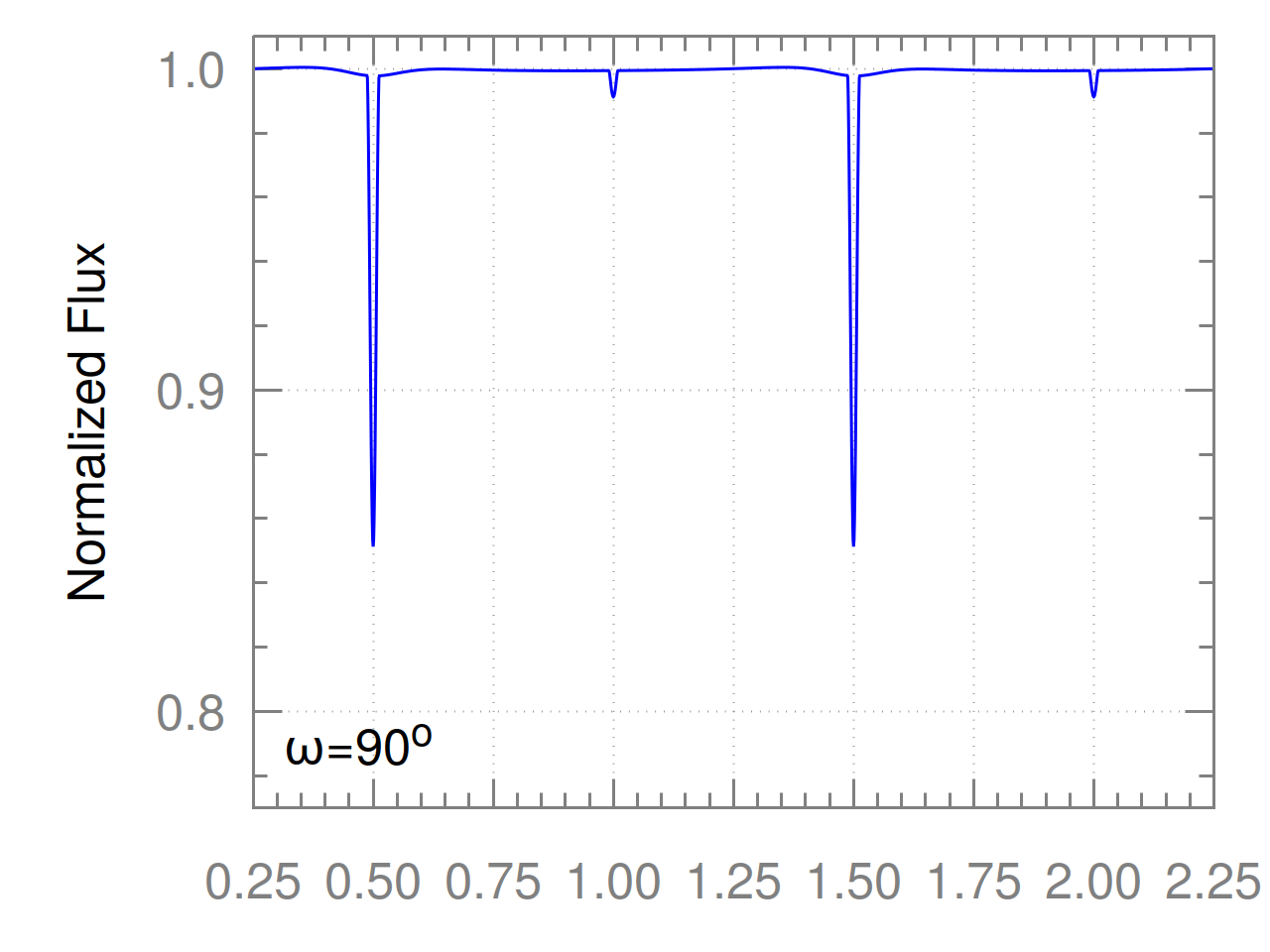}\includegraphics[width=5.5 cm]{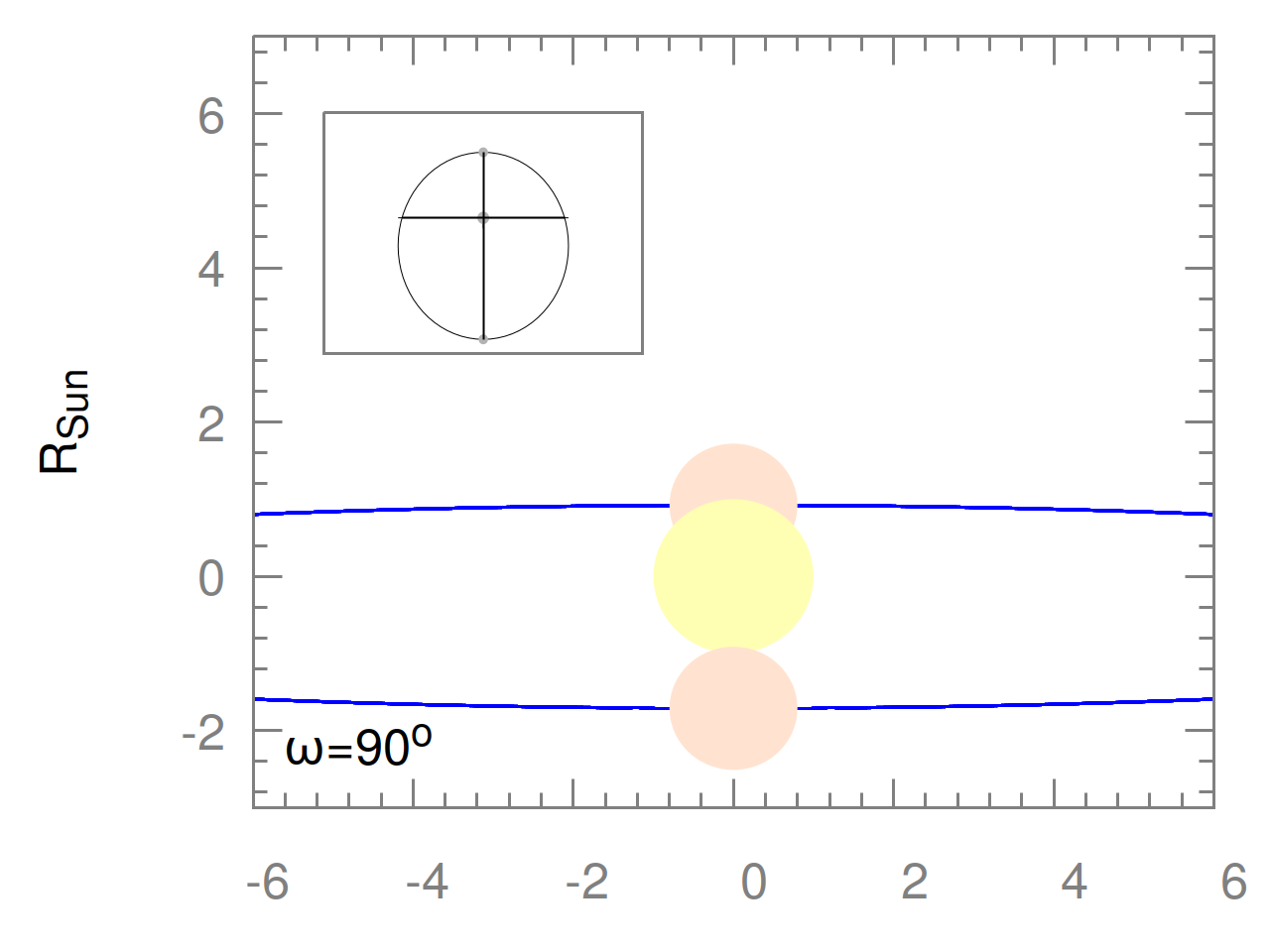}
\includegraphics[width=5.5 cm]{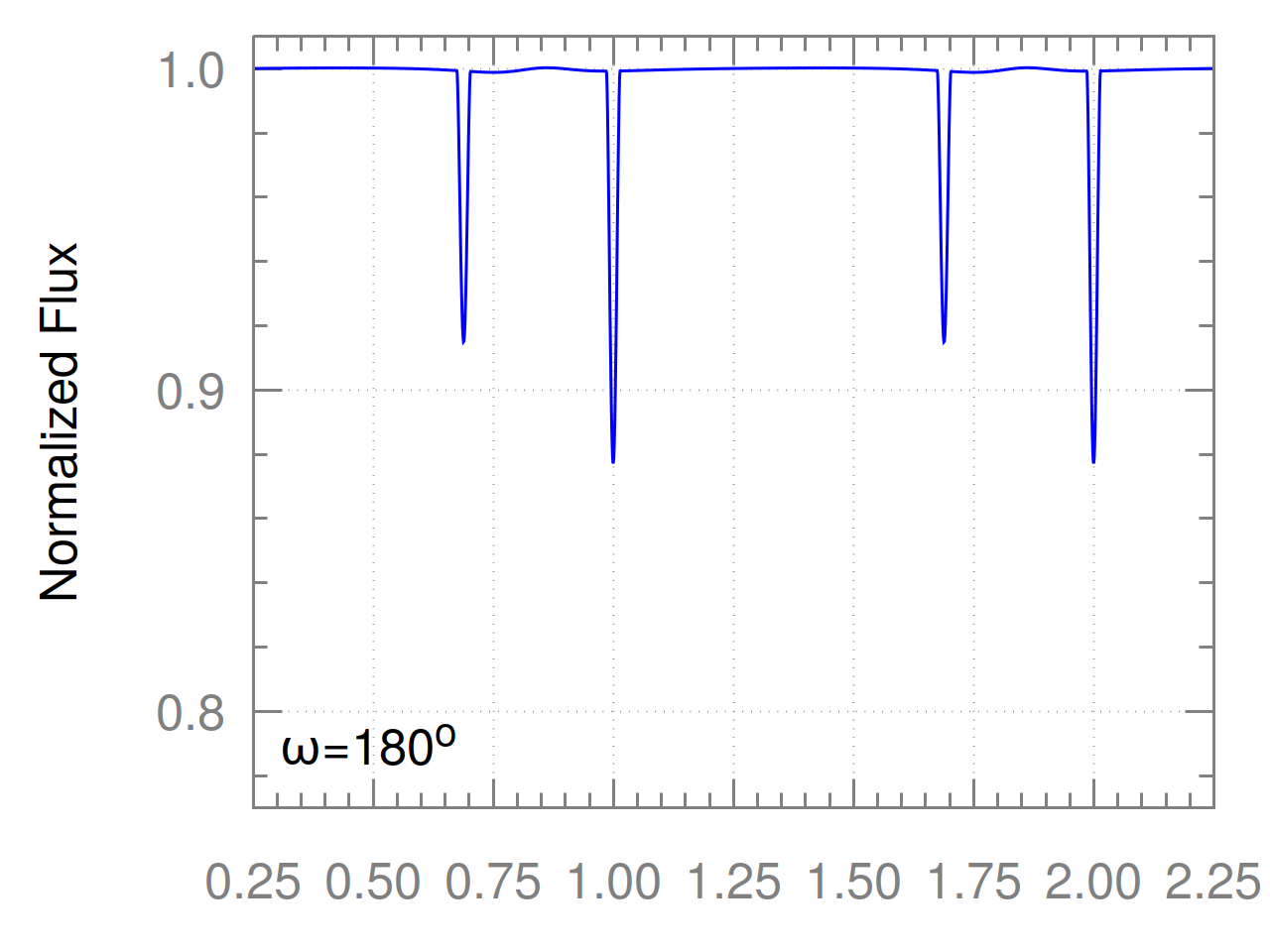}\includegraphics[width=5.5 cm]{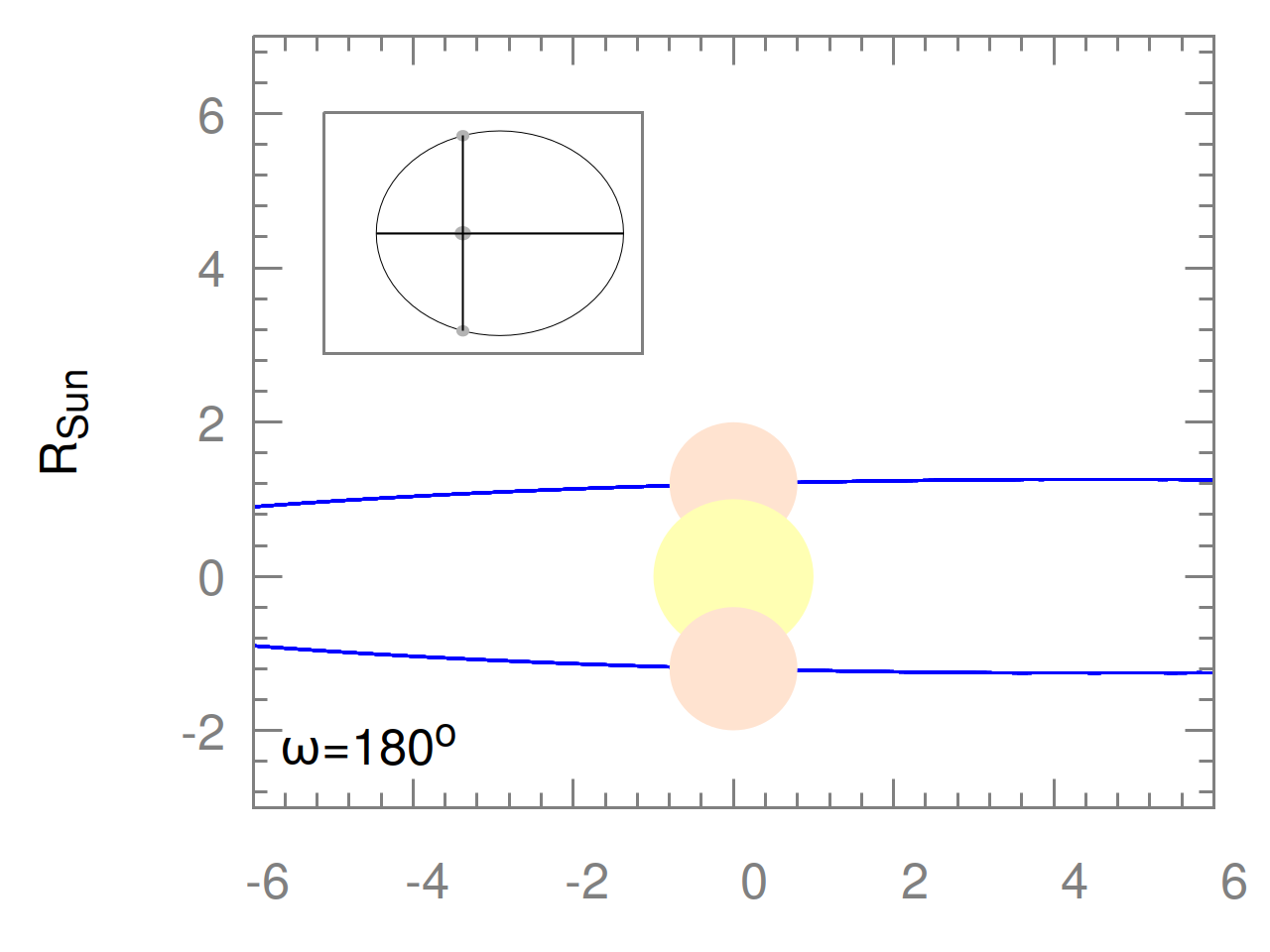}
\includegraphics[width=5.5 cm]{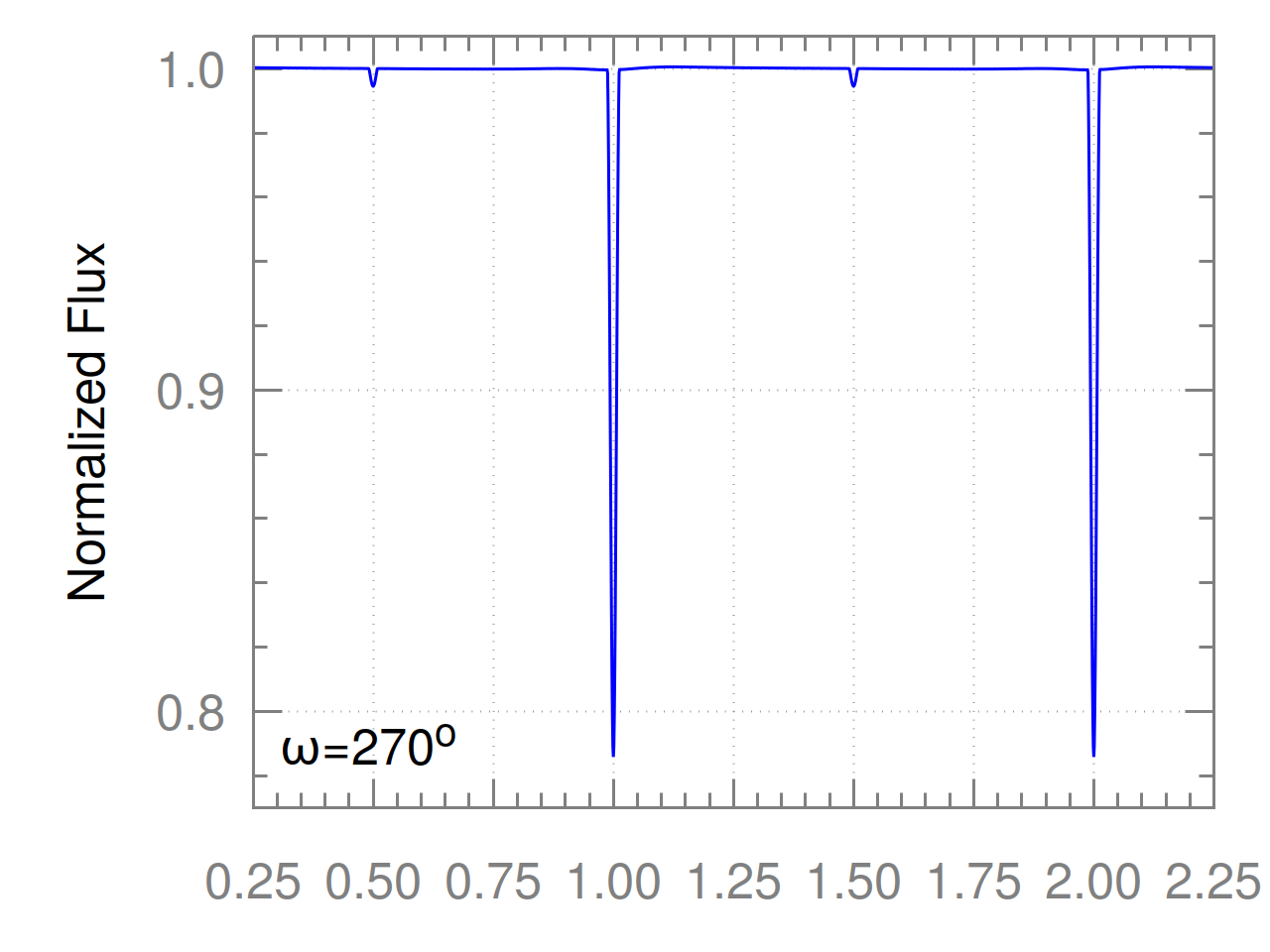}\includegraphics[width=5.5 cm]{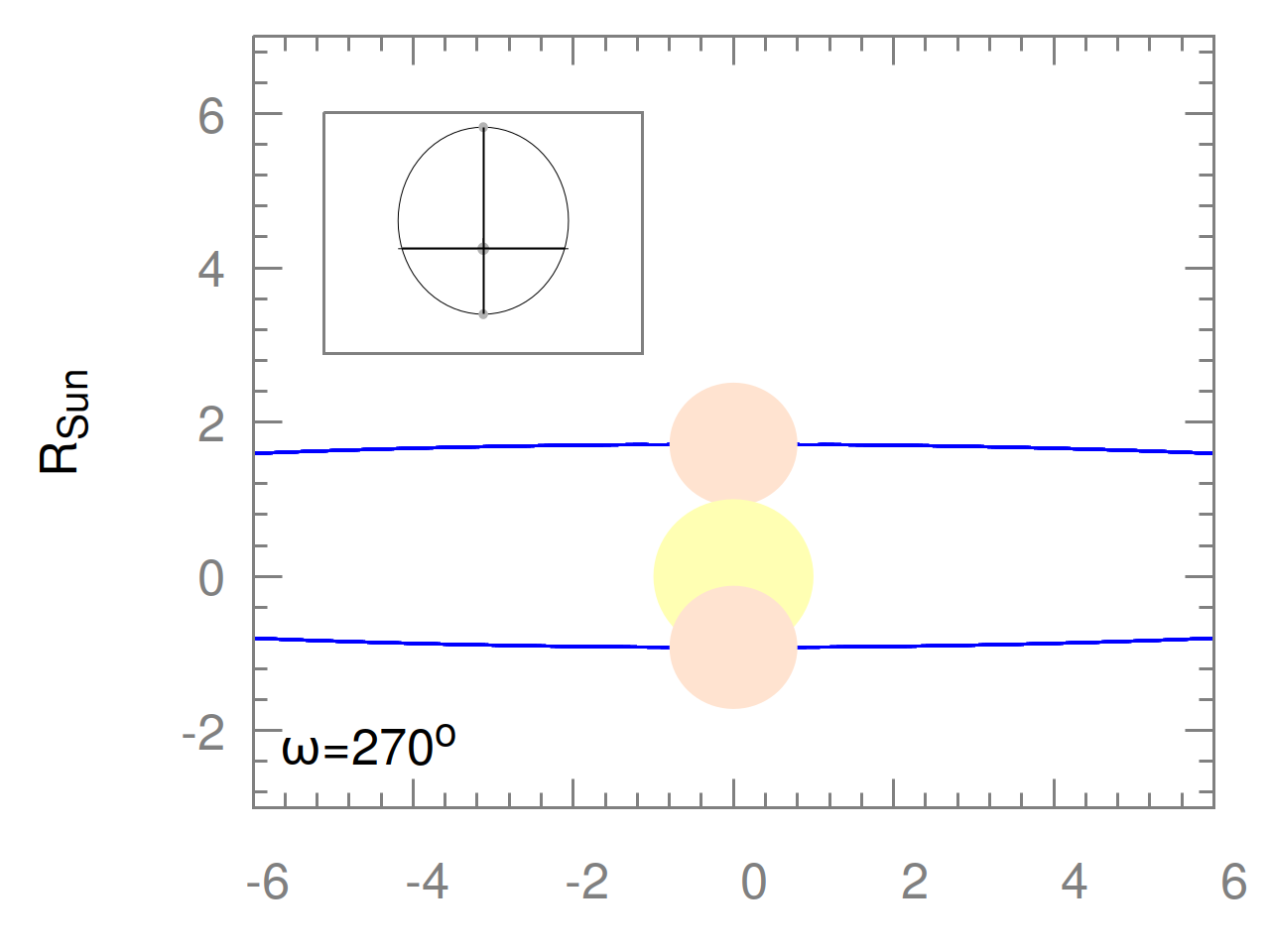}
\caption{Illustration of the effect of the orientation of the orbit on the properties of the eclipses. For illustration purposes we generated light curves for a hypothetical binary with parameters: $m_\mathrm{Aa}=1.00\,\mathrm{M}_\odot$, $m_\mathrm{Ab}=0.80\,\mathrm{M}_\odot$; $R_\mathrm{Aa}=1.03\,\mathrm{R}_\odot$, $m_\mathrm{Ab}=0.799\,\mathrm{R}_\odot$;  $T_\mathrm{Aa}=5878$\,K, $T_\mathrm{Ab}=5274$\,K; $P=5.00$\,d; $e=0.30$; $i=85^\circ$. The only difference amongst the four rows is the orientation of the apsidal line of the ellipse, which are $\omega=0^\circ$, $90^\circ$, $180^\circ$ and $270^\circ$, respectively. The left panels represent the synthetic light curves, which in all four cases are phased to the inferior conjunction of the less massive secondary, while the right panels show the corresponding geometry, as seen from the Earth, and also as it would be seen from the pole of the orbit (small, inserted figures). (In the case of the insert plots, the direction of the Earthly observer is along the negative Y axis. Note also that the secondary revolves counter-clockwise in all the orbit plots.)  As one can see on the light curve plots, the depth ratio of the primary and secondary eclipses depend strongly on the orientation of the ellipse. When the ellipse seen along the minor axis ($\omega=0^\circ$, $180^\circ$) the eclipsed disk areas are similar during the inferior and superior conjunctions and, hence, the ratio of the eclipse depths -- similar to the circular orbit case -- primarily depend only upon the ratio of the surface brightnesses. On the other hand, when the ellipse is seen along its major axis ($\omega=90^\circ$ and $270^\circ$), the eclipsed surface area is much larger during the periastron-eclipse than during the apastron-eclipse. In the current illustration, due to the relatively large eccentricity of $e=0.3$, the difference in the eclipsed surface areas is so large that the eclipse depths reverse.}
\label{fig:apsis_illustration}
\end{figure}  
     
\begin{figure}
\includegraphics[width=7 cm]{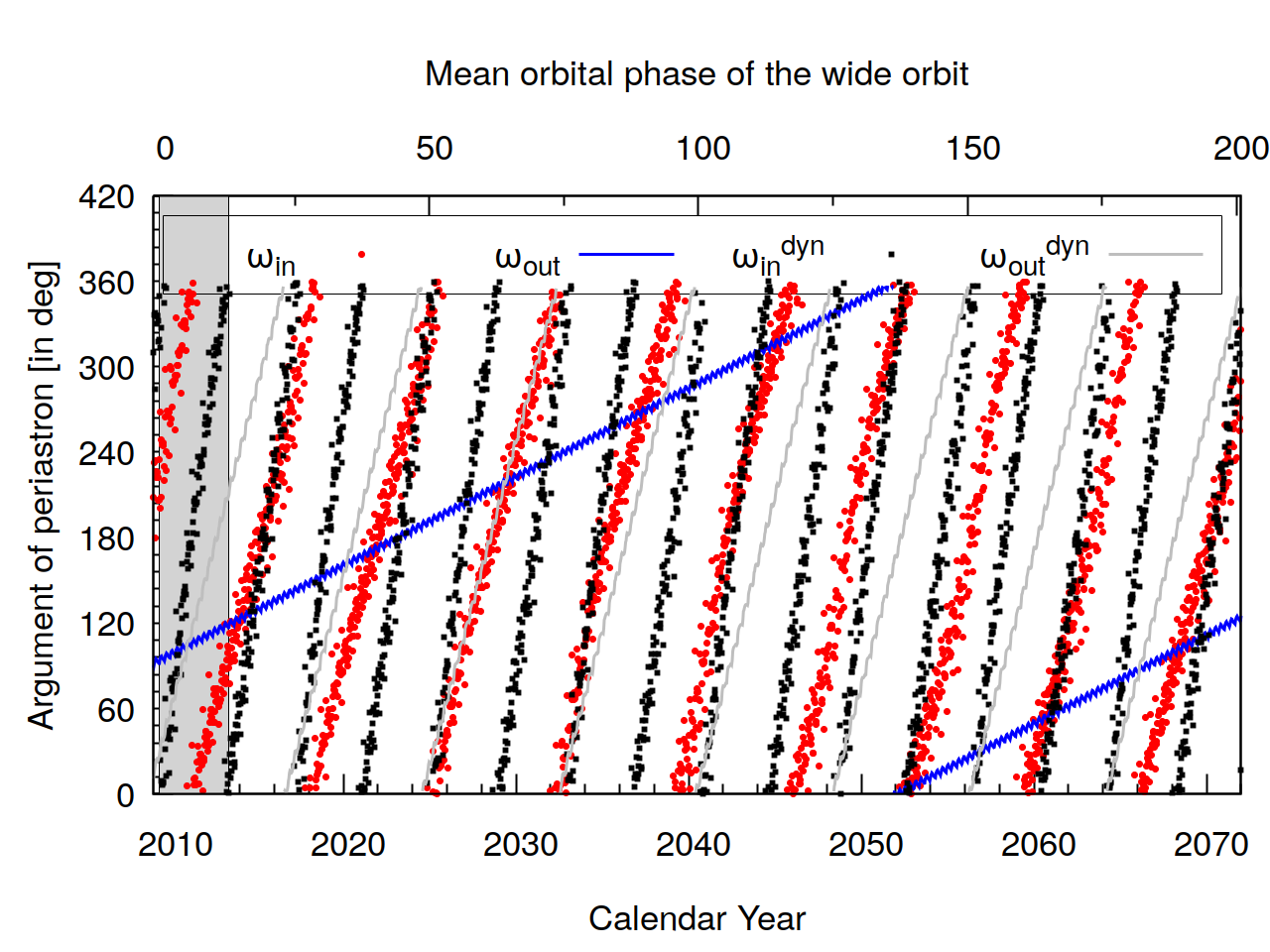}\includegraphics[width=7 cm]{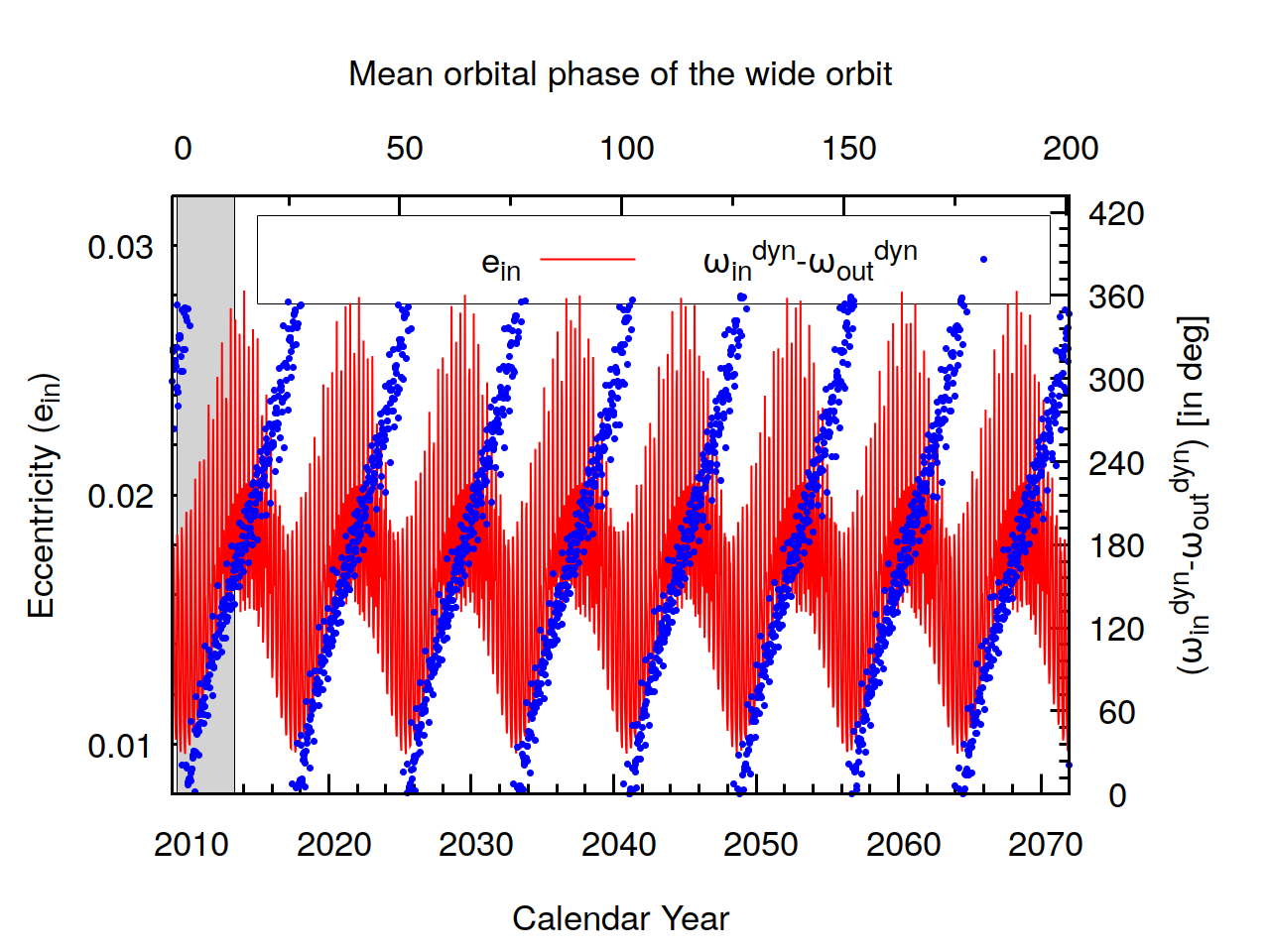}
\caption{The same plots as in Fig.~\ref{fig:K9714358elements}, but for KIC~5771589. See text for further details.}
\label{fig:K5771589elements}
\end{figure}  

\subsection{KIC~5771589}

As mentioned above in Sect.~\ref{sect:selection}, the two noteworthy observables of this triple system are the very rapid apsidal motion and the very quickly varying eclipse depths. Despite our best efforts, we were unable to model the variations of the eclipse depths perfectly. In general, as one can see in  Fig.~\ref{fig:K5771589lcs}, {\sc Lightcurvefactory} returns the main properties of the EDVs in the case of both the \textit{Kepler} and the \textit{TESS} data. The problem, however, shows up in the second half of the \textit{Kepler} data where our models fail to reproduce the correct depth ratios of the secondary vs primary eclipses. In other words, while the \textit{Kepler} observations suggest that the ratio of the depths of the secondary and primary eclipses remains near constant during the four years of the observations, our model fits tend to equalize the depths of the two kinds of eclipses toward the end of the \textit{Kepler} observations. 

To understand the origin of this problem, one should keep in mind that for an eccentric ($e\neq0$) EB viewed not exactly edge-on ($i\neq90^\circ$}), the ratio of the eclipse depths, in addition to being a function of the ratio of the surface brightnesses, also depends strongly upon the position of the observable argument of periastron. This is so because, for the eclipse that occurs closer to periastron, the projected separation of the two stellar disks will be smaller than for the other kind of eclipse (closer to the apastron). Thus, a larger portion of the surface of the eclipsed star will be occulted. Naturally, the more eccentric the orbit, and the more grazing the eclipse, the more significant this effect may be. And, naturally, one can reach a situation where the eclipse closer to apastron disappears entirely, as was observed recently e.g., in the case of KIC~5731312 \citep{borkovitsetal22b}. Note, we illustrate graphically these effects for a hypothetical eccentric EB in Fig.~\ref{fig:apsis_illustration}.

In the case of the current system, both the position of the argument of periastron and the inclination have varied during the four years of \textit{Kepler} observations. As we mentioned above, but it can also be seen directly from the ETV curve (middle left panel of Fig.~\ref{fig:ETVs}), the inner orbital ellipse made a bit more than half revolution during the \textit{Kepler} observations.  According to our photodynamical model, the observable argument of periastron attained a value of ($\omega=270^\circ$) around the very beginning of the Year 2010 (circa 2\,455\,200, see also the left panel of Fig.~\ref{fig:K5771589elements}). This means that during that time, the primary eclipses occurred during the periastron passages, while the secondaries  occurred at the apastron points, thereby causing the largest differences in the sizes of the eclipsed areas (bottom row of Fig.~\ref{fig:apsis_illustration}). Then, toward the end of the \textit{Kepler} era, the situation reversed. During the last days of the perfectly operating three gyroscopes on the \textit{Kepler} spacecraft, the observable argument of periastron reached $\omega=90^\circ$, i.e., the last \textit{Kepler}-observed primary eclipses occurred around the apastron points, while the secondaries were near periastron (second row of Fig.~\ref{fig:apsis_illustration}). This is the reason why the secondary eclipse depths tend to be much closer to the primary ones by the end of the \textit{Kepler} data in our photodynamical models. We note, however, that the same argument can be made purely from the observations, without the use of our photodynamical model results and, hence, one cannot claim this discrepancy as a failure of the model. This is so, because the ETV curves show in themselves that the system was seen from the direction of the major axis at the very beginning and the end of the \textit{Kepler} observations. (This can be deduced from the fact, that the primary and secondary curves intersect each other only when the orbital ellipse is seen from the direction of the major axis, i.e., when one of the eclipses occurs at periastron, while the other at apastron.) Then, the fact, that the secondary (blue) ETV curve is located below the primary (red) one during the entire \textit{Kepler} observations reveals that during this interval, the secondary eclipses were ``in a hurry'' in contrast to the primary ones (or, in other words, the secondary eclipses occurred before photometric phase of 0.5). This means that the periastron passage happened in between the primary and the secondary eclipses (i.e., $\omega$ had values in between $270^\circ$ [or, $-90^\circ$] and $90^\circ$, reaching $\omega=0^\circ$ at the mid-time  -- see the top row of Fig.~\ref{fig:apsis_illustration}). But, even if one would assume retrograde apsidal motion, which would mean that the role of the apastrons and periastrons were exchanged, this cannot explain why the observed ratios of the secondary to primary eclipse depths would have immunity to the apsidal revolution. 

In our understanding, the only possibility for explaining the near constant depth ratios during half an apsidal cycle is that the true observable inclination of the system should be closer to $i=90^\circ$ than our findings of $i_\mathrm{in}=86.05^\circ\pm0.09^\circ$ would suggest (since the more edge-on an eccentric EB is viewed, the smaller the apsidal-phase-dependent difference of the eclipsed disk areas and, for orbits viewed perfectly edge-on, this effect entirely disappears). This fact, however, would require more contaminating light in the system. The main possibilities to fulfill this requirement are as follows: (i) the system has a fourth, more distant, undetected component; (ii) the tertiary is more luminous than would be inferred from the \texttt{PARSEC} stellar isochrones that we utilized or, (iii) the outer mass ratio ($q_\mathrm{out}$) could be larger and, hence, the tertiary would be more massive (and, consequently brighter) relative to the binary members than our solution suggests. Regarding this last possibility, we made efforts to find acceptable solution with higher $q_\mathrm{out}$ but all of our trials failed. So, in what follows, we discuss our findings in the imperfect case, where the  (non-)variation of the eclipse depth ratios is imperfectly modelled. Despite this fact, however, we believe that our findings, to be discussed below, are basically correct.

According to our joint photodynamical solution, in KIC~5771589 the most massive star is the third, distant component, with $m_\mathrm{B}=1.02\pm0.01\,\mathrm{M}_\odot$, though, the primary of the inner binary is only slightly less massive ($m_\mathrm{Aa}=0.94\pm0.01\,\mathrm{M}_\odot$), while its binary companion has a lower mass of $m_\mathrm{Ab}=0.73\pm0.01\,\mathrm{M}_\odot$. Here we note that, despite the lack of RV data, our solutions yield masses with surprisingly small uncertainties of only $1-2\%$. Due to the problematic nature of the light curve solution discussed above, however, we are not convinced that these small uncertainties are realistic. The quantity which is certainly more robust is the outer mass ratio, being $q_\mathrm{out}=0.612\pm0.003$. This value is substantially lower than what was inferred from the previous analytic ETV fits of \citet{borkovitsetal15,borkovitsetal16}. One should keep in mind, however, that those former fits were based only on the \textit{Kepler} data (long before the launch of the \textit{TESS} space telescope) and, because of the extremely tight nature of the current system, as the authors noted, the analytical ETV models they used were somewhat inadequate.

Our solution suggests a metal deficient ($[M/H]\approx-0.50$), old ($\log\tau=9.83\pm0.02$), and slightly evolved system. According to the accepted solution, the massive tertiary is clearly on its way toward the giant branch, having $\log g_\mathrm{B}=3.7\pm0.1$. One should keep in mind, however, that in the absence of third-body eclipses, the direct effects of the tertiary on the joint, complex solutions (apart from the SED fitting) manifest themselves only in the outer mass ratio (via timings) and the contaminated light (via the eclipse depths). Hence, any further statements about the astrophysical condition of the tertiary depend strongly upon the pre-assumption that the whole system formed and evolved in a coeval manner. Our solution gives a distance of $d=709_{-55}^{+33}$\,pc, which differs substantially from the Gaia EDR3-based value of $d_\mathrm{EDR3}=870\pm100$\,pc. This fact might again suggest that the system could have an additional, fourth stellar component (not considered in the photodynamical solution), and the high value of the RUWE parameter (11.34) indeed indicates that the Gaia astrometric solution is probably significantly influenced by the multiplicity of the system. 

Turning to the readily observable long-term or, secular perturbations in this triple, the rapid, dynamically forced apsidal motion is evident. The doubly averaged secular theory predicts apsidal motion periods of the inner and outer orbits to be ($P_\mathrm{apse}=11.22\pm0.04$ and $51.7\pm0.2$\,yrs, respectively). From the fact that the $\sim4$-yr-long \textit{Kepler} dataset covers a bit more than half of an apsidal cycle, one can infer directly that the quadrupole model, again, gives an incorrect result for this system. This can also be seen in the left panel of Fig.~\ref{fig:K5771589elements} which shows that the true apsidal motion period is about $P_\mathrm{apse}^\mathrm{meas}\approx6.3$\,yr. On the other hand, however, note that the outer apsidal motion period according to our numerical integration is $P_\mathrm{apse}^\mathrm{meas}\approx58.1$\,yr which is, again, in much better agreement with the theoretical quadrupole value.

The significance of the octupole order perturbations in the variations of the inner eccentricity is, again, nicely demonstrated in the right panel of Fig.~\ref{fig:K5771589elements}. As one can see, the cyclic variations of $e_\mathrm{in}$ in between $\sim0.01$ and $\sim0.02$ have exactly the same period as for the quantity $\omega_\mathrm{in}^\mathrm{dyn}-\omega_\mathrm{out}^\mathrm{dyn}$, which, in the current system varies between $0^\circ$ and $360^\circ$ with a period of $\sim8$\,yrs.

\begin{figure}
\includegraphics[width=7cm]{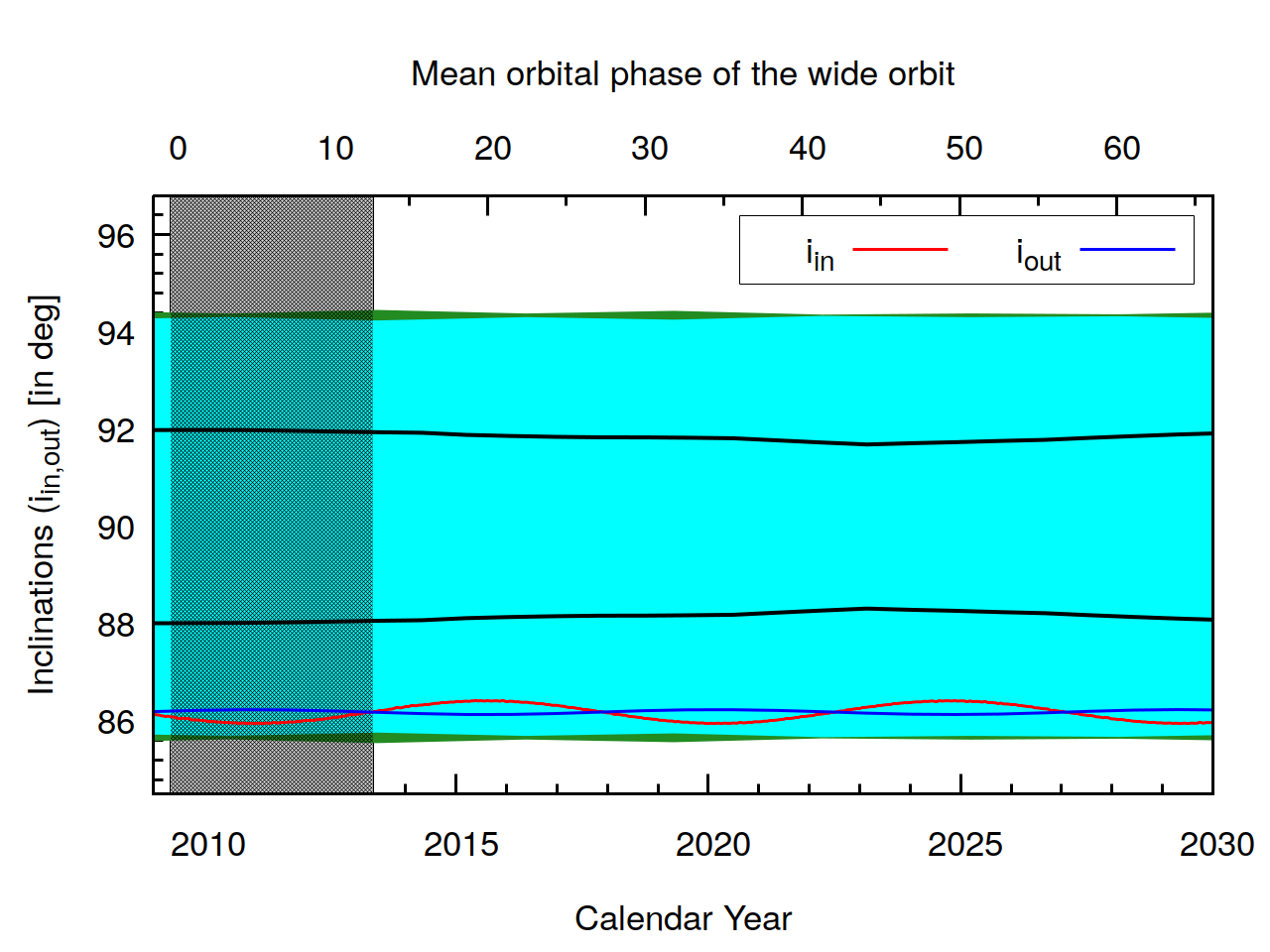}\includegraphics[width=7cm]{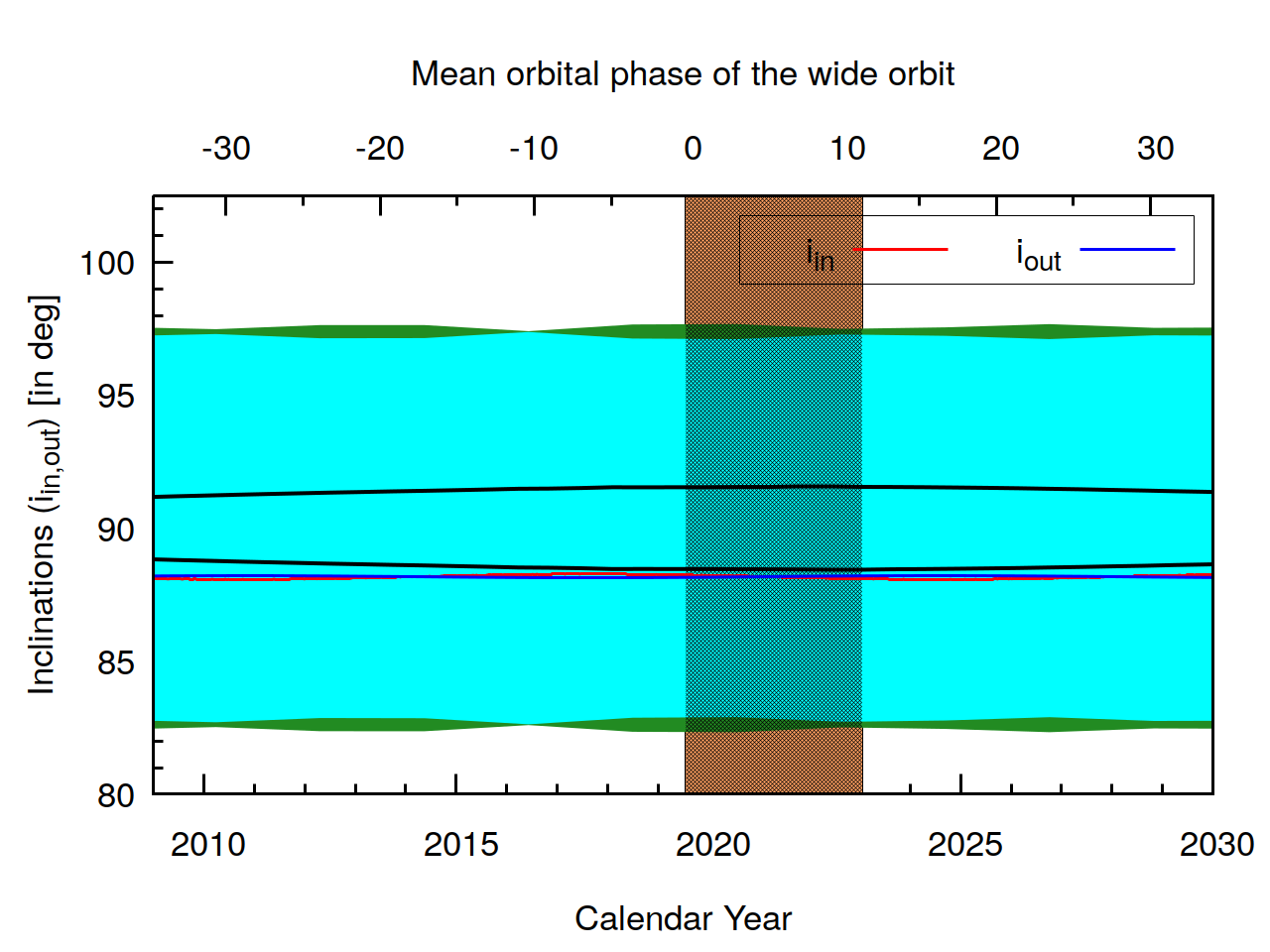}
\caption{Variations of the observable inclinations of the inner and outer orbital planes ($i_\mathrm{in}$ -- red; $i_\mathrm{out}$ -- blue, respectively) of KIC~5771589 (left) and TIC~219885468 (right). The green area represents the domain where the inner binary produces at least one eclipse during a revolution, while the cyan area covers that inclination domain where both eclipses are present. Furthermore, the nearly horizontal. black lines denote the lower and upper limits of that domain of the outer inclination ($i_\mathrm{out}$), where third-body eclipses might occur. As one can see, in the case of KIC~5771589, the inner inclination ($i_\mathrm{in}$) shows small amplitude oscillations near the border of the EB domain, implying that the eclipses are grazing. Moreover, the blue curve ($i_\mathrm{out}$) remains continuously below the third-body eclipse inclination limit, hence, no third-body eclipses can occur. In contrast to this, in the case of TIC~219885468 the small amplitude oscillations of $i_\mathrm{in}$ are located quite deeply in the (cyan/green) EB domain, indicating that the eclipses are quite deep and, hence, the eclipse depths are insensitive to such small inclination variations. Note also that this is an {\it almost} triply eclipsing system, as the outer inclination is just below the triply eclipsing limit.}
\label{fig:K5771589T219885468incl}
\end{figure}

The most interesting feature of KIC~5771589, however, is its continuous eclipse depth variations. What may be surprising, therefore, at first sight is the fact that according to our results the inner and outer orbital planes are almost coplanar. The non-zero mutual inclination, which is the origin of the nodal precession and, hence, the eclipse depth variations is very small, being only $i_\mathrm{mut}=0.29^\circ\pm0.03^\circ$. The question naturally arises as to how is it possible that such a small mutual inclination can cause such a readily observable eclipse depth variation? The answer lies in the extreme grazing nature of the eclipses, as is shown in the left panel of Fig.~\ref{fig:K5771589T219885468incl}. 

\begin{figure}
\includegraphics[width=14cm]{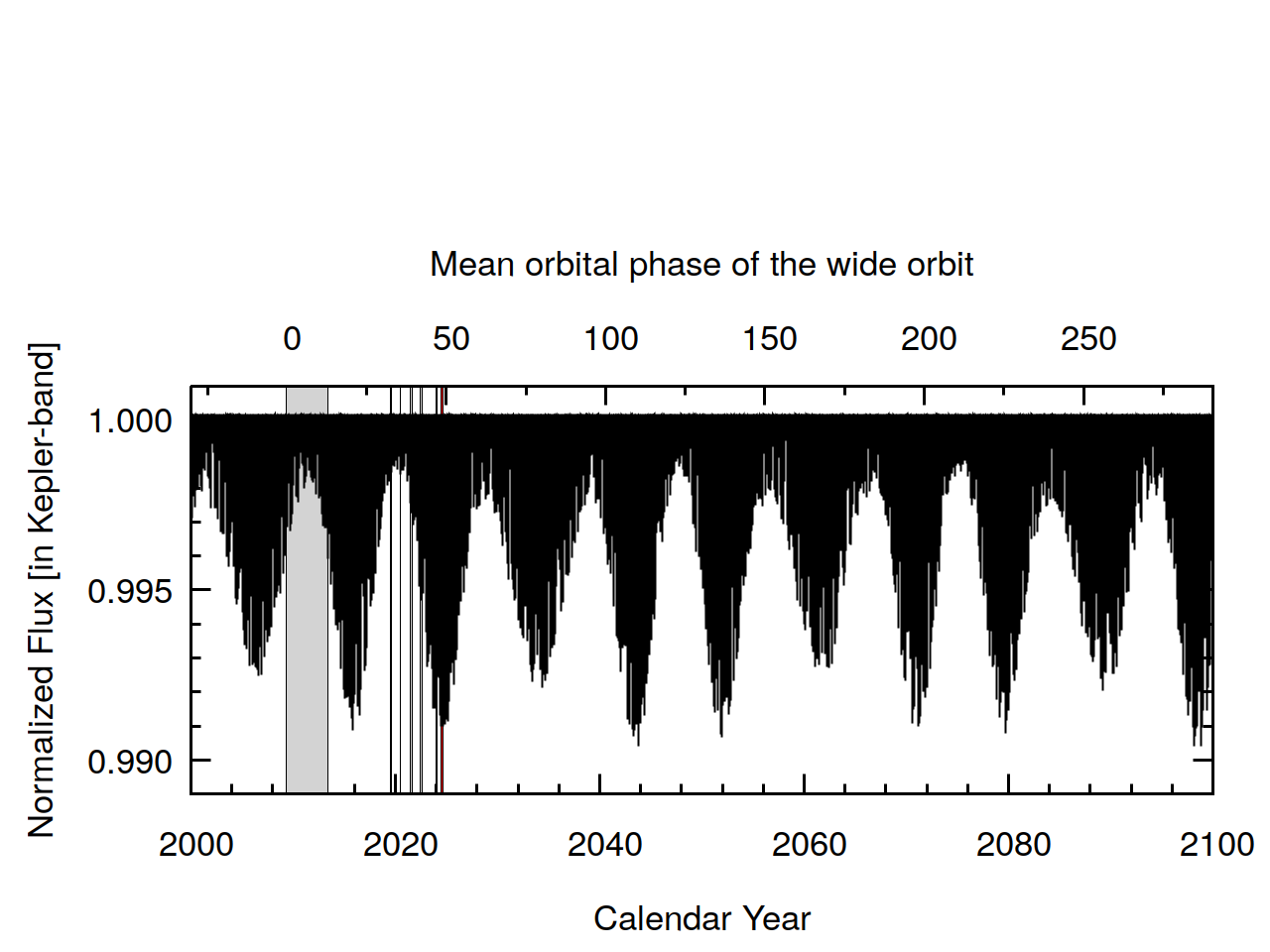}
\caption{Simulated light curve of KIC~5771589 in \textit{Kepler}'s photometric band for all of the 21st century. The gray, shaded area, as before, represents the interval of the \textit{Kepler} observations, while the other vertical lines stand for the past and scheduled future observations of \textit{TESS}.}
\label{fig:K5771589lc100yr}
\end{figure}

Finally, in Fig.~\ref{fig:K5771589lc100yr} we plot the predicted eclipse depth variations for the present century. As one can see, for the expected near-future observations of \textit{TESS} (in Sectors 74, 80, 81) the inner inclination will be larger and, hence, one can predict somewhat deeper and better detectable eclipses. The numerical integrations give a nodal regression period of $P_\mathrm{node}^\mathrm{meas}\approx9.2$\,yr which is in perfect accord with the theoretical value tabulated in Table~\ref{tab: syntheticfit_KIC5771589}.

\subsection{TIC~219885468}

Our third triple is a good counterexample in regard to both (i) the octupole perturbations driving apsidal motion and eccentricity variations, and (ii) the remarkable eclipse depth variations in an almost flat system.

\begin{figure}
\includegraphics[width=7cm]{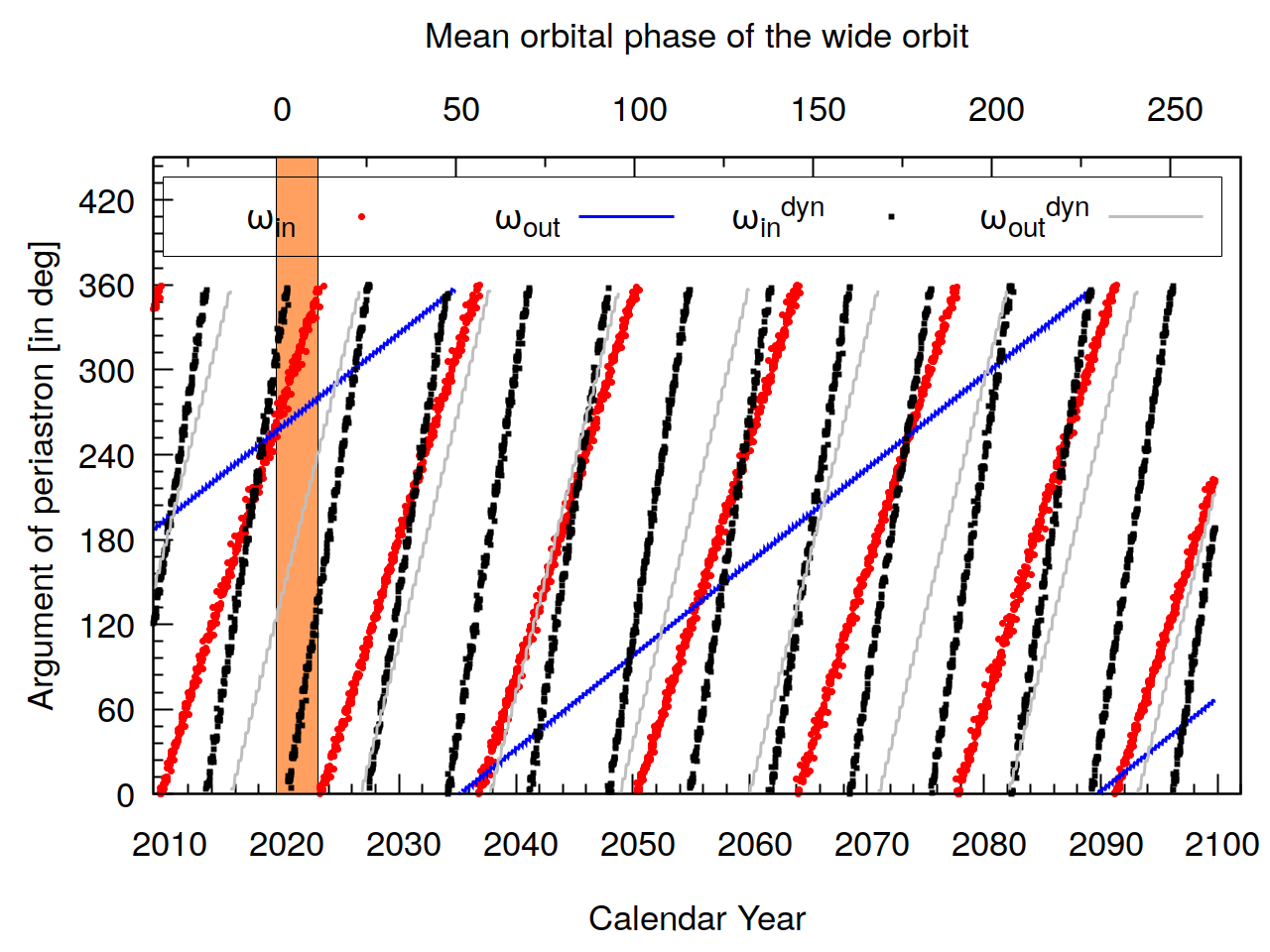}\includegraphics[width=7cm]{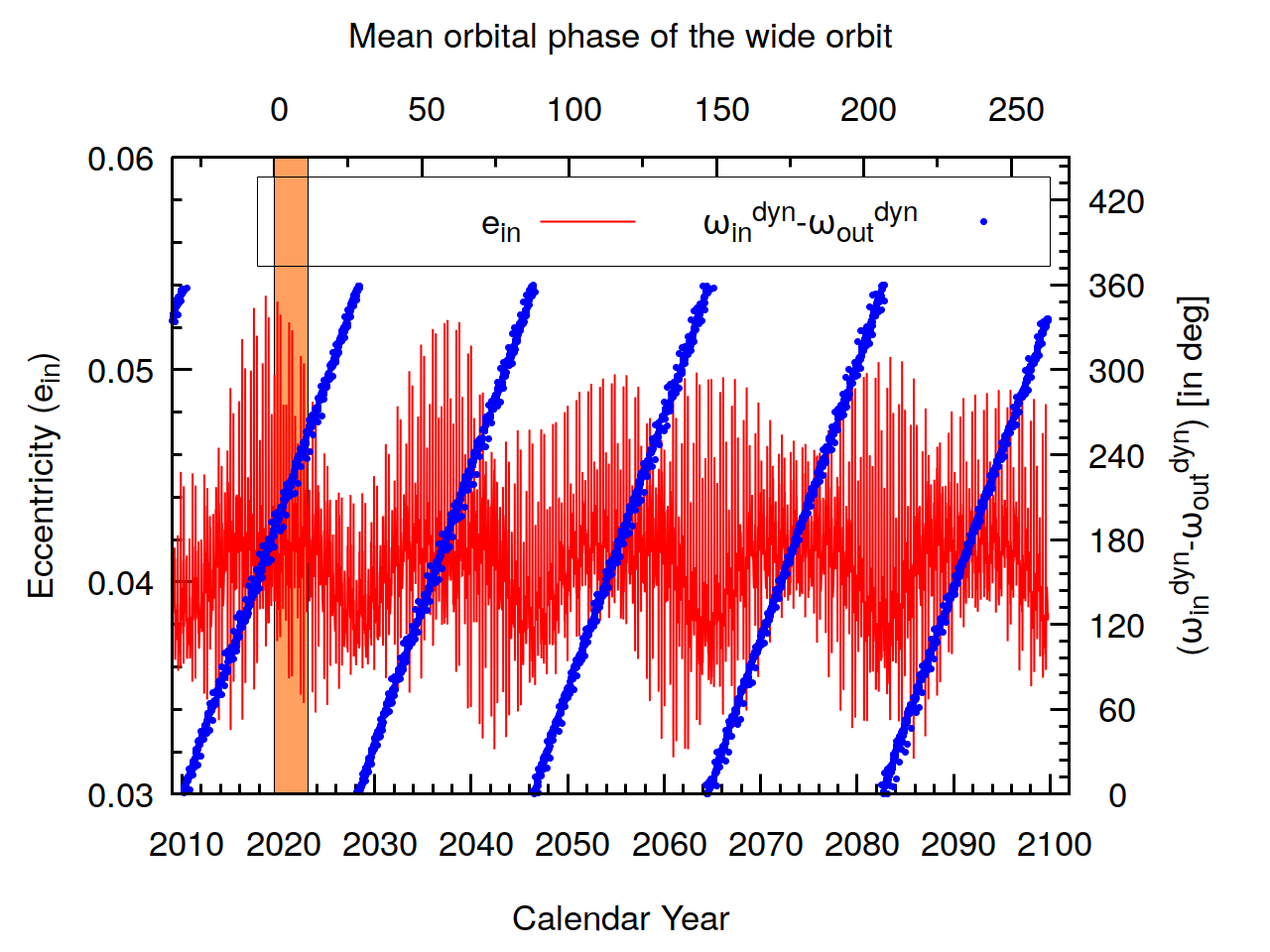}
\caption{The same plots as in Fig.~\ref{fig:K9714358elements}, but for TIC~219885468. The filled brown area represents the interval in between the beginning of the very first (S14) and the end of the momentarily last (S60) NCVZ observations of \textit{TESS}. See text for further details.}
\label{fig:T219885468elements}
\end{figure}

Both the tightness and the compactness of this system is in between our previous two example triples. Moreover, similar to the other two triples, the inner orbit has a small eccentricity ($e_\mathrm{in}=0.0423\pm0.0006$), while the outer orbit is moderately eccentric ($e_\mathrm{out}=0.3903\pm0.0007$). The outer mass ratio, again, makes this triple quite similar to KIC~9714358 ($q_\mathrm{out}=0.319\pm0.002$ vs. $q_\mathrm{out}=0.265\pm0.003$ for the TIC and the KIC systems, respectively), while the mutual inclination is almost identical to that of the EDV system KIC~5771589 ($i_\mathrm{mut}=0.25^\circ\pm0.14^\circ$ vs. $i_\mathrm{mut}=0.29^\circ\pm0.03^\circ$, respectively). The similarities, however, end at this point. From a dynamical point of view, the main difference between the present and the two KIC systems is that the inner EB of TIC~219885468 consists of two twin stars ($q_\mathrm{in}=0.988\pm0.017$) and, hence, the octupole (and any other even order) perturbation terms can be neglected. This is demonstrated nicely in the two panels of Fig.~\ref{fig:T219885468elements}, where it can be readily seen that both the inner and outer ellipses rotate with constant rates, as is expected for coplanar systems in the quadrupole theory. Moreover, the theoretically calculated quadrupole apsidal advance rates ($P_\mathrm{apse,in}^\mathrm{theo}=19.49\pm0.09$\,yr and $P_\mathrm{apse,out}^\mathrm{theo}=51.2\pm0.1$\,yr) agree quite well with the apsidal motion periods `measured' via numerical integration ($P_\mathrm{apse,in}^\mathrm{meas}\approx13.6$\,yr and $P_\mathrm{apse,out}^\mathrm{meas}\approx54.5$\,yr). 

The agreement is even better in the nodal regression period being $P_\mathrm{node}^\mathrm{meas}\approx13.9$\,yr vs. $P_\mathrm{node}^\mathrm{theo}=14.09\pm0.07$\,yr. Turning to this latter effect, despite the similar amplitude of the precession cone, in the case of TIC~219885468 one cannot detect any EDVs. The two panels of Fig.~\ref{fig:K5771589T219885468incl} illustrate nicely the substantial differences amongst the eclipse geometries of the two EBs. While in the case of KIC~5771578 there are grazing eclipses, in TIC~219885468 the eclipses are much deeper. Hence, the areas of the eclipsed surfaces are less sensitive to such small inclination variations. 

Regarding the astrophysical parameters of the three stars, the twin EB members are found to be late F/early G-type, slightly evolved MS stars ($\log\tau=9.49\pm0.06$) with masses of $m_\mathrm{Aa}=1.19\pm0.03\,\mathrm{M}_\odot$ and $m_\mathrm{Ab}=1.18\pm0.04\,\mathrm{M}_\odot$ and, effective temperatures of $T_\mathrm{Aa}=6\,410\pm40$\,K and $T_\mathrm{Ab}=6\,400\pm40$\,K. The less massive third component is a K dwarf with $m_\mathrm{B}=0.75\pm0.02\,\mathrm{M}_\odot$ and $T_\mathrm{B}=4\,885\pm50$\,K. The inferred distance of the triple star agrees with the Gaia EDR3 distance within two sigma ($d=1\,150\pm20$\,pc vs $d_\mathrm{EDR3}=1\,111\pm13$\,pc).

%%%%%%%%%%%%%%%%%%%%%%%%%%%%%%%%%%%%%%%%%%
\section{Conclusions}
\label{sec:conclusions}

We investigated three such EBs which form the inner pairs of TCHT stellar systems. Two systems (KICs~9714358 and 5771589) were observed with the \textit{Kepler} space telescope in its 4-yr-long prime mission, while the third target (TIC~219885468) is located in the NCVZ of \textit{TESS}. Besides the evident medium-period third-body perturbations, the ETVs of all three EBs show rapid dynamically driven apsidal motion, and one of the three targets (KIC~5771589) also exhibits cyclic EDVs. We carried out complex photodynamical analyses for the three systems analyzing jointly their \textit{Kepler} and \textit{TESS} light curves (naturally, for TIC~219885468 only the \textit{TESS} light curve was used), ETV curves, composite SEDs and, in the case of KIC~9714358, APOGEE-2 RV data as well. We determined reliable, robust astrophysical and orbital parameters for all three systems and their constituent stars. 

In our study, however, we focused mainly on the dynamical properties, and higher order third-body perturbations of the systems. We found clear evidence that in the case of the almost perfectly coplanar KIC~9714358 the inner and outer orbital ellipses are oriented in the same directions and while, on average, they rotate evenly with the same apsidal revolution rate (with a period of $P_\mathrm{apse}^\mathrm{meas}\approx78$\,yr), the axis of the inner orbit librates around this average direction with an oscillation period of $P_\mathrm{apse, in}^\mathrm{osc}\approx30$\,yr and with an amplitude of $\sim15^\circ$. The variations of the inner eccentricity of KIC~9714358 show the same period as the apsidal oscillations, but with a phase shift of $\sim0.25$ with respect to the libration. We connect this behaviour to the octupole perturbations; however, we leave the quantitative investigation of this issue for a future study. We expect, however, that the near-future \textit{TESS} observations will provide us with additional observational evidence for this behaviour.

Regarding KIC~5771589, despite the evident cyclic EDVs of this system (detected both in the \textit{Kepler} and \textit{TESS} eras), we found, somewhat surprisingly, that the triple was almost coplanar, with a mutual inclination of $i_\mathrm{mut}=0.29^\circ\pm0.03^\circ$. We explain this unusual finding with the grazing nature of the eclipses, as the depths of grazing eclipses are much more sensitive to even very small variations in the observable inclination. This triple is the second tightest \textit{Kepler}-triple, and also this EB has the second shortest apsidal motion period. Despite this fact, we did not find observational evidence for the operation of the octupole-order perturbations, however, our numerical integration did reveal that they are present. The weakness of the octupole effects in the present system could be in accord with the fact that the relative strength of the octupole order effects relative to the quadrupole ones is much lower than in KIC~9714358.

Finally, TIC~219885468 was selected as a counterexample to the two KIC sources. This triple, in both compactness and tightness, is very similar to the other two systems, but the nearly equal primary and secondary eclipse depths suggested that the two stars of the inner EB should be similar in mass and, hence, one cannot expect octupole order third-body perturbations in this system. While our analysis justified this pre-assumption, finding $q_\mathrm{in}=0.988\pm0.017$ revealed another aspect in which TIC~219885468 serves as counterexample. The mutual inclination of the system was found to be almost identical to that of the former system ($i_\mathrm{mut}=0.25^\circ\pm0.14^\circ$ vs. $0.29^\circ\pm0.03^\circ$). In the current system, however, the eclipses are deep and, hence, they are insensitive to such small variations in inclination.

As our primary aim was to detect observationally the higher order secular perturbations, we did not investigate the past and future evolution of these three triple systems as this question is beyond the focus of the present paper. Currently all three systems are stable in the sense of the semi-empirical dynamical stability limit of \citet{mardlingaarseth01}. This fact, however, does not imply that the systems also remain stable in the distant future. Such investigations may also be the part of a future study.

\dataavailability{The \textit{Kepler} and \textit{TESS} observations were provided by NASA and obtained either from the Kepler Eclipsing Binary Catalog at \url{http://keplerebs.villanova.edu/} or the Mikulski Archive for Space Telescopes (MAST) at \url{https://mast.stsci.edu/portal/Mashup/Clients/Mast/Portal.html}. The RV data used for the analysis of KIC~9714358 were provided by the DR 16 of the APOGEE-2 project, and it was downloaded from the Vizier site at \url{http://vizier.cfa.harvard.edu/viz-bin/VizieR?-source=III/284}.}

\acknowledgments{The authors are very grateful for Prof. S. A. Rappaport for his extensive linguistic corrections. This project has received funding from the HUN-REN Hungarian Research Network. This paper includes data collected by the \textit{Kepler} and \textit{TESS} missions and obtained from the MAST data archive at the Space Telescope Science Institute (STScI). Funding for the Kepler mission is provided by the NASA Science Mission Directorate. Funding for the TESS mission is provided by the NASA Explorer Program. STScI is operated by the Association of Universities for Research in Astronomy, Inc., under NASA contract NAS5--26555. We used the Simbad service operated by the Centre des Donn\'ees Stellaires (Strasbourg, France) and the ESO Science Archive Facility services (data obtained under request number 396301). This research has made use of NASA's Astrophysics Data System.}

\conflictsofinterest{The authors declare no conflict of interest.} 

%% Optional
%\sampleavailability{Samples of the compounds ... are available from the authors.}

%%%%%%%%%%%%%%%%%%%%%%%%%%%%%%%%%%%%%%%%%%
%\end{paracol}
%%%%%%%%%%%%%%%%%%%%%%%%%%%%%%%%%%%%%%%%%%
% To add notes in main text, please use \endnote{} and un-comment the codes below.
%\begin{adjustwidth}{-5.0cm}{0cm}
%\printendnotes[custom]
%\end{adjustwidth}
%%%%%%%%%%%%%%%%%%%%%%%%%%%%%%%%%%%%%%%%%%
\reftitle{References}

\end{document}